\documentclass[prd%
               ,twocolumn%
               ,showpacs%
               ,floatfix%
               ,preprintnumbers%
               ,superscriptaddress%
               ,amsmath%
               ,amssymb%
	       ,nofootinbib
               ]{revtex4}
\usepackage[dvips]{color,graphicx}
\usepackage{dcolumn}
\usepackage{bm}
\definecolor{ref}{rgb}{1.0,0.0,0.0}
  \makeatletter
  \def\mathcomposite{%
     \@ifstar
        {\def\@mathcomposite@option{%
            \baselineskip\z@skip\lineskiplimit-\maxdimen}%
         \@mathcomposite}%
        {\let\@mathcomposite@option\offinterlineskip
         \@mathcomposite}}
  \def\@mathcomposite{%
     \@ifnextchar[\@@mathcomposite{\@@mathcomposite[0]}}
  \def\@@mathcomposite[#1]#2#3#4{%
     #2{\mathchoice
        {\@mathcomposite@{#1}{#3}{#4}\displaystyle{1}}%
        {\@mathcomposite@{#1}{#3}{#4}\textstyle{1}}%
        {\@mathcomposite@{#1}{#3}{#4}%
         \scriptstyle\defaultscriptratio}%
        {\@mathcomposite@{#1}{#3}{#4}%
         \scriptscriptstyle\defaultscriptscriptratio}}}
  \def\@mathcomposite@#1#2#3#4#5{%
     \vcenter{\m@th\@mathcomposite@option
        \dimen@\f@size\p@\dimen@#1\dimen@\dimen@#5\dimen@
        \divide\dimen@ 18
        \edef\@mathcomposite@skipamount{\the\dimen@}%
        \ialign{\hfil$#4##$\hfil\cr
           #2\crcr
           \noalign{\vskip\@mathcomposite@skipamount}%
           #3\crcr}}}
  \makeatother
\renewcommand{\agt}{\mathcomposite{\mathrel}{>}{\sim}}
\renewcommand{\alt}{\mathcomposite{\mathrel}{<}{\sim}}
\newcommand{\Sqrt}[1]{\sqrt{\mathstrut #1}}

\def\bfm#1{\mbox{\boldmath $#1$}}
\def\bfsm#1{\mathstrut\mbox{\scriptsize{\boldmath $#1$}}\mathstrut}

\newcommand{\De}{\Delta}
\newcommand{\ep}{\varepsilon}
\newcommand{\eps}{\epsilon}

\newcommand{\dsp}{\displaystyle}

\newcommand{\Eqn}[1]{Eq.~(\ref{#1})}  
\newcommand{\Fig}[1]{FIG.~{\ref{#1}}}
\newcommand{\e}{{\rm e}}  
\newcommand{\beq}{\begin{equation}}
\newcommand{\eeq}{\end{equation}}
\newcommand{\ba}{\begin{array}}
\newcommand{\ea}{\end{array}}
\newcommand{\bea}{\begin{eqnarray}}
\newcommand{\eea}{\end{eqnarray}}
\newcommand{\bal}{\begin{align}}  
\newcommand{\eal}{\end{align}}
\newcommand{\bi}{\begin{itemize}}  
\newcommand{\ei}{\end{itemize}}
\newcommand{\ben}{\begin{enumerate}}  
\newcommand{\een}{\end{enumerate}}

\newcommand\hide[1]{}

\newcommand{\tr}{\mbox{tr}}

\renewcommand{\Re}{{\rm Re}\,}
\renewcommand{\Im}{{\rm Im}\,}
\newcommand{\ie}{{i.e.}}

\newcommand{\ds}[1]{
  \setbox0=\hbox{\ensuremath{#1}}
  \hbox to\wd0{\hbox to0pt{\hbox to\wd0{\hss/\hss}\hss}\box0}}

\newcommand{\MeV}{\,{\rm MeV}}

\newcommand{\g}{\mathcal G}
\newcommand{\btem}{\bibitem}
\newcommand{\bra}[1]{\langle#1\vert}

\newcommand{\braket}[1]{\left\langle #1\right\rangle}

\begin{document}
\preprint{YITP-07-07, BARI-TH/07-563}
\title{BCS/BEC crossover in Quark Matter\\
and Evolution of its Static and Dynamic properties\\
{\it -- from the atomic unitary gas to color superconductivity --}
}
\author{Hiroaki Abuki}
\email[E-mail:~]{abuki@yukawa.kyoto-u.ac.jp}
\affiliation{Yukawa Institute for Theoretical Physics, 
Kyoto University, Kyoto 606-8502, Japan}
\affiliation{I.N.F.N., Sezione di Bari, I-70126 Bari, Italia}
\date{\today}

\begin{abstract}
We study the evolution of dynamic properties of the BCS/BEC
 (Bose-Einstein Condensate) crossover in a relativistic superfluid as
 well as its thermodynamics.
We put particular focus on the change in the soft mode dynamics
 throughout the crossover, and find that three different effective
 theories describe it; these are, the time-dependent
 Ginzburg-Landau (TDGL) theory in the BCS regime, the Gross-Pitaevskii
 (GP) theory in the BEC regime, and the relativistic Gross-Pitaevskii
 (RGP) equation in the relativistic BEC (RBEC) regime.
Based on these effective theories, we discuss how the physical nature of
 soft mode changes in the crossover.
We also discuss some fluid-dynamic aspects of the crossover using these
 effective theories with particular focus on the shear viscosity. 
In addition to the study of soft modes, we show that the ``quantum
 fluctuation'' is present in the relativistic fermion system,
 which is in contrast to the usual Nozi{\`e}res--Schmit-Rink (NSR)
 theory.
We clarify the physical meaning of the quantum fluctuation, and find
 that it drastically increases the critical temperature in the weak
 coupling BCS regime.
\end{abstract}


\maketitle

\section{Introduction}
Quantum ChromoDynamics (QCD) is expected to exhibit a surprisingly rich
phase structure at finite temperature and/or density. 
The lattice calculations strongly support the conjecture that the
nuclear matter undergoes a phase transition to the color-deconfined 
quark-gluon plasma (QGP) phase
at some critical temperature \cite{Karsch:1994hm,Karsch:2003jg}.
The QGP is a longstanding theoretical issue since the discovery of the
asymptotic freedom of QCD \cite{Gross:1973ju} and is now being searched
experimentally in the Relativistic Heavy Ion Collider (RHIC)
\cite{Arsene:2004fa}.

The QGP was originally thought as the weakly coupled plasma and its
early conjectured signature \cite{Matsui:1986dk} was based on this
observation, \ie, the strong suppression of heavy $q\bar{q}$ bound
states above $T_c$.
However, the lattice calculations of mesonic spectral functions using
the Maximum Entropy Method (MEM) show that the mesonic bound states
persist well above the critical temperature \cite{Umebo}.
Moreover, the RHIC data of the collective flow together with the
theoretical analyses using the parton cascade
simulation \cite{Molnar:2001ux} and the hydrodynamics
\cite{Teaney:2003kp} suggests the strongly correlated plasma where the
partonic cross section is almost $50$ times larger than its
perturbative estimate.
The recent lattice calculation of the shear viscosity to entropy ratio
\cite{Nakamura:2004sy} in the purely gluonic plasma also shows the
value which is a factor 2-3 smaller than its perturbative estimate
\cite{Arnold:2000dr}; the lattice value is rather close to
the minimum bound $(1/4\pi)$ speculated in \cite{Kovtun:2004de} using the
string theory together with the AdS/CFT correspondence
\cite{Policastro:2001yc}.

These theoretical and experimental results show that the characteristics
of the QGP are, (i) the existence of the quasi-bound states above $T_c$
and (ii) an almost perfect fluidity with small sound-attenuation
length \cite{Asakawa:2006tc}.
Some theoretical efforts have been devoted to understanding such a
strongly coupled QGP
\cite{Shuryak:2003ty,Simonov:2005jj,Castorina:2005tm,He:2005bd} in which
there exist two temperatures, one for the deconfining transition or
chiral-restoration ($T_c$) and the other for the vanishing of 
resonances ($T^*>T_c$).
The existence of the mesonic correlation above the chiral-restoration
temperature $T_c$ was suggested earlier \cite{Hatsuda:1985eb} using
the NJL model analysis \cite{Hatsuda:1994pi,Klevansky:1992qe}.

The baryonic matter is also expected to be deconfined to quark matter
when it is compressed, which was conjectured in \cite{Collins:1974ky} or
even earlier \cite{Itoh:1970uw}.
Moreover, a variety of color superconducting phases at high baryon
density is suggested \cite{Bailin:1983bm,Iwasaki:1994ij,reviews}.
It is now widely accepted that the ground state of quark matter is the
Color-Flavor-Locked (CFL) phase \cite{Alford:1998mk} at extremely
high density where the gap and critical temperature can be
studied by the perturbative Dyson-Schwinger type equations
\cite{Son:1998uk,Schafer:1999jg,Brown:1999aq,Schafer:1999fe}.
In contrast, which phase is realized at relatively low density relevant
to compact stars still remains a matter of debate 
because there is no general scheme.
However, recent extensive studies based on effective models have
revealed that there may arise a surprisingly rich variety of non-BCS
exotic phases
\cite{Alford:2003fq,Fukushima:2004zq,Abuki:2004zk,%
Ruster:2005jc,Ciminale:2006sm,Kitazawa:2006zp,Fukushima:2006su,He:2006tn}
once the kinematical constraints such as the neutrality with respect to
the color and electric charges are incorporated.
Due to a relatively large strange quark mass, these {\em kinematic
effects} plays an important role there.
Such a {\em stressed pairing} is attracting broad interest not only from
the QCD community but also from the audience of the atomic polarized
Fermi gas \cite{Carlson:2005kg,Son:2005qx,Gubankova:2006gj,Mannarelli:2006hr}.

Another (and perhaps more direct) source of difficulty in
investigating the pairing at low density is the presence of {\em
dynamical effects} due to its {\em strong coupling} nature.
As the system goes towards lower density, the gauge coupling grows and
some strong coupling effects beyond the mean field approximation (MFA)
come to play an important role as pointed out earlier
by evaluating the Cooper pair correlation length {\em within} the MFA
\cite{Matsuzaki:1999ww,Abuki:2001be}.
Some theoretical efforts to examine the strong coupling effects
beyond the MFA were then made with a particular focus on
the precursory soft mode
\cite{Kitazawa:2001ft,Kitazawa:2003cs,Kitazawa:2005vr},
where non-vanishing diquark correlation and the {\em pseudogap} above
the critical temperature are reported.
Along with this line, the limit temperature $T^*$ where the resonances
decouple from the spectrum is evaluated both for the chiral and diquark
channels in \cite{He:2005bd}.

Such existence of two temperature scales at strong coupling may 
be naturally understood in the scenario of the crossover from the
BCS pairing to the Bose-Einstein condensate (BEC), \ie, the 
BCS/BEC crossover \cite{eagles,leggett,nozieres}.
The BCS/BEC crossover has been the longstanding theoretical idea
first discussed in the context of a theory of superconductivity in low
concentration systems such as SrTiO${}_3$ doped with Zr \cite{eagles}. 
It has also been introduced to understand the isotropic nature of
elementally excitation in ${}^3$He superfluid where the spin fluctuation
results in strong attraction in ${}^3P$ channel leading to the
non-$s$-wave pairing \cite{leggett}.
Their work was extended to finite temperature in \cite{nozieres}
where the fluctuation about the MFA was inevitably included within the
gaussian approximation.
The basic concept of the BCS/BEC crossover is the following;
in the weak coupling, the system exhibits the BCS superconductivity
due to the attraction and the large density of state at the Fermi
surface, while in the strong coupling, the {\em composite bosons} are
formed at $T^*$ prior to their condensation to the bosonic
zero mode at $T_c<T^*$.
Although the symmetry breaking pattern is the same in both sides
and therefore there is no sharp phase boundary in between, 
the mechanism of the condensation is completely different 
in the sense that the former has a dynamical origin while the latter has
a rather kinematical origin; in the BEC side, the short range quantum
effects are taken only into the structure of composite boson.

So far, the BCS/BEC crossover has been widely discussed in various
contexts including not only the liquid $^3$He \cite{leggett} or the
high $T_c$ superconductivity \cite{highTc}, but also the nuclear matter
\cite{Lombardo:2001ek}, and the magnetically trapped alkali atom systems
\cite{ohashi02}.
Although the crossover scenario in color superconductivity was conjectured
in the literature \cite{Matsuzaki:1999ww,Abuki:2001be,Fukushima:2004bj},
the first explicit investigation of this problem was given very
recently in \cite{Nishida:2005ds} where it is shown that the
relativistic fermion system exhibits the two-step crossovers; one is the
ordinary BCS/BEC crossover but with some unconventional behaviour of the
critical temperature brought about by relativistic effects, and the
other is the crossover from the BEC to the relativistic BEC (RBEC)
\cite{kapusta} where the critical temperature increases up to the order
of Fermi energy. 
Recently the existence of the RBEC phase has been confirmed at
$T=0$ in the Leggett's flamework and the evolution of collective
modes are reported \cite{He:2007kd}.
Investigation of such a relativistic crossover is also performed in
a boson-fermion model \cite{Nawa:2005sb,Deng:2006ed} which together with
a chemical equilibrium condition enables us to describe the crossover
thermodynamics with a boson mass or chemical potential controlled by hand.
A possible importance of the diquark BEC in QCD phase diagram at low
density is demonstrated in a model calculation \cite{Rezaeian:2006yj}.
The BCS/BEC crossover is also discussed in \cite{Babaev:1999iz} with possible
relevance to the chiral transition using the nonlinear sigma model. 
Moreover, in recent work \cite{Castorina:2005tm}, it is discussed 
with special emphasis on the chiral {\em pseudogap} phase above $T_c$.

As above, the BCS/BEC crossover is discussed in the QCD context to
give an insight into the strong coupling nature of the quark-gluon
plasma in either low or high density.
However, how the transport and hydrodynamic properties change throughout
the crossover remains unclarified; this is interesting
because it may provide some unified view on the two apparently
independent aspects of QGP, \ie, (i) the survival of the bound states
above $T_c$ and (ii) the perfect liquidity.

It is also worth mentioning that the hydrodynamic properties of the
unitary Fermi gas have recently attracted considerable attention
both experimentally \cite{Thomas,Grimm-collective,Thomas-breakdown,%
Thomas-2transitions,OHara,Salomon-intE}
and theoretically \cite{Gelman:2004fj,Schaefer:2007pr,Son:2005tj}.
The unitary Fermi gas is the intermediate of the BCS/BEC crossover,
where the $s$-wave scattering length diverges.
Such strongly coupled system can be created in the atomic traps
with the external magnetic field fine tuned to the Feshbach resonance
\cite{regal04}. 
It provides us with a theoretical challenge to describe such a unique
many body system which is {\em universal} in the sense that there is no
intrinsic dimensionful parameter except for temperature and density
\cite{Nishida:2006br}.
Understanding such system also may shed light on the physics 
of high $T_c$ superconductivity \cite{highTc}, and possibly, the QGP.

The aim of this paper is to understand how the transport properties
of the relativistic fermion system, as well as its thermodynamics,
change throughout the BCS/BEC crossover.
In order to do this, we first formulate the relativistic
Nozi{\`e}res--Schmitt-Rink (NSR) framework extending our earlier work
\cite{Nishida:2005ds} to fermions having SU$(2)_{\rm F}$ flavor and
SU$(3)_{\rm c}$ color with a little refined regularization scheme.
We then derive the effective theory for soft modes and study how this
effective theory changes as the system moves from BCS to (R)BEC regime.
Based on this effective theory, we will discuss some of fluid dynamic
aspects of the crossover with particular focus on the shear viscosity
and its ratio to the entropy which serves as a measure of
``perfect liquidity''.

The rest of this paper is organized as follows:
In Sec.~\ref{intro}, we formulate the Nozi\`eres--Schmit-Rink theory for
a relativistic fermion system.
It turns out that, in contrast to the nonrelativistic system,
the relativistic fermion system has an additional source of
fluctuation, \ie, the ``quantum fluctuation'' which we have just ignored
in our previous study \cite{Nishida:2005ds}.
Our framework is not restricted to quark matter, and may be used for the
other relativistic fermion systems such as a possible neutrino
superfluid trapped in compact stars at the early stage of their thermal
evolution \cite{Kapusta:2004gi}.
In Sec.~\ref{thermodynamics}, we discuss the static part of problem,
\ie, thermodynamics, the pair size, the spectrum, etc.
In Sec.~\ref{softmodes}, we derive the effective theory for soft modes
and study how it evolves with the crossover.
In addition, we discuss the shear viscosity to entropy ratio based on
the effective theory.
In Sec.~\ref{densquan}, we investigate the density dependence of the
crossover.
We also study how the ``quantum fluctuation'' affects the crossover.
In Sec.~\ref{concluding}, we make concluding remarks and perspectives.

\section{Formulation}\label{intro}
In this section, we extend the Nozi\`eres--Schmit-Rink theory
so as to apply it to study relativistic fermions interacting via
a point attraction.
In Sec.~\ref{nozie}, we derive the thermodynamic potential up to the
gaussian fluctuation about the MFA.
In Sec.~\ref{specs}, we introduce an intuitive representation of
the thermodynamic potential in terms of the spectral density.
In Sec.~\ref{phsh}, we introduce the in-medium phase shift which
enables us to handle easily the fluctuation effect to the number
conservation.
In Sec.~\ref{1d}, we discuss the Thouless criterion and present how
to renormalize it using the low energy scattering parameter, the
scattering length. 
In Sec.~\ref{1e}, we give the renormalization scheme for the 
dynamic pair susceptibility.
In Sec.~\ref{1f}, we end up with a couple of basic equations, \ie, the
number conservation, and the Thouless criterion.
\subsection{Nozi\`eres--Schmit-Rink theory for a relativistic fermion
  system}\label{nozie}
We start with the following four-Fermi model inspired by the one-gluon
exchange (OGE) \cite{Nishida:2005ds},
\beq
 {\mathcal L}=\bar{q}(i\ds{\partial}+\ds{\mu}-m)q%
		     -\frac{g^2}{2}\sum_{a=1}^8%
		     \bar{q}T^a\gamma_\mu q%
		     \,\bar{q}T^a\gamma^\mu q
		     \label{eq:lag}.
\eeq
Here quark has $N_c=3$ colors and $N_f=2$ flavors, and $T^a=\lambda^a/2$
is the Gell-Mann matrix for color.
At high density, we expect the pair formation in the $J^P=0^+$,
and the color-flavor triplet channel \cite{Bailin:1983bm}, \ie,
\beq
 (P_\eta)_{ab}^{ij}=iC\gamma_5\eps^{ij}\eps_{abc}.
\eeq
The indices $\{i,j,\cdots\}$ represent flavors and $\{a,b,\cdots\}$
indicate colors. 
Fierz transformation allows us to re-write the Lagrangian
\Eqn{eq:lag} as follows.
\beq
\ba{rcl}
 {\mathcal L}&=&\bar{q}(i\ds{\partial}+\ds{\mu}-m)q%
                     +\frac{G}{4}\sum_{\eta=1}^3[iq^tP_\eta
		     q][i\bar{q}\bar{P}_\eta \bar{q}^t]\\[2ex]
             & &+\frac{G}{4}\sum_{\eta=1}^3[iq^tP_\eta\gamma_5
		     q][i\bar{q}\bar{P}_\eta\gamma_5\bar{q}^t]\cdots\\[2ex]
\ea
\label{eq:lag2},
\eeq
with $G=\frac{N_c+1}{4N_c}g^2$. 
Though we find the attraction in the scaler $q\bar{q}$ channel, we shall
ignore this channel throughout this paper; consequently, our framework
is not reliable in the vicinity of the chiral-restoration density.
By introducing a spinor doublet $Q=(q,\bar{q})^t$, we write
\beq
\ba{rcl}
{\mathcal L}&=&\dsp\frac{1}{2}\bar{Q}\left(%
            \ba{cc}
            i\ds{\partial}+\ds{\mu}-m & 0 \\
            0 & i\ds{\partial}^t-\ds{\mu}^t+m \\
	    \ea
            \right)Q\\[2ex]
           & &\dsp+\frac{G}{4}\bar{Q}\left(%
              \ba{cc}
              0 & 0\\
              P_\eta& 0 \\
              \ea
              \right)Q%
              \bar{Q}\left(%
              \ba{cc}
              0 & \bar{P}_\eta\\
              0 & 0 \\
              \ea
              \right)Q.
\ea
\label{lagNG}
\eeq
Here, we have defined $\bar{P}_\eta=\gamma_0(P_\eta)^\dagger\gamma_0$.
Then the partition function is
\beq
  Z=\int{\mathcal D}Q\exp\biggl[%
  -\int_0^\beta d\tau\int d\bfm{x}{\mathcal L}_E(Q)\biggr],
\eeq
where ${\mathcal L}_E=-{\mathcal L}|_{t\to-i\tau}$ is the 
Lagrangian density in the Euclid space, which is periodic in
the imaginary time interval $[0,\beta]$.
Introducing the Hubberd-Stratnobich fields
\beq
\ba{rcl}
 \De_\eta(\tau,x)&=&\frac{G}{2}\biggl\langle\bar{Q}\left(%
              \ba{cc}
              0 & 0\\
              P_\eta& 0 \\
              \ea
              \right)Q\biggr\rangle,\\[1ex]
 \De^*_\eta(\tau,x)&=&\frac{G}{2}\biggl\langle\bar{Q}\left(%
              \ba{cc}
              0 & \bar{P}_\eta\\
              0 & 0 \\
              \ea
              \right)Q\biggr\rangle,\\[1ex]
\ea
\eeq
and integrating out quark fields, we obtain
\beq
\ba{rcl}
  Z&=&\dsp\int{\mathcal D}\De{\mathcal D}\De^*%
    \exp\left[-\int d\tau d\bfm{x} \left(%
    \frac{|\De(-i\tau,\bfm{x})|^2}{G}\right)\right]\\[2ex]
   & &\dsp\times\exp\left[\frac{1}{2}{\rm log}\,%
      \underset{(x,y)}{\rm Det}%
      \left(\ba{cc}
             S_{Fx,y}^{-1} & \bar{P}_\eta\De(x)\delta_{x,y}\\
             P_\eta\De^*(x)\delta_{x,y} & \bar{S}_{Fx,y}^{-1}
             \ea\right)\right].\\[2ex]
\ea
\eeq
Here the bare propagator is defined by
\beq
\ba{rcl}
  S_{Fx,y}^{-1}&=&(i\ds{\partial}+\ds{\mu}-m)\delta_{x,y},\\[2ex]
  \bar{S}_{Fx,y}^{-1}&=&\gamma_5CS_{Fy,x}^{-1}C\gamma_5.
\ea
\label{ferp}
\eeq
$\delta_{x,y}$ 
is the delta function antiperiodic in the imaginary time $\tau$;
in momentum space,
\beq
  \delta_{x,y}=T\sum_n\int\frac{d\bfm{p}}{(2\pi)^3}%
                  e^{-i\omega_n(\tau_x-\tau_y)+i\bfm{\scriptstyle p}%
                  \cdot(\bfsm{x}-\bfsm{y})},
\eeq
with $\omega_n$ being the fermionic Matsubara-frequency.
Introducing the following notations for the Nambu-Gor'kov propagator and
self energy,
\beq
\ba{rcl}
  \bfm{S}_{Fx,y}^{-1}&=&\left(\ba{cc}%
                  S_{Fx,y}^{-1} & 0\\
                  0   & \bar{S}_{Fx,y}^{-1}
                  \ea\right),\\[3ex]
  -\bfm{\Sigma}_{x,y}&=&\left(\ba{cc}%
                  0   &   \bar{P}_\eta\De_\eta(x)\delta_{x,y}\\
                  P_\eta\De^*_\eta(x)\delta_{x,y}   &   0
                  \ea\right),\\[2ex]
\ea
\eeq
we can write the partition function in the following
form where the fluctuation contribution is factorized:
\beq
  Z(\mu,T)\equiv e^{-\beta\Omega(\mu,T)}=Z_0(\mu,T) Z_{\rm fluc}(\mu,T),
\eeq
where $Z_{\rm fluc}$ is the part coming from the fluctuation:
\beq
\ba{rcl}
   Z_{\rm fluc}&=&\dsp\int{\mathcal D}\De{\mathcal D}\De^*%
    \exp\left[-\int d\tau d\bfm{x} \left(%
    \frac{|\De_\eta(-i\tau,\bfm{x})|^2}{G}\right)\right]\\[2ex]
   & &\dsp\times\exp\left[\frac{1}{2}{\rm log}\,%
      \underset{(x,y)}{\rm Det}%
      \left(\delta_{x,y}-%
      \sum_z\bfm{S}_{Fx,z}\bfm{\Sigma}_{z,y}\right)\right].\\[2ex]
\ea
\label{Z}
\eeq
$\sum_z$
is a shorthand notation of $\int_0^\beta d\tau_z\int d\bfm{z}$.
The free partition function $Z_0$ is $e^{-\beta\Omega_0}$ 
with $\Omega_0$ being the thermodynamic potential for free 
(massive) quarks:
\beq
\ba{rcl}
 \dsp\frac{\Omega_0}{V}%
 &=&\dsp-2N_fN_c\,T\sum_{\alpha=\pm}\int\!\!\!%
     \frac{d\bfm{p}}{(2\pi)^3}%
     \log\left(1+e^{-\eps_{p\alpha}/T}\right),\\[1ex]
\ea
\label{eq:renthouless}
\eeq
where $\eps_{p\pm}=\Sqrt{m^2+p^2}\mp\mu$, and the vacuum fluctuation
which has no dependence on $(\mu,T)$ is suppressed.

Up to quadratic order in $(\Delta,\Delta^*)$ (the gaussian
approximation), we have
\begin{widetext}
\beq
 Z_{\rm fluc}\equiv e^{-\beta\Omega_{\rm fluc}}%
 =\prod_{\eta,N,\bfsm{P}}\int d\De_{\eta}(i\Omega_N,\bfm{P})%
  d\De^*_{\eta}(i\Omega_N,\bfm{P})%
  \exp\left[-\frac{T}{V}%
  \left(\frac{1}{G}-\chi_{\mu,T}(i\Omega_N,\bfm{P})\right)%
  |\De_\eta(i\Omega_N,\bfm{P})|^2\right],
\eeq
where $\Omega_N$ and ${\bfm{P}}$ denote the {\em bosonic}
 Matsubara-frequency and momentum. 
The {\em pair correlation function} $\chi_{\mu,T}(i\Omega_N,\bfm{P})$ 
 at one-loop level is defined by \cite{Kitazawa:2001ft,Kitazawa:2005vr}
\beq
\ba{rcl}
 \dsp\chi_{\mu,T}(i\Omega_N,\bfm{P})%
 &=&\dsp 2T\sum_n\int\!\!\!\frac{d\bfm{q}}{(2\pi)^3}%
    \tr\Big[S_F(i\omega_n+i\Omega_N,\bfm{q}+\bfm{P})%
    S_F(-i\omega_n,-\bfm{q})\Big]\\[2ex]
 &=&\dsp-2\int\!\!\!\frac{d\bfm{q}}{(2\pi)^3}%
    \bigg(1+\frac{m^2+\bfm{q}\cdot(\bfm{q}+\bfm{P})}%
    {E_{\bfsm{q}}E_{\bfsm{q}+\bfsm{P}}}\bigg)%
    \frac{1-f_F(E_{\bfsm{q}+\bfsm{P}}-\mu)%
           -f_F(E_{\bfsm{q}}-\mu)}%
     {i\Omega_N+2\mu-E_{\bfsm{q}+\bfsm{P}}-E_{\bfsm{q}}}\\[2ex]
 & &\dsp-4\int\!\!\!\frac{d\bfm{q}}{(2\pi)^3}%
    \bigg(1-\frac{m^2+\bfm{q}\cdot(\bfm{q}+\bfm{P})}%
    {E_{\bfsm{q}}E_{\bfsm{q}+\bfsm{P}}}\bigg)%
    \frac{-f_F(E_{\bfsm{q}+\bfsm{P}}-\mu)%
           +f_F(E_{\bfsm{q}}+\mu)}%
     {i\Omega_N+2\mu-E_{\bfsm{q}+\bfsm{P}}+E_{\bfsm{q}}}\\[2ex]
 & &\dsp+2\int\!\!\!\frac{d\bfm{q}}{(2\pi)^3}%
    \bigg(1+\frac{m^2+\bfm{q}\cdot(\bfm{q}+\bfm{P})}%
    {E_{\bfsm{q}}E_{\bfsm{q}+\bfsm{P}}}\bigg)%
    \frac{1-f_F(E_{\bfsm{q}+\bfsm{P}}+\mu)%
           -f_F(E_{\bfsm{q}}+\mu)}%
     {i\Omega_N+2\mu+E_{\bfsm{q}+\bfsm{P}}+E_{\bfsm{q}}},\\[2ex]
\ea
\label{eq:dynamic}
\eeq
\end{widetext}
where $S_F(\omega_n,\bfm{p})$ is the Fourier transform of
the fermion propagator $S_{Fx,y}$, \ie, the inverse of \Eqn{ferp}.
Integrating out the gaussian fluctuation leads to 
the following expression for thermodynamic potential.
\beq
  \Omega(\mu,T)=\Omega_0(\mu,T)+\Omega_{\rm fluc}(\mu,T),\\[2ex]
\eeq
with the gaussian fluctuation being defined by
\beq
\ba{rcl}
 \Omega_{\rm fluc}&=&\dsp d_BT\sum_{N:\mbox{\scriptsize even}}%
   \int\!\!\!\frac{d\bfm{P}}{(2\pi)^3}\log%
   \left[\frac{1}{G}-\chi_{\mu,T}(i\Omega_N,\bfm{P})\right].\\[3ex]
  &&\dsp-d_B\int_{-\infty}^{\infty}d\Omega%
     \int\!\!\!\frac{d\bfm{P}}{(2\pi)^3}\log%
     \left[\frac{1}{G}-\chi_{0}(i\Omega,\bfm{P})\right],\\[2ex]
\ea
\label{effpot1}
\eeq
where $d_B=\frac{N_c(N_c-1)}{2}$ is the number of ``flavors''
of boson and $\chi_0\equiv\chi_{\mu=0,T=0}$; we have subtracted the
``vacuum'' contribution at $\mu=T=0$, which in general leads 
a $G$-dependent fermion mass and wavefunction renormalization through a
Hartree and higher order terms \cite{tokumitsu}.
Some notes are in order here.
(i) The fluctuation is symmetric under $\mu\leftrightarrow-\mu$ as it
should be. This is guaranteed by the charge conjugation property
$\chi_{-\mu,T}(-i\Omega_N,\bfm{P})|=\chi_{\mu,T}(i\Omega_N,\bfm{P})$.
(ii) The above formula is analytic above some $T_c$ below which the
system undergoes the transition to pairing phase. 

The dynamic pair susceptibility $\Gamma(\omega,\bfm{P})$ can
be obtained by the analytic continuation of pair correlation,
\Eqn{eq:dynamic}, to the real $\omega$-axis:
\beq
 \Gamma_{\mu,T}^{-1}(\omega,\bfm{P})=%
 \frac{1}{G}-\chi_{\mu,T}(\omega+i\delta,\bfm{P}).
\label{dyp}
\eeq

\subsection{Thermodynamic potential in terms of spectral
  density}\label{specs}
We introduce a spectral density at some coupling strength
$G=\mathcal{G}$ by
\beq
 -\frac{1}{1/\g-\chi_{\mu,T}(i\Omega_N,\bfm{P})}%
 =\int_{-\infty}^{\infty}\frac{d\omega}{\pi}%
 \frac{\rho^{\g}_{\mu,T}(\omega,\bfm{P})}{i\Omega_N-\omega}.
\eeq
It is clear that the spectral density is related to the imaginary part
of the dynamic pair susceptibility at $\g$.
\beq
 \rho^{\g}_{\mu,T}(\omega,\bfm{P})=%
  {\Im}\Gamma_{\mu,T}^{\g}(\omega,\bfm{P})%
\eeq
To express the thermodynamic potential in terms of
the spectral density, we differentiate and integrate 
\Eqn{effpot1} with respect to $G$. We obtain
\beq
  \ba{rcl}
  \Omega_{\rm fluc}&=&
  \dsp d_B\int_0^{G}\frac{d\g}{\g^2}%
   \int_{-\infty}^{\infty}\frac{d\omega}{\pi}\frac{d\bfm{P}}{(2\pi)^3}%
   {T}\sum_{N}\frac{\omega\rho^\g_{\mu,T}(\omega,\bfm{P})}%
   {\Omega_N^2+\omega^2}\\[3ex]
   & &\dsp-(\mbox{$T=\mu=0$ part})
\ea
\eeq
Because of the identity
$\rho_{-\mu,T}^\g(\omega,\bfm{P})=-\rho_{\mu,T}^\g(-\omega,\bfm{P})$,
the thermodynamic potential still possesses the charge conjugation
symmetry.
By performing the Matsubara-summation, we obtain
\beq
   \ba{rcl}
   \Omega_{\rm fluc}&=&%
   \dsp-d_B\int_0^{G}\!\frac{d\g}{\g^2}%
      \int_{-\infty}^{\infty}\!\frac{d\omega}{\pi}\frac{d\bfm{P}}{(2\pi)^3}%
      \tilde{f}_B(\omega)\rho^\g_{\mu,T}(\omega,\bfm{P})\\[2ex]
   &&\dsp-d_B\int_0^{G}\!\frac{d\g}{\g^2}%
      \int_{-\infty}^{\infty}\!\frac{d\omega}{\pi}\frac{d\bfm{P}}{(2\pi)^3}%
      \frac{\eps(\omega)}{2}{\it\Delta}\rho^\g_{\mu,T}%
      (\omega,\bfm{P}).
\ea
\eeq
where $\tilde{f}_B(\omega)=f_B(\omega)+\theta(-\omega)$ 
with $f_B(\omega)=1/(e^{\beta\omega}-1)$ being the bose distribution
function.
We have defined the in-medium spectral shift 
${\it\Delta}\rho^{\g}_{\mu,T}(\omega,\bfm{P})=\rho^{\g}_{\mu,T}%
(\omega,\bfm{P})-\rho_0^{\g}(\omega,\bfm{P})$
with $\rho_0^{\g}(\omega,\bfm{P})=\rho_{0,0}^\g(\omega,\bfm{P})$ being
the spectral density at $\mu=T=0$.
We denote the first term by $\Omega_{\rm NSR}$, and the second term
by $\Omega_{\rm qfl}$ hereafter. 
The total fluctuation is then the sum of these two pieces.
\beq
  \Omega_{\rm fluc}=\Omega_{\rm NSR}+\Omega_{\rm qfl},
\eeq
where
\beq
 \Omega_{\rm NSR}=\dsp-d_B\int_0^{G}\!\frac{d\g}{\g^2}%
      \int_{-\infty}^{\infty}\!\frac{d\omega}{\pi}\frac{d\bfm{P}}{(2\pi)^3}%
      \tilde{f}_B(\omega)\rho^\g_{\mu,T}(\omega,\bfm{P}),
\eeq
and
\beq
 \Omega_{\rm qfl}=-d_B\int_0^{G}\!\frac{d\g}{\g^2}%
      \int_{-\infty}^{\infty}\!\frac{d\omega}{\pi}\frac{d\bfm{P}}{(2\pi)^3}%
      \frac{\eps(\omega)}{2}{\it\Delta}\rho^\g_{\mu,T}%
      (\omega,\bfm{P}).
\eeq
$\Omega_{\rm NSR}$
is the Nozi\`eres--Schmit-Rink correction to the
thermodynamic potential and roughly corresponds to the thermal
fluctuation, while $\Omega_{\rm qfl}$ can be regarded as the quantum
fluctuation which is ignored in the nonrelativistic
Nozi\`eres--Schmit-Rink theory.
If we take the limit $T\to0$, the former goes to zero, but the 
latter remains:
\beq
   \ba{l}
   -d_B\int_0^{G}\!\frac{d\g}{\g^2}%
      \int_{-\infty}^{\infty}\!\frac{d\omega}{\pi}\frac{d\bfsm{P}}{(2\pi)^3}%
      \frac{\eps(\omega)}{2}\left[\rho^\g_{\mu,0}%
      (\omega,\bfm{P})-\rho^\g_0(\omega,\bfm{P})\right].\\[2ex]
\ea
\eeq
$\Omega_{\rm qfl}$
may play an important role at low temperatures, 
but we shall ignore this contribution for a while
because 
(i) the fluctuation itself is less significant in the weak coupling 
(low $T_c$) regime, and 
(ii) the charge conjugation symmetry is maintained even if $\Omega_{\rm
qfl}$ is ignored.
We will come back to this problem in Sec.~\ref{densquan} where
we examine how large this ``quantum correction'' is.

\subsection{Thermodynamic potential in terms of phase shift}%
\label{phsh}
It is sometimes more convenient to express the thermodynamic potential
in terms of the {\em in-medium phase shift} rather than the spectral
density. It may be defined by the integration of the spectral density:
\beq
\ba{rcl}
  \dsp\int_{0}^{G}\!\frac{d\g}{\g^2}\rho^\g_{\mu,T}(\omega,\bfm{P})
  &=&\dsp\frac{i}{2}\log\left(\frac{\frac{1}{G}-%
    \chi_{\mu,T}(\omega+i\delta,\bfm{P})}%
    {\frac{1}{G}-\chi_{\mu,T}(\omega-i\delta,\bfm{P})}\right)\\[4ex]
  &\equiv&\delta_{\mu,T}(\omega,\bfm{P}),
\ea
\eeq
\ie, the argument of the dynamic pair susceptibility
\beq
\frac{\frac{1}{G}-\chi_{\mu,T}(\omega\pm i\delta,\bfm{P})}%
{\big|\frac{1}{G}-\chi_{\mu,T}(\omega,\bfm{P})\big|}%
=e^{\mp i\delta_{\mu,T}(\omega,\bfsm{P})}.
\eeq
The phase shift possesses the charge conjugation 
$\delta_{-\mu,T}(-\omega,\bfm{P})=-\delta_{\mu,T}(\omega,\bfm{P})$.
Using this phase shift, we can express $\Omega_{\rm NSR}$ and
$\Omega_{\rm qfl}$ as
\beq
  \ba{rcl}
  \Omega_{\rm NSR}%
  &=&\dsp\!\!-d_B\int_{-\infty}^{\infty}\!\frac{d\omega}{\pi}%
     \frac{d\bfm{P}}{(2\pi)^3}%
     \tilde{f}_B(\omega)\delta_{\mu,T}(\omega,\bfm{P}),\\[3ex]
  \Omega_{\rm qfl}&=&\dsp\!\!-d_B\int_{-\infty}^{\infty}\!\frac{d\omega}%
     {\pi}\frac{d\bfm{P}}{(2\pi)^3}%
     \frac{\eps(\omega)}{2}{\it\Delta\delta_{\mu,T}}(\omega,\bfm{P}),\\[1ex]
\ea
\eeq
where the in-medium shift of the phase shift is defined by
${\it\Delta}\delta_{\mu,T}(\omega,\bfm{P})%
\equiv\delta_{\mu,T}(\omega,\bfm{P})-\delta_{0}(\omega,\bfm{P})$
with $\delta_0(\omega,\bfm{P})\equiv\delta_{0,0}(\omega,\bfm{P})$ 
being the phase shift in vacuum.
$\Omega_{\rm NSR}$
is exactly of the same form as that derived by the
Nozi{\`e}res--Schmitt-Rink \cite{nozieres}.

\subsection{Thouless criterion and the renormalization of the coupling}
\label{1d}
In the weak coupling regime, the Thouless condition
determines the critical temperature.  
This condition ensures the divergence of the long wavelength limit of
the dynamic pair susceptibility at the critical temperature for
a given chemical potential $\mu$: 
\beq
   \Gamma^{-1}_{\mu,T_c}(0,\bfm{0})%
   =\frac{1}{G}-\chi_{\mu,T_c}(0,\bfm{0})=0.
\label{eq:th1}
\eeq
$\chi_{\mu,T}(0,\bfm{0})$
can be calculated as
\beq
   \chi_{\mu,T_c}(0,\bfm{0})=2\int\!\!\frac{d\bfm{q}}{(2\pi)^3}%
   \frac{\tanh\frac{E_q-\mu}{2T_c}}{E_q-\mu} + (\mu\rightarrow-\mu).
\eeq
which is quadratically divergent. 
Accordingly, the critical temperature should depend on the cutoff
as 
\beq
T_c=\Lambda f(m/\Lambda,\mu/\Lambda,G\Lambda^2),\label{eq:tcbare}
\eeq
where $f$ is some function.
We can reduce the cutoff dependence via the partial renormalization
of the coupling using the low energy information about the
scattering $T$-matrix.
We here consider the fermion-fermion 
scattering $(f(1)+f(2)\to f(3)+f(4))$ with $(1,2,3,4)$ labeling the
momentum, color, flavor and helicity of each quark, as
$(\mbox{``1''}=(\bfm{p},a,i,h_1),\,\mbox{``2''}=(-\bfm{p},b,j,h_2)),%
\,(\mbox{``3''}=(\bfm{k},c,k,h_3),\,\mbox{``4''}=(-\bfm{k},d,l,h_4))$.
Then the on-shell $T$-matrix can be parameterized as
\beq
   T(12\rightarrow34)%
   =T(\bfm{p},\bfm{k})\big(\Gamma_{12}\Gamma_{34} 
   - (3\leftrightarrow 4)\big),
\eeq
where $\Gamma_{12}$ is defined by
\beq
 \Gamma_{12}=\frac{\ep_{ab}}{\Sqrt{2}}\frac{\ep_{ij}}{\Sqrt{2}}%
 \frac{\sigma^3_{h_1h_2}}{\Sqrt{2}}.
\eeq
$\sigma^3$ 
is the Pauli matrix. $T(\bfm{p},\bfm{k})$ can be evaluated with the
Lippman-Schwinger equation as
\beq
  T(\bfm{p},\bfm{k})=\frac{-G}{1-G\chi_{0}(2E_p+i\delta,\bfm{0})}%
  \equiv -\Gamma_0(2E_p,\bfm{0}).
\eeq
Here, $\Gamma_0$ is the dynamic pair susceptibility, \Eqn{dyp},
at $\mu=T=0$, \ie, $\Gamma_0(\omega,\bfm{P})=\Gamma_{0,0}(\omega,\bfm{P})$.
Using the definition of the phase shift function introduced in the
previous section, we find the following expression for the scattering
amplitude $f(\bfm{p},\bfm{k})=-\frac{\bar{m}}{2\pi}T(\bfm{p},\bfm{k})$
with ($\bar{m}=m/2$) being the reduced mass
\beq
  f(\bfm{p},\bfm{k})=\frac{m}{4\pi}%
  ||\Gamma_0(2E_p,\bfm{0})||e^{i\delta_{0}(2E_p,\bfsm{0})}.
\eeq
At sufficiently low energy $p=k\ll m$, one can make use of
$2E_p\cong 2m+\frac{p^2}{m}$ in the above formula, and the result should
be matched with the general form of the low-energy expansion of scattering
amplitude, \ie,
\beq
\ba{rcl}
  &&f(\bfm{p},\bfm{k})=\frac{e^{i\delta}\sin\delta}{p}=\frac{1}{p
  \cot\delta -ip}\sim\frac{1}{-\frac{1}{a_s} + \frac{1}{2}r_e p^2 - ip},
\ea
\eeq
where $a_s$ is the $s$-wave scattering length and $r_e$ is the effective
range.
We find the scattering length is quadratically divergent as
\beq
\ba{rcl}
\dsp-\frac{m}{4\pi a_s}&=&\dsp\frac{1}{G}-2\int\!\!\frac{d\bfm{q}}{(2\pi)^3}%
 \left(\frac{1}{E_q-m}+\frac{1}{E_q+m}\right),\\[1ex]
\ea
\label{eq:renom}
\eeq
From now, we use the renormalized coupling $G_R$ instead of $a_s$
itself for notational simplicity; that is defined by
\beq
  \frac{1}{G_R}\equiv\frac{m}{4\pi a_s}.
\eeq
\Eqn{eq:renom} can be written as
\beq
 -\frac{1}{G_R}=\frac{1}{G}-\chi_{0}(2m,\bfm{0}).
\eeq
Using this renormalized coupling, the gap equation can be casted into
the following form:
\beq
\ba{rcl}
  \dsp-\frac{1}{G_R}&=&\dsp2\int\!\!\frac{d\bfm{q}}{(2\pi)^3}%
   \biggl(\frac{\tanh\frac{E_q-\mu}{2T_c}}{E_q-\mu}-\frac{1}{E_q-m}\\[2ex]
   &&\qquad\qquad\qquad+\,(\mu,m\rightarrow-\mu,-m)\biggl).
\ea
\label{eq:thr}
\eeq
It can be easily seen that this integral still has a weak logarithmic
divergence.
This is in contrast to the nonrelativistic case where the UV divergence
can be completely taken away because of the quadratic momentum
dependence of the single fermion excitation, $p^2/2m$, 
in the energy denominator.
$T_c$ 
is now parameterized by
\beq
   T_c=m f_R(\mu/m,G_R m^2;\ln(\Lambda/m)).
\label{eq:sc}
\eeq
$f_R$
is again some unknown function.
The apparent cutoff dependence seems to be smaller than the
\Eqn{eq:tcbare}.

\Eqn{eq:thr} can also be formally written as
\beq
 -\frac{1}{G_R}-\chi^{\rm Ren}_{\mu,T_c}(0,\bfm{0})=0,
\eeq
with the renormalized pair correlation function
\beq
 \chi_{\mu,T}^{\rm Ren}(\omega,\bfm{P})%
 \equiv\chi_{\mu,T}(\omega,\bfm{P})-\chi_0(2m,\bfm{0}).
\eeq

We next discuss the bound state equation in vacuum $\mu=T=0$ in terms of
the renormalized coupling.
The bound state or the resonance pole in two fermion scattering can be
determined by
\beq
  -\frac{1}{G_R}-\chi^{\rm Ren}_0(\omega,\bfm{0})=0,
\eeq
where 
$\chi^{\rm Ren}_0(\omega,\bfm{P})\equiv\chi^{\rm Ren}_{0,0}(\omega,\bfm{P})$. 
Because $\chi_0^{\rm Ren}(2m,0)=0$, if $-1/G_R>0$ then
the resonance pole is located at $|\omega|>2m$. 
Otherwise it is located at $|\omega|<2m$, which corresponds to
the {\em stable bound state} because $\Im\chi(|\omega|<2m,\bfm{0})=0$.
Then the critical coupling $G_R=G_{0}$ for the zero binding is given 
by the condition 
\beq
-\frac{1}{G_{0}}-\chi^{\rm Ren}_0(2m,\bfm{0})=-\frac{1}{G_{0}}=0, 
\label{eq:G0}
\eeq
\ie, $1/a_s=0$ which is sometimes called as the unitary limit.
If the coupling is above this critical coupling ($1/G_R>0$) and if 
the bound state is sufficiently loosely bound so that the pole $\omega$
satisfies $0<4m^2-\omega^2\ll m^2$, then the bound state pole can be
approximated model-independenty by the unitary condition; that is
\beq
 |\omega|\cong 2m-\frac{1}{ma_s^2}=2m-\frac{16\pi^2}{m^3G_R^2}.
\eeq
The binding energy is given as usual by $1/ma_s^2$.
In our relativistic case, we have another typical coupling $1/G_{c}$
which is stronger than the unitary coupling:
When the coupling approaches this critical point $1/G_R\to 1/G_{c}-0$, 
the bound state becomes massless. 
This critical coupling satisfies the following condition,
\beq
  -\frac{1}{G_{c}}=\chi_{0}^{\rm Ren}(0,\bfm{0})%
  =-|\chi_{0}^{\rm Ren}(0,\bfm{0})|.
\label{criC}
\eeq
When the attraction is increased beyond this $G_c$,
the vacuum becomes unstable against the formation of $qq$-condensate 
which leads to the Majorana mass gap in the single quark excitation
even in vacuum.

It is also noteworthy that because of the identity
\beq
\chi^{\rm Ren}_{\mu,0}(\omega,\bfm{0})=\chi^{\rm
Ren}_{0}(\omega+2\mu,\bfm{0}),
\eeq
the Thouless condition $(-1/G_R-\chi^{\rm Ren}_{\mu,T}(0,\bfm{0})=0)$
rather determines the chemical potential $\mu$ in the 
region $-1/G_R\ll 0$ and $2(m-\mu)\gg T$ where 
the Pauli-blocking gives only a minor effect 
($\chi^{\rm Ren}_{\mu,T}(0,\bfm{0})\sim\chi^{\rm Ren}_{\mu,0}(0,\bfm{0})$).
Such conditions are actually realized in the nonrelativistic BEC
regime, where $\mu$ should be about one half of bound state energy at
rest, but with small $T$-dependent entropic correction \cite{melo93}.

Before closing this section, let us briefly discuss the $m\to0$ limit.
Although we have introduced the renormalized coupling $G_R$ via
$a_s$, the low energy information about $f(\bfm{p},\bfm{k})$
assuming $p\ll m\ne0$, if we extend the definition of $G_R$ 
by \Eqn{eq:thr}, the succeeding discussion can be applied even
to the $m=0$ case. 
By defining an appropriate function $f'_R$, \Eqn{eq:sc} can be casted
into $T_c=\mu f'_R(m/\mu,G_R\mu^2,\ln(\Lambda/\mu))$ in which
it is rather clear that it has a smooth massless limit.
Also in this limit, $1/G_0=1/G_c=0$ as is clear from their definitions,
\Eqn{eq:G0} and \Eqn{criC} combined with $\chi^{\rm Ren}_0(0,0)=0$.

\subsection{Renormalization of the dynamic pair susceptibility}
\label{1e}
In this section, we polish up the argument in the previous section.
By doing this, it turns out that we only need a regularization of
the vacuum (and also real) part of the dynamic pair susceptibility.

First we notice that the imaginary part of
$\Gamma_{\mu,T}^{-1}(\omega,\bfm{P})$
is finite while the real part is divergent.
Thus, we only look at the real part
${\Re}\Gamma_{\mu,T}^{-1}(\omega,\bfm{P})$
leaving the explicit formula of the imaginary part in the Appendix.
The real part of the pair dynamic susceptibility can be decomposed
into two part:
\beq
  \Re\Gamma^{-1}_{\mu,T}(\omega,\bfm{P})%
  =\Re\Gamma^{-1}_{0}(\omega+2\mu,\bfm{P})-\Re\chi^{\rm
  mat}_{\mu,T}(\omega,\bfm{P}),
\eeq
where we have defined the ``pair susceptibility'' 
in the vacuum $(\mu=T=0)$ as
\beq
\ba{rcl}
  \Gamma^{-1}_{0}(z,\bfm{P})&\equiv&\dsp\frac{1}{G}-\chi_0(z,\bfm{P}),\\[2ex]
\ea
\eeq
with $\chi_0$ being $\chi_{\mu=0,T=0}$.
The explicit forms of $\chi_0$ and $\chi^{\rm mat}_{\mu,T}$ can be read
from \Eqn{eq:dynamic},
\begin{widetext}
\beq
\ba{rcl}
  \Re\chi_{0}(z,\bfm{P})%
 &=&\dsp-2\int\!\!\!\frac{d\bfm{q}}{(2\pi)^3}%
    \bigg(1+\frac{m^2+\bfm{q}\cdot(\bfm{q}+\bfm{P})}%
    {E_{\bfsm{q}}E_{\bfsm{q}+\bfsm{P}}}\bigg)%
    \left[\frac{\mathcal P}{z-E_{\bfsm{q}+\bfsm{P}}-E_{\bfsm{q}}}%
    -\frac{\mathcal P}{z+E_{\bfsm{q}+\bfsm{P}}+E_{\bfsm{q}}}\right],\\[2ex]
 \Re\chi_{\mu,T}^{\rm mat}(\omega,\bfm{P})%
 &=&\dsp+4\int\!\!\!\frac{d\bfm{q}}{(2\pi)^3}%
    \bigg(1+\frac{m^2+\bfm{q}\cdot(\bfm{q}+\bfm{P})}%
    {E_{\bfsm{q}}E_{\bfsm{q}+\bfsm{P}}}\bigg)%
    \left[\frac{{\mathcal P}\,f_F(E_{\bfsm{q}}-\mu)}%
     {\omega+2\mu-E_{\bfsm{q}+\bfsm{P}}-E_{\bfsm{q}}}%
    -\frac{{\mathcal P}\,f_F(E_{\bfsm{q}}+\mu)}%
    {\omega+2\mu+E_{\bfsm{q}+\bfsm{P}}+E_{\bfsm{q}}}\right]\\[2ex]
 & &\dsp-4\int\!\!\!\frac{d\bfm{q}}{(2\pi)^3}%
    \bigg(1-\frac{m^2+\bfm{q}\cdot(\bfm{q}+\bfm{P})}%
    {E_{\bfsm{q}}E_{\bfsm{q}+\bfsm{P}}}\bigg)%
    \left[\frac{{\mathcal P}\,f_F(E_{\bfsm{q}}+\mu)}%
     {\omega+2\mu-E_{\bfsm{q}+\bfsm{P}}+E_{\bfsm{q}}}%
    -\frac{{\mathcal P}\,f_F(E_{\bfsm{q}}-\mu)}%
     {\omega+2\mu-E_{\bfsm{q}}+E_{\bfsm{q}+\bfsm{P}}}\right].\\[2ex]
\ea
\eeq
\end{widetext}
From these formulas, we can see 
that $\chi^{\rm mat}_{\mu,T}(\omega,\bfm{P})$ is finite while the vacuum
part $\chi_0(z,\bfm{P})$ is quadratically divergent.
In the same way as in the previous section, we can reduce the
divergence in $\Gamma^{-1}_0$ by the renormalization of the attractive
coupling (see \Eqn{eq:renom}).
\beq
\ba{rcl}
  \Re\Gamma^{-1}_0(\omega+2\mu,\bfm{P})&=&%
  \dsp\frac{1}{G}-\chi_0(\omega+2\mu,\bfm{P})\\[2ex]
  &\to&\dsp-\frac{1}{G_R}-\chi_0^{\rm Ren}(\omega+2\mu,\bfm{P})\\[2ex]
  &\equiv&\Re\Gamma^{-1}_{0\rm Ren}(\omega+2\mu,\bfm{P}),
\ea
\eeq 
with $\chi_0^{\rm Ren}(\omega+2\mu,\bfm{P})%
  =\chi_0(\omega+2\mu,\bfm{P})-\chi_0(2m,\bfm{0})$.
The momentum integral in the right hand side is again logarithmically
divergent.
We arrived at the following renormalized pair susceptibility:
\beq
\ba{rcl}
  \big[\Gamma^{{\rm Ren}}_{\mu,T}(\omega,\bfm{P})\big]^{-1}%
  &=&\Re\Gamma^{-1}_{0\rm Ren}(\omega+2\mu,\bfm{P})\\[2ex]
  &&-\Re\chi^{\rm mat}_{\mu,T}(\omega,\bfm{P})\\[2ex]
  &&-i\,\Im\chi_{\mu,T}(\omega+i\delta,\bfm{P}).\\[2ex]
\ea
\eeq
We note here again that only the first term has a logarithmic
divergence. The renormalized Thouless condition which we have 
derived in the previous section can be expressed in the compact form:
\beq
  \big[\Gamma^{\rm Ren}_{\mu,T_c}(0,\bfm{0})\big]^{-1}=0.
\label{thl}
\eeq
This is exactly the same as \Eqn{eq:renthouless} if the
matter part of the integral in \Eqn{eq:renthouless}, 
the piece being proportional to $f_F$, is evaluated without
a cutoff $\Lambda$.

\subsection{Number conservation}\label{1f}
Using the renormalized dynamic pair susceptibility discussed in the
previous section, the in-medium spectral density and phase shift can be
defined by
\beq
\ba{rcl}
  \rho^{\rm Ren}_{\mu,T}(\omega,\bfm{P})&=&%
  \Im\Gamma^{\rm Ren}_{\mu,T}(\omega,\bfm{P}),\\[2ex]
  \delta^{\rm Ren}_{\mu,T}(\omega,\bfm{P})&=&%
  {\rm Arg}\left[\Gamma^{\rm Ren}_{\mu,T}(\omega,\bfm{P})\right].
\ea
\eeq
The total density $N_{\rm tot}(\mu,T)$ is given by the
derivative of the thermodynamic potential with respect to $\mu$:
\beq
\ba{rcl}
  N_{\rm tot}(\mu,T)&=&N_{\rm MF}(\mu,T)\\[2ex]
  &&\dsp+d_B\int_{-\infty}^{\infty}%
  \!\!\frac{d\omega}{\pi}\frac{d\bfm{P}}{(2\pi)^3}%
  \tilde{f}_B(\omega)\frac{\partial\delta_{\mu,T}^{\rm
  Ren}(\omega,\bfm{P})}{\partial\mu},
\ea
\label{eq:number}
\eeq
where
$N_{\rm MF}(\mu,T)=-\frac{\partial\Omega_0(\mu,T)}{\partial\mu}$
is the quark number density from free quarks. 
This contribution can be further decomposed into two parts, \ie,
the contribution from quarks, and that from antiquarks, \ie,
\beq
  N_{\rm MF}(\mu,T)=N_q(\mu,T)-N_{\bar{q}}(\mu,T),
\eeq
with
\beq
\ba{rcl}
 N_q(\mu,T)&=&\dsp
 2N_cN_f\int\!\!\frac{d\bfm{q}}{(2\pi)^3}f_F(E_q-\mu),\\[2ex]
 N_{\bar{q}}(\mu,T)&=&\dsp
 2N_cN_f\int\!\!\frac{d\bfm{q}}{(2\pi)^3}f_F(E_q+\mu).
\ea
\eeq
The second term in \Eqn{eq:number} represents the quark number density
from pair correlation, which we denote by
\beq
 N_{\rm NSR}(\mu,T)=d_B\int_{-\infty}^{\infty}%
  \!\!\frac{d\omega}{\pi}\frac{d\bfm{P}}{(2\pi)^3}%
  \tilde{f}_B(\omega)\frac{\partial\delta_{\mu,T}^{\rm
  Ren}(\omega,\bfm{P})}{\partial\mu}.
\eeq
In order to give an intuitive interpretation of this formula,
we extract some specific contributions from this expression.
We first note the identity
\beq
\ba{rcl}
 \dsp\frac{\partial\delta_{\mu,T}^{\rm Ren}(\omega,\bfm{P})}{\partial\mu}%
 &=&\dsp-\rho_{\mu,T}^{\rm Ren}(\omega,\bfm{P})%
 \frac{\partial\Re\big[\Gamma_{\mu,T}^{\rm
 Ren}(\omega,\bfm{P})^{-1}\big]}{\partial\mu}\\[2ex]
 &&\dsp-\Re\Gamma_{\mu,T}^{\rm Ren}(\omega,\bfm{P})%
 \frac{\partial\Im\big[\Gamma_{\mu,T}^{\rm
 Ren}(\omega,\bfm{P})^{-1}\big]}{\partial\mu}.\\[2ex]
\ea
\eeq
We focus on the first term because the second term plays only
minor role as long as $(\mu,T)$ satisfies the Thouless condition. 
The number density from the first part is
\beq
\ba{rcl}
 d_B\int_{-\infty}^{\infty}%
  \!\!\frac{d\omega}{\pi}\frac{d\bfsm{P}}{(2\pi)^3}%
  \tilde{f}_B(\omega)\rho_{\mu,T}^{\rm Ren}(\omega,\bfm{P})
  \frac{\partial\Re[-\Gamma_{\mu,T}^{\rm
  Ren}(\omega,\bfsm{P})^{-1}]}{\partial\mu}.
\ea
\label{eq:spec}
\eeq
If the attractive coupling is strong enough $(1/G_R\agt0)$,
we have a stable bound boson (antiboson) pole in the spectral function.
\beq
\ba{rcl}
  \left[\rho^{\rm Ren}_{\mu,T}(\omega,\bfm{P})\right]_{\mbox{\scriptsize
  pole part}}%
  &=&Z_B^{\mu,T}\delta(\omega+2\mu-E_{B\bfsm{P}}^{\mu,T})\\[2ex]
  & &-Z_{\bar{B}}^{\mu,T}%
  \delta(\omega+2\mu+E_{\bar{B}\bfsm{P}}^{\mu,T}),\\[2ex]
\ea
\eeq
where $Z_B^{\mu,T}$, $E_{B\bfsm{P}}^{\mu,T}$ 
($Z_{\bar{B}}^{\mu,T}$, $E_{\bar{B}\bfsm{P}}^{\mu,T}$)
are the wavefunction renormalization and the bound state energy for
boson (antiboson). $E_{B\bfsm{P}}^{\mu,T}$
and $E_{\bar{B}\bfsm{P}}^{\mu,T}$
are defined by the following bound state pole conditions:
\beq
\ba{rcl}
  &&%
  \Gamma_{\mu,T}^{\rm Ren}(E_{B\bfsm{P}}^{\mu,T}-2\mu,\bfm{P})^{-1}=0,\\
  &&%
  \Gamma_{\mu,T}^{\rm Ren}(-E_{B\bfsm{P}}^{\mu,T}-2\mu,\bfm{P})^{-1}=0.
\ea
\eeq
Imaginary parts should vanish for stable bound states.
The wavefunction renormalizations are calculated by
\beq
\ba{rcl}
 &&\dsp Z_B^{\mu,T}=\frac{\pi}{\Big|\partial\Gamma^{\rm
 Ren}_{\mu,T}(\omega,\bfsm{P})^{-1}\big/%
 \partial\omega\Big|_{\omega=E_{BP}^{\mu,T}-2\mu}},\\[3ex]
 &&\dsp Z_{\bar{B}}^{\mu,T}=\frac{\pi}{\Big|\partial\Gamma^{\rm
 Ren}_{\mu,T}(\omega,\bfsm{P})^{-1}\big/%
 \partial\omega\Big|_{\omega=-E_{\bar{B}P}^{\mu,T}-2\mu}}.\\[2ex]
\ea
\eeq
Substituting these expressions to \Eqn{eq:spec}, we find
the following bound state contributions to $N_{\rm NSR}(\mu,T)$
\beq
\ba{rcl}
  &&\dsp d_B\int\!\!\frac{d\bfm{P}}{(2\pi)^3}%
  \Biggl[2-\frac{\partial E_{B\bfsm{P}}^{\mu,T}}{\partial\mu}\Biggr]%
  \tilde{f}_B(E_{B\bfsm{P}}^{\mu,T}-2\mu)\\[3ex]
  &&\dsp-d_B\int\!\!\frac{d\bfm{P}}{(2\pi)^3}%
  \Biggl[2+\frac{\partial
  E_{\bar{B}\bfsm{P}}^{\mu,T}}{\partial\mu}\Biggr]%
  \tilde{f}_B(E_{\bar{B}\bfsm{P}}^{\mu,T}+2\mu).\\[2ex]
\ea
\label{effectivenumber}
\eeq
We denote the first term by $N_B(\mu,T)$ and the second term by 
$-N_{\bar{B}}(\mu,T)$.
Note that the ``effective'' quark number charge of (anti)boson
slightly deviates from $(-)2$, which is caused by the Pauli-blocking
effect.
The remaining contribution to $N_{\rm NSR}(\mu,T)$ comes from
unstable pair correlation such as the $qq$-continuum excitation or the
Landau damping.
We define these contribution by
\beq
  N_{\rm un}(\mu,T)\equiv N_{\rm NSR}(\mu,T)-N_{B}(\mu,T)+N_{\bar{B}}(\mu,T).
\eeq

\begin{table*}[t]
 \begin{tabular}{|r||c|c|c|}
  \hline
  &Weak coupling BCS ($G_R^{-1}\ll0$) & Crossover regime & BEC
  ($0\ll G_R^{-1}\ll G_{c}^{-1}$) \\ \hline
  \multicolumn{1}{|l||}{\bf\sl (A) Thouless criterion~~} & $T_c\sim E_F
  \mathstrut e^{-\frac{\pi}{2k_F|a_s|}}$ 
  & $(\mu_c,T_c)$ & 
  $2\mu_c\sim 2m-\frac{1}{m a_s^2}$\\ \hline
  \multicolumn{1}{|l||}{\bf\sl (B) Number equation~~} & $\mu_c\sim E_F$ 
  & $(\mu_c,T_c)$ & 
  $T_c\sim\frac{k_F^2}{\pi^{1/3}m}\left(\frac{2}{3(N_c-1)%
  \zeta_{2/3}}\right)^{2/3}$ \\ \hline
  \multicolumn{1}{|l||}{\bf\sl Physical gap at $T=0$~~} 
  &$2\Delta_0\sim {2}{\pi}e^{-\gamma_{\rm E}}(=3.53)T_c$ 
  & $\sim2\Sqrt{\De_0^2+(m-\mu_c)^2}$ & 
  $2m-2\mu_c$\\
  \hline
 \end{tabular}
 \caption[]{The roles of two basic equations. $E_F=\Sqrt{k_F^2+m^2}$ is
 the Fermi energy, and the $s$-wave scattering length is related to the
 renormalized coupling by $a_s=m G_R/4\pi$.}
 \label{tab:1}
\end{table*}
The chemical potential $\mu_c$ may be determined by the quark
number conservation:
\beq
 N_cN_f\frac{k_F^3}{3\pi^2}\equiv N_{\rm tot}(\mu_c,T)=N_{\rm
 MF}(\mu_c,T)+N_{\rm NSR}(\mu_c,T),
\label{numeq}
\eeq
where the Fermi momentum $k_F$ is introduced as a parameter which
controls quark number density; we fix the total charge of the system
as $N_{\rm tot}=N_cN_f\frac{k_F^3}{3\pi^2}$.
In the weak coupling, this equation actually determines the
chemical potential $\mu_c\sim E_F\equiv\Sqrt{k_F^2+m^2}$, while in the
strong coupling region, it rather determines $T_c$ because
the Thouless condition determines $\mu_c$. 

In the following, we derive some analytic approximations for $T_c$
in the strong coupling. 
As noted, the Thouless condition in the strong coupling is essentially
the bound state (Bethe-Salpeter) equation in vacuum where $2\mu_c$ plays
a role of boson mass $M_B=2\mu_c<2m$.
Then the fermionic contribution to the number density gives
only a minor contribution and instead the bound state contributions
dominate the total density. 
Therefore we approximate the number equation \Eqn{numeq} as
\beq
\ba{rcl}
  \dsp\frac{N_{\rm tot}}{2d_B}=\int\frac{d\bfsm{P}}{(2\pi)^3}%
  \Big[f_B(E_{B\bfsm{P}}^{\mu_c,T}-2\mu_c)
  -f_B(E_{\bar{B}\bfsm{P}}^{\mu_c,T}+2\mu_c)\Big].\\[1ex]
\ea
\label{apps}
\eeq
We further approximate this for the following two cases,
\ie, (1) nonrelativistic case and (2) relativistic case.

\vspace*{1ex}
\noindent
{\bf (1) Nonrelativistic BEC;}~this means that the boson 
mass $2\mu_c$ is much larger than the boson's kinetic 
energy $N_{\rm tot}^{2/3}/4\mu_c$ or $T_c$, \ie
\beq
2\mu_c\gg N_{\rm tot}^{2/3}/4\mu_c.
\eeq
This may allow us to approximate boson's dispersion as
\beq
 E_{B\bfsm{P}}^{\mu_c,T}\sim 2\mu_c+\frac{P^2}{4\mu_c}.
\label{appd}
\eeq
Then the boson pole in \Eqn{apps} dominates
the integral because of the bose-enhancement factor.
We have
\beq
  \frac{N_{\rm tot}}{2 d_B}=\left(\frac{\mu_c
  T}{\pi}\right)^{3/2}\zeta(3/2).
\eeq
Therefore, if the temperature becomes smaller than the critical value
\beq
  T<T_{\rm BEC}^{\rm NR}
  =\frac{k_F^2}{\pi^{1/3}\mu_c}\left(\frac{N_f}{N_c-1}%
  \frac{1}{3\zeta(3/2)}\right)^{2/3},
\eeq
the macroscopic fraction has to be condensed in the $P=0$ mode
to maintain the number conservation;
that is the classical BEC for the nonrelativistic bose gas.
The critical temperature is of order $N_{\rm tot}^{2/3}/4\mu_c\,%
(\ll 2\mu_c)$ which also justifies our approximation \Eqn{appd}.

\vspace*{1ex}
\noindent
{\bf (2) Relativistic BEC;}~there is another situation which was
first considered in \cite{kapusta}; that is the case
\beq
  2\mu_c\ll N_{\rm tot}^{1/3}.
\label{appr}
\eeq
In this case, we have to take care both of the boson and antiboson
contributions in \Eqn{apps}. We may approximate the boson and antiboson
dispersion by $\Sqrt{4\mu_c^2+P^2}$. 
Then, we obtain the following approximate formula in the totally same
way as \cite{kapusta},
\beq
\ba{rcl}
\dsp
  \frac{N_{\rm tot}}{2d_B}%
  =\frac{2\mu_cT^2}{3}.\\[2ex]
\ea
\eeq
Then the critical temperature to the relativistic BEC (RBEC) is given by
\beq
  T_{\rm BEC}^{\rm RL}=\frac{k_F}{\pi}%
  \Sqrt{\frac{k_F}{\mu_c}\frac{N_f}{2(N_c-1)}}.
\eeq
The critical temperature is of order $(N_{\rm tot}/2\mu_c)^{1/2}$
which is much larger than the boson mass $2\mu_c$.

\vspace*{2ex}
\noindent
{\bf The crossover regime:~}
If the coupling is in the intermediate range such that the system
is in the crossover regime between the BCS and the BEC, the coupled set
of the Thouless condition and the number equation determines $(\mu_c,T_c)$.
Thus, our basic equations to determine $(\mu_c,T_c)$ in the whole region
of the attractive coupling are
\begin{widetext}
\beq
\ba{lrcl}
 \mbox{\bf (A) Thouless condition:~}&0&=&\Re\left[\Gamma^{\rm
 Ren}_{\mu_c,T_c}(0,\bfm{0})^{-1}\right],\\[2ex]
 \mbox{\bf (B) Number conservation:~}
 &\dsp N_c N_f\frac{k_F^3}{3\pi^2}%
 &=&N_{\rm MF}(\mu_c,T_c)%
 \dsp+\int_{-\infty}^{\infty}%
  \!\!\frac{d\omega}{\pi}\frac{d\bfm{P}}{(2\pi)^3}%
  \tilde{f}_B(\omega)\frac{\partial{\rm Arg}\big[\Gamma_{\mu_c,T_c}^{\rm
  Ren}(\omega,\bfm{P})\big]}{\partial\mu_c}. 
\ea
\label{eq:numcon}
\eeq
\end{widetext}
We summarized the roles of the Thouless criterion and the number equation
in the BCS and BEC regimes in TABLE.~\ref{tab:1}.
It may be worth noting that the physical gap at $T=0$ for the charge
neutral excitation is always handled by the Thouless condition,
\ie, by the low energy (long wavelength) limit of the dynamic pair
susceptibility.
We will see that, in the crossover regime, or in the ultra strong
coupling regime, these two equations are strongly coupled, giving rise
to a complex behaviour of $(\mu_c,T_c)$ as a function of the
renormalized coupling $G_R$.

\section{Thermodynamics of the BCS-to-BEC crossover}\label{thermodynamics}
Here, we discuss how the thermodynamic character of the system
changes throughout the crossover.
In the numerical calculation, we set $N_c=3$, $m/\Lambda=0.2$
and fix the fermion number density at $k_F=0.2m$.
\subsection{The BCS, BEC, and RBEC phases}
\begin{figure}[tp]
  \includegraphics[width=0.45\textwidth,clip]{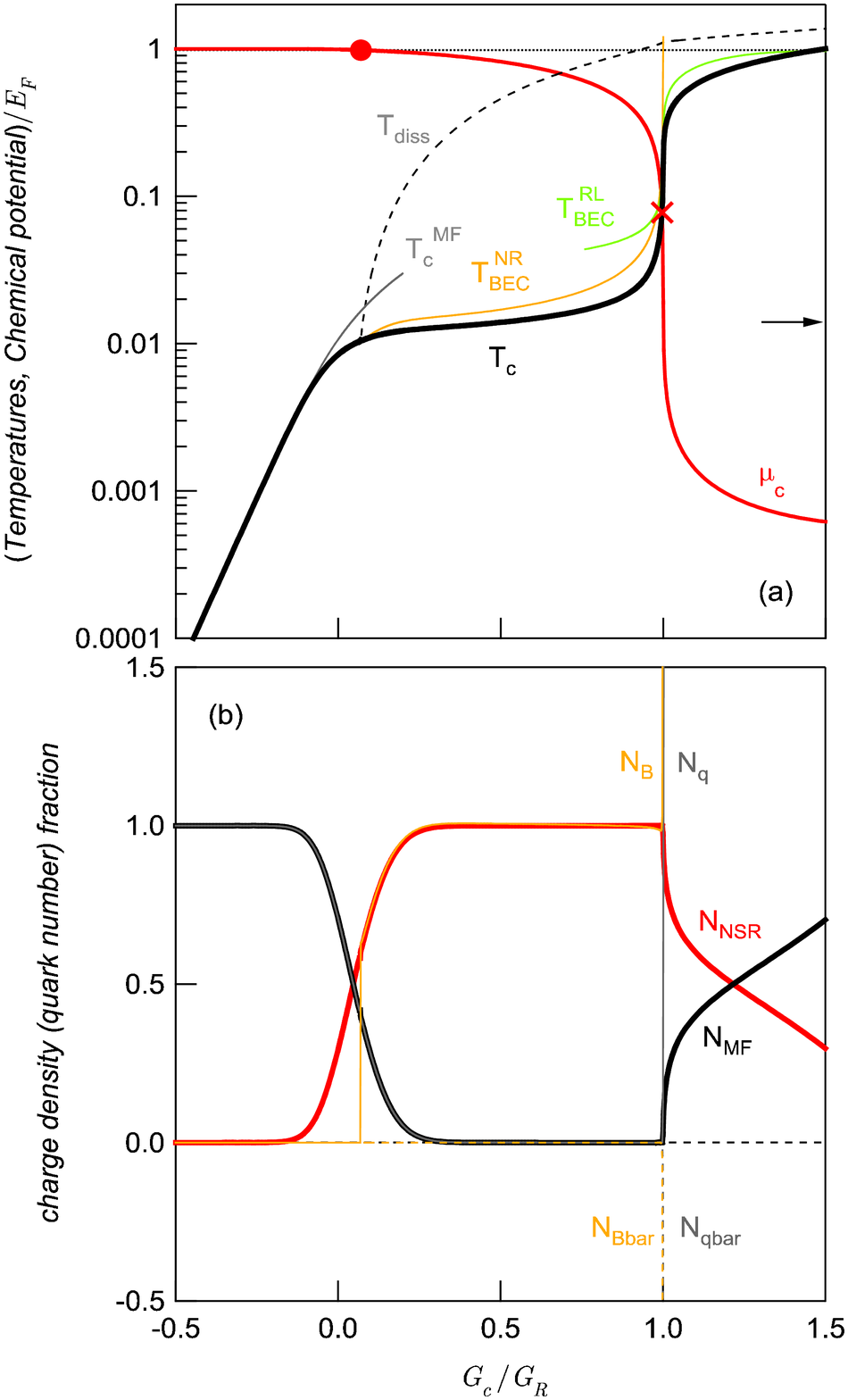}
  \caption[]{%
  {\bf(a)}~~The critical temperature (chemical potential) $T_c$
 ($\mu_c$) as a function of attractive coupling $G_c/G_R$.
  The dissociation temperature of pre-formed pair is also
  depicted by dashed line.
 {\bf(b)}~~The quark number fractions.
  }
  \label{Tcmuc}
\end{figure}

\begin{figure}[tp]
  \includegraphics[width=0.45\textwidth,clip]{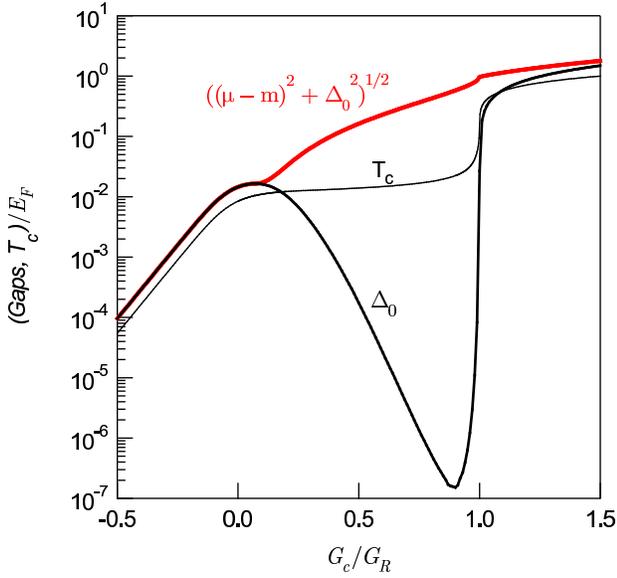}
  \caption[]{%
  The zero temperature gap parameter $\De_0$, and $T_c$ as
  a function of $G_R^{-1}$. 
  The mean field gap parameter $\De_0$ is calculated at
  $(\mu=\mu_c,T=0)$.
  The physical gap in the single quark excitation at $T=0$,
  $\Sqrt{\De_0^2+{\rm max.}(m-\mu_c,0)^2}$, 
  is also shown by the bold red line.
  }
  \label{gapvsTc}
\end{figure}

\Fig{Tcmuc}(a) shows how the critical temperature and chemical
potential ($T_c$ and $\mu_c$) change along with the attractive
coupling $G_c/G_R$; we use $G_c$, the critical coupling defined by
\Eqn{criC}, as a normalization.
Figure (b) shows the quark number contents.
We can clearly see that there are three distinct regions,
which we have called the BCS, BEC and RBEC phases in our previous paper
\cite{Nishida:2005ds}.
The critical temperature in the BCS region is well-approximated by the
mean field result ($T_c^{\rm MF}$; thin grey line) where $T_c^{\rm MF}$
and $\mu_c^{\rm MF}$ are determined without the fluctuation effect to
the number conservation, \ie, the result in which the last term of (B)
of \Eqn{eq:numcon} is ignored.
In this regime, the weak coupling universal relation between the zero
temperature gap $\De_0$ and $T_c$, \ie, $T_c\sim0.567\De_0$ is well
realized as we can see in \Fig{gapvsTc}.


\begin{figure}[tp]
  \includegraphics[width=0.45\textwidth,clip]{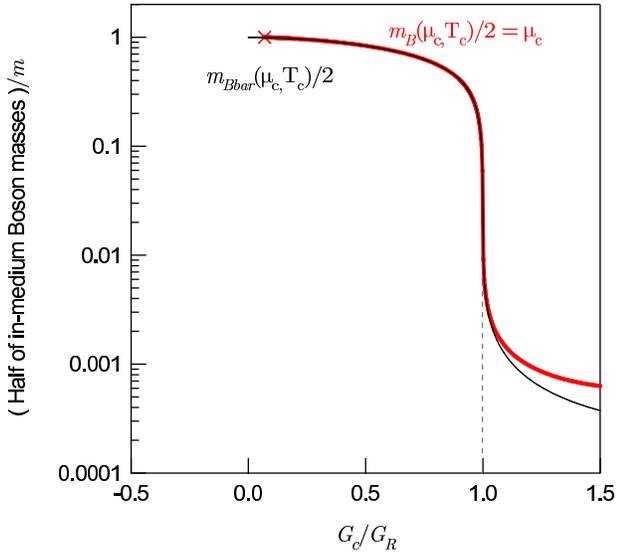}
  \caption[]{%
  The in-medium boson and antiboson masses $(M_B,M_{\bar{B}})$ as
  a function of attraction $G_R^{-1}$.
  The cross indicates the point $G_c/G_R\cong0.07$ where the bound
  boson forms, \ie, $M_B(\mu_c,T_c)=2\mu_c=2m$ holds at this point.
  }
  \label{bosonmass}
\end{figure}

As $G_c/G_R$ grows, $T_c$ gradually deviates from the mean field result
and for $G_c/G_R\agt 0.07$ the bound state formation takes place and the
system goes into the BEC regime.
The bound state formation in medium $(k_F\ne0)$ always takes place
at a stronger coupling than $G_c/G_R=0$, the unitary coupling.
This is because the Bethe-Salpeter kernel is smeared the Pauli-blocking
effect at finite density. 
\Fig{bosonmass} also shows this fact.
The bold line shows the half of boson mass $M_B(\mu_c,T_c)/2\equiv\mu_c$
while the thin line represents the half of antiboson 
mass $M_{\bar{B}}(\mu_c,T_c)/2$ as a function of $G_R$.
We can clearly see that the antiboson forms prior to the formation of
boson.
The antiboson forms almost $G_c/G_R\sim0$ where the bound state forms in
vacuum, while boson does not form up to a little stronger 
attraction $G_c/G_R=0.07$; 
this means that the Pauli-blocking by on-shell fermions actually
prevents the formation of boson.

In the BEC regime, the growth of $T_c$ as a function of $G_c/G_R$ is
suppressed because the increase of attraction is mainly used to reduce
the in-medium bound state mass.
Because the in-medium boson mass changes, temperature does not saturate
to the ideal BEC temperature of boson with mass $M_B=2m$ which is
indicated by the arrow in the figure; 
this is in contrast to the nonrelativistic calculations
\cite{nozieres,melo93,Haussmann}. $T_c$ in this region is well
approximated by the nonrelativistic
ideal BEC temperature ($T_{\rm BEC}^{\rm NR}$) of the boson
with mass $2\mu_c$,
\beq
  T_{\rm BEC}^{\rm
  NR}=k_F\frac{k_F}{\mu_c}\frac{1}{\pi^{1/3}}\left(\frac{1}{3%
  \zeta_{2/3}}\right)^{2/3}.
\eeq
This is shown by the thin orange line in figure (a).
It is remarkable that this expression has no explicit dependence
on $G_R$; there is only the implicit dependence on $G_R$
thorough $2\mu_c$, \ie, the in-medium boson mass. 
This fact expresses that the condensation in this regime has 
a rather kinematical origin.
The factor $N_c-1$ in the above expression also
indicates the condensation in this regime depends on the kinematical
degrees of freedom; the factor comes from the fact there are $N_c$
fermions in the system while we have $d_B=N_c(N_c-1)/2$ bosons belonging
to the $\bar{\bf 3}_c$ representation of SU$(3)$.
This is in contrast to $T_c^{\rm MF}$ in the BCS region where we find
no parametric dependence on $N_c$.
It is also noteworthy that the universal relation $T_c/\De_0=0.567$
is significantly violated as seen in \Fig{gapvsTc}; $T_c/\De_0$
grows up to $\sim10^5$.
However, the physical gap in the single quark excitation at $T=0$,
$\Sqrt{\De_0^2+\theta(m-\mu_c)(m-\mu_c)^2}$,
is always comparable to $T_c$.

As the attraction, $G_c/G_R$, is further increased beyond the unity,
the system goes into the RBEC phase where $T_c$ exceeds the
in-medium boson mass $2\mu_c$.
The critical temperature in this regime is well approximated by the
ideal BEC temperature $T_{\rm BEC}^{\rm RL}$ of a relativistic bose
gas:
\beq
  T_{\rm BEC}^{\rm
  RL}=\frac{k_F}{\pi}\Sqrt{\frac{k_F}{2\mu_c}}.
\eeq
Because the in-medium boson mass is smaller than the critical
temperature, there are a lot of bosons and antibosons as shown
in Fig.~\ref{Tcmuc} (b).
We notice that unlike the ideal BEC of the elementary boson gas
\cite{kapusta}, there are a lot of (anti)fermions as well as
(anti)bosons in the RBEC regime.
This is the characteristic feature of the composite boson system, 
where there exists the competition between free energy and entropy;
two fermion state is more favorable than one boson state in terms of
entropy, but is less favorable in terms of free energy.
This consists one reason why we find a relatively large deviation
between the real $T_c$ and $T_{\rm BEC}^{\rm RL}$ in the RBEC regime.

\begin{figure}[tp]
  \includegraphics[width=0.45\textwidth,clip]{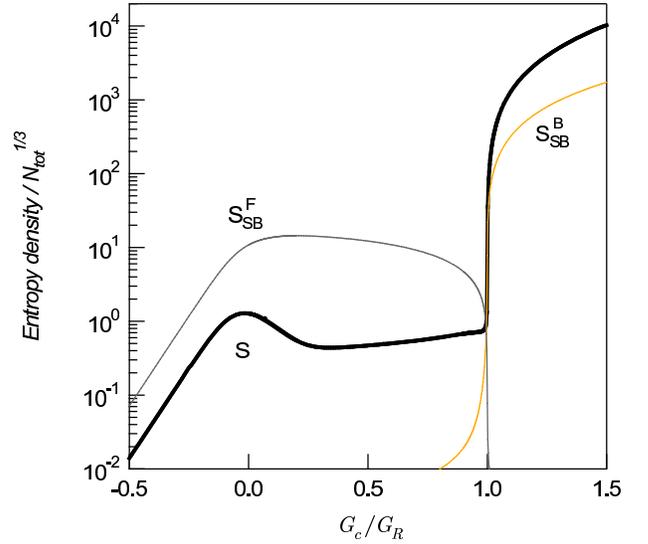}
  \caption[]{%
  The entropy par quark number charge as a function of $G_R^{-1}$.
  $S_F=N_c N_f(\frac{\mu_c^2 T_c}{3}+\frac{7\pi^2T_c^3}{45})$
  is the entropy for free massless quark gas and
  $S_B=N_c(N_c-1)\frac{2\pi^2T_c^3}{45}$
  is the Stephan-Boltzmann entropy for $N_c(N_c-1)/2$-``flavored'' boson
  gas.
  }
  \label{entropy}
\end{figure}

To confirm the above-discussed physical picture, we show in
\Fig{entropy}, how the total entropy behaves going from the BCS to the
RBEC.
We can see that the entropy in the BCS regime is suppressed from the
free massless fermion result because of the pairing correlation and
finite mass.
When the system goes into the BEC region, the entropy takes a nearly
constant value. 
This is because the increase of the coupling mainly affects the internal
structure (the binding) of the boson and the kinetic degrees of freedom
controlling $T_c$ does not change.
When the system goes into the RBEC phase, the entropy rapidly increases
because there appear a lot of kinetic degrees of freedom,
$\{q,\bar{q},B,\bar{B}\}$. 
Since plenty of $(q,\bar{q})$ are present in addition to $(B,\bar{B})$,
the Stephan-Boltzmann entropy for the ideal bose gas underestimates the
total entropy.

\subsection{Character change of Cooper pair and Deformation of Fermi surface}
\begin{figure*}[tp]
  \includegraphics[width=0.99\textwidth,clip]{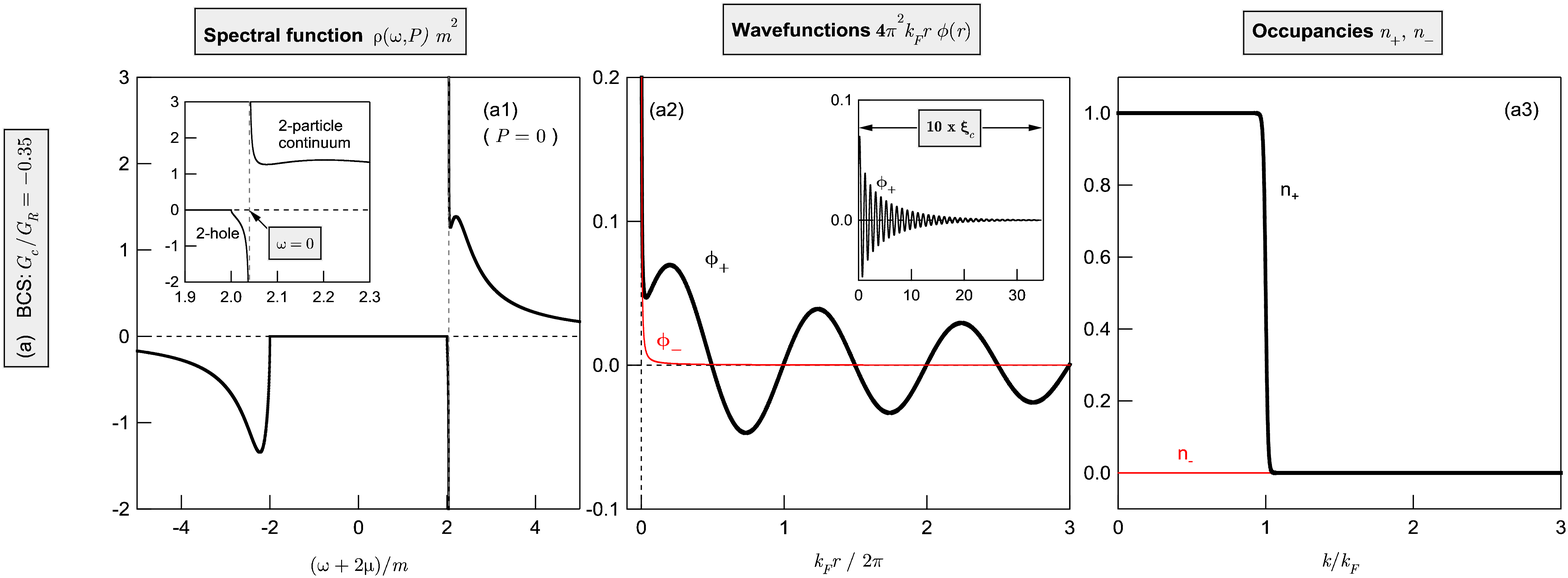}
  \includegraphics[width=0.99\textwidth,clip]{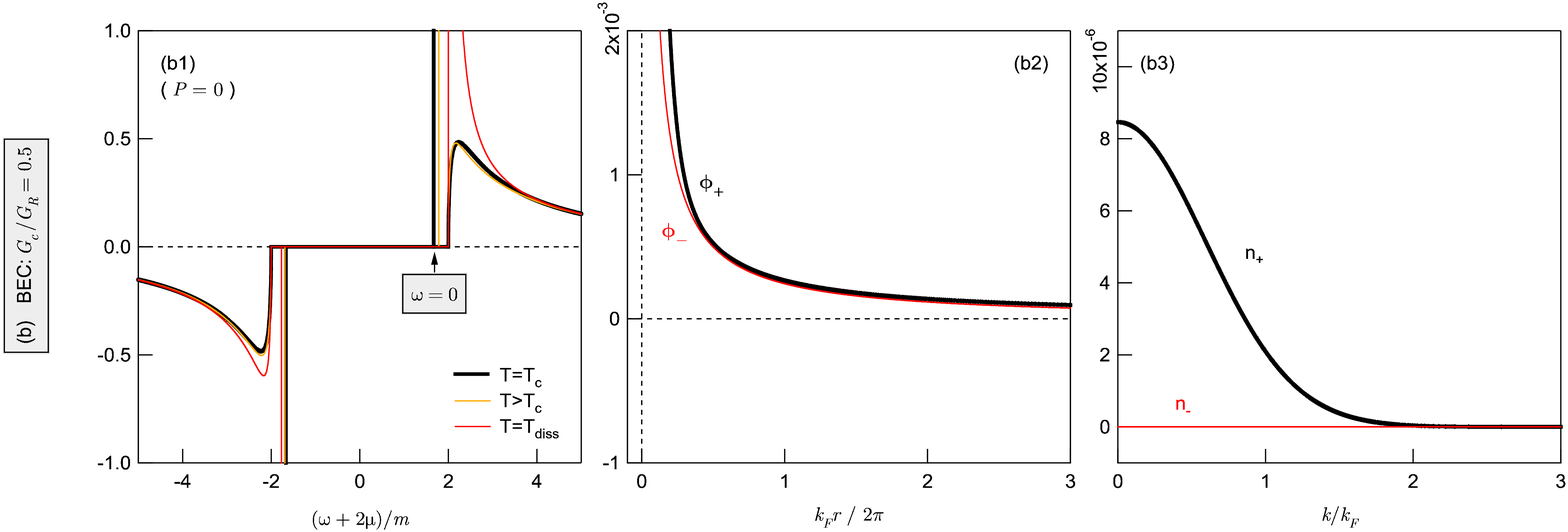}
  \includegraphics[width=0.99\textwidth,clip]{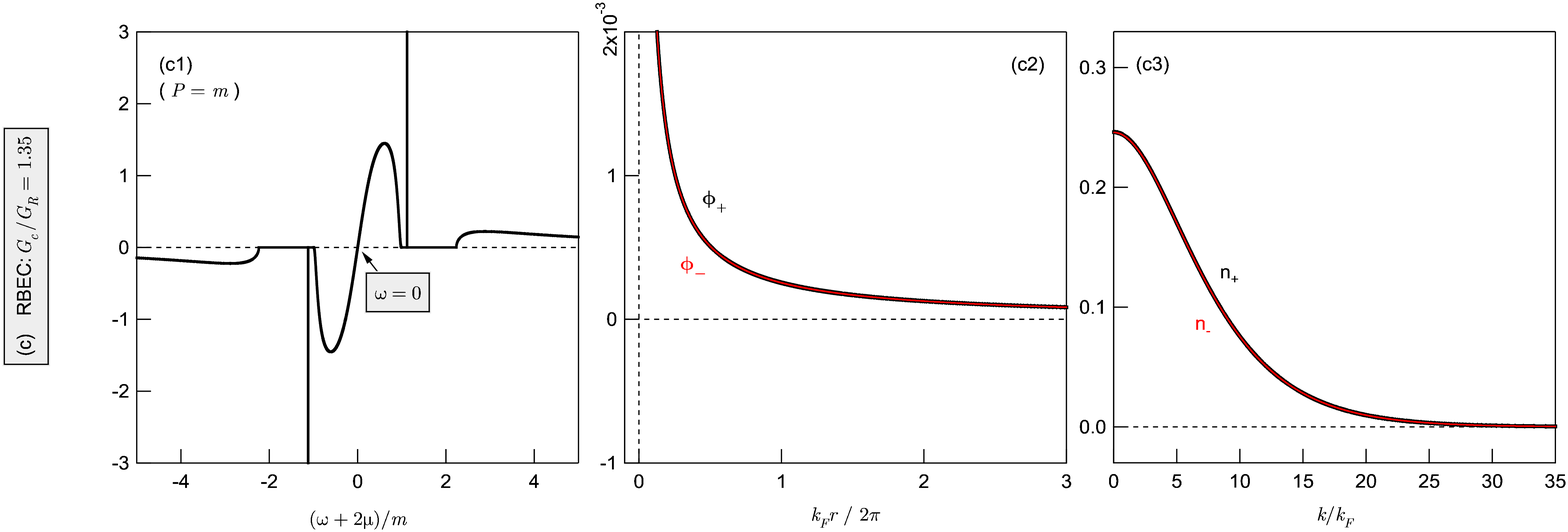}
  \caption[]{%
  The spectral density as a function of $\omega+2\mu$ (left panel),
  wavefunctions in the real space (middle), and the occupation numbers
  as a function of momentum $k$ (right panel). 
  From top to bottom, the attractive coupling increases
  as $G_c/G_R=-0.35$ (a), $G_c/G_R=0.5$ (b), and $G_c/G_R=1.35$ (c); the
  system evolves as the BCS (a), the BEC (b), and the RBEC state (c).
  The inset of (a1) is an enlargement near $\omega=0$, and
  the inset of (a2) shows the behaviour of the wavefunction at large
  length scale.
  }
  \label{fspecs}
\end{figure*}

To obtain a more intuitive insight into the three regions, we here study
the character change of Cooper pair wavefunction, the spectral density, and
fermion occupation numbers.
In \Fig{fspecs}, we show the spectral density [(a); left panel], 
the Cooper pair wavefunction [(b); middle panel], and the
(anti)fermion occupancy [(c); right panel].
From top to bottom, the renormalized coupling grows as $G_c/G_R=-0.35$
(BCS), $G_c/G_R=0.5$ (BEC), and $G_c/G_R=1.35$ (RBEC).

We first discuss how the spectral density evolves 
from the weak coupling BCS to the strong coupling RBEC phase.
Figure (a1) shows the spectral density $m^2\rho^{\rm
Ren}_{\mu_c,T_c}(\omega,\bfm{0})$ in the BCS phase.
We can see that the Thouless ``zero'' of the inverse susceptibility
at $\omega=0$ is located above the 2-particle ``decay'' (and 2-hole
``absorption'') continuum threshold, \ie, $0>2m-2\mu_c$. 
This means that the pair fluctuation with finite momentum in the BCS
state decays into two quarks.
We confirm this later by constructing the low energy effective theory
for the fluctuating pair field.
\Fig{fspecs} (b1) shows the spectral density in the BEC state.
In contrast to the BCS case, the Thouless singularity
at $\omega=0$ is realized as a bound state (boson) isolated pole.
This is because $\omega=0$ is located below the 2-particle continuum,
\ie, $0<2m-2\mu_c$.
There exists another singularity in the region $\omega+2\mu<0$
corresponding to the in-medium antiboson pole.
We note, however, that the thermodynamic quantity should be expressed 
by the integral of the combination 
$\tilde{f}_B(\omega)\times\rho_{\mu_c,T_c}^{\rm Ren}(\omega,\bfm{P})$,
so the antiboson pole gives only an exponentially suppressed
contribution to the thermodynamic quantities like entropy,
heat capacity, etc.
If we go higher temperature $T>T_c$ with fixing $\mu=\mu_c$, then the
boson pole gradually shifts to higher $\omega$ because the bound state
dissociates thermally (see the thin orange line in the figure).
As we increase the temperature, the pole eventually gets absorbed into
the 2-particle continuum at the dissociation 
temperature $T=T_{\rm diss}$. 
The spectral density at $T=T_{\rm diss}$ is depicted in the figure by
the thin red line.
Interestingly enough, the antiboson pole still survives at this
temperature; this is attributed to the fact that the Pauli-blocking
effect is less significant in the antiboson sector than in the boson
sector.
Finally, we show the spectral 
density $m^2\rho^{\rm Ren}_{\mu_c,T_c}(\omega,P=m)$ in the RBEC phase in
\Fig{fspecs} (c1).
Because $\mu_c\ll T_c$ in this regime, the spectral function is almost
antisymmetric with respect to $\omega\leftrightarrow-\omega$.
In the region $-P<\omega+2\mu<P$, the decay of the pair fluctuating mode
by absorbing thermally excited antiquarks (the Landau damping
process) is kinematically allowed.
Because of this, the spectral density takes non-zero value in this
region.
In the RBEC phase, the temperature is so high ($T_c\gg\mu_c$) that
this Landau damping process gives a significant contribution to the
unstable content $N_{\rm un}$ in $N_{\rm NSR}(\mu_c,T_c)$. 
This is one of the characteristic features in the relativistic fermion
system.

Next we study how the internal structure of the Cooper pair changes
throughout the crossover.
Let us first begin with the definition of the Cooper pair ``wavefunction''. 
We may define it from the expression of the correlation functions for
$T\le T_c$ \cite{Abuki:2003ut}:
\beq
\ba{rcl}
&&\braket{q^a_i(0,\bfm{x})q^b_j(-i0^-,\bfm{y})}\\[2ex]
&=&+(\tau_2\lambda_2)^{ab}_{ij}\int%
\frac{d\bfsm{q}}{(2\pi)^3}e^{i\bfsm{q}%
\cdot(\bfsm{x}-\bfsm{y})}\phi_+(q)\Delta\gamma_5C\Lambda_+(\hat{q})\\[2ex]
&&+(\tau_2\lambda_2)^{ab}_{ij}\int%
\frac{d\bfsm{q}}{(2\pi)^3}e^{i\bfsm{q}%
\cdot(\bfsm{x}-\bfsm{y})}\phi_-(q)\Delta\gamma_5C\Lambda_-(\hat{q}),\\[2ex]
\ea
\eeq
where the quark-quark pair wavefunction $\phi_+$ and the
antiquark-antiquark pair wavefunction $\phi_-$ are defined by
\beq
\phi_\pm(q)=\tanh\left(\frac{\eps_{q\pm}}{2T}\right)%
\frac{1}{2\eps_{q\pm}},
\eeq
$\eps_{q\pm}\equiv\Sqrt{(E_q\mp\mu)^2+\De^2}$
is the quasi-(anti)quark energy. 
These wavefunction in $q$-space can be Fourier-transformed to the real
space:
\beq
\phi_\pm(r)=\frac{1}{2\pi^2r}\int_0^{\infty}dq%
\frac{q\sin(qr)}{2\eps_{q\pm}}\tanh\left(\frac{\eps_{q\pm}}{2T}\right).
\eeq
Unlike the nonrelativistic situation, this integral does not converge
even for $r\ne0$; the integrand oscillates with $\sin(q r)$ as $q\to\infty$. 
Evaluating the oscillating integral by the sharp cutoff $\Lambda$ is
inappropriate, and therefore we take the following prescription to
regulate this integral. 
By integrating by parts, we obtain
\beq
\ba{rcl}
\phi_\pm(r)&=&-\frac{1}{2\pi^2 r^2}\left[\frac{q\cos(qr)}{2\eps_{q\pm}}%
 \tanh\left(\frac{\eps_{q\pm}}{2T}\right)\right]_{\scriptsize q\to\infty}\\
&&+\frac{1}{2\pi^2r^2}\int_0^{\infty}dq%
\cos(qr)\frac{\partial}{\partial
q}\left[\frac{q}{2\eps_{q\pm}}\tanh\left(\frac{\eps_{q\pm}}{2T}\right)\right].\\[2ex]
\ea
\label{wf}
\eeq
The second term is conversing while there remains a oscillating
uncertainty in the first term.
However, if we adopted the smooth cutoff scheme, the first term
should vanish. Thus we may define the second term of \Eqn{wf}
as the regularized Cooper pair wavefunction:
\beq
\ba{rcl}
\phi_\pm(r)&=&+\frac{1}{2\pi^2r^2}\int_0^{\infty}dq%
\cos(qr)\frac{\partial}{\partial
q}\left[\frac{q}{2\eps_{q\pm}}\tanh\left(\frac{\eps_{q\pm}}{2T}\right)\right].\\[2ex]
\ea
\label{wff}
\eeq
Using the regularized Cooper pair function, we can evaluate the
Cooper pair size (Coherence length) \cite{melo97,Abuki:2001be}:
\beq
 \xi_c^2(T)=\frac{\bra{\phi_+(\bfm{r})}\bfm{r}^2\phi_+(\bfm{r})\rangle}%
 {\langle\phi_+(\bfm{r})|\phi_+(\bfm{r})\rangle}%
 =-\frac{\bra{\phi_+(\bfm{q})}\nabla_{\bfsm{q}}^2\phi_+(\bfm{q})\rangle}%
 {\langle\phi_+(\bfm{q})|\phi_+(\bfm{q})\rangle_{\scriptsize\Lambda}}.
\eeq
\Fig{fspecs}(b1) shows the Cooper pair wavefunctions $4\pi^2 k_F
r\phi_\pm(r)$ at $T=T_c$ in the BCS phase ($G_c/G_R=-0.35$).
We can see that the wave function $\phi_+(r)$ oscillates with
a period $\sim2\pi/k_F$ and decays exponentially
$\e^{-\frac{r}{\pi\xi_p}}$
with the Pippard length
\beq
  \xi_p=\frac{k_F}{E_F}\frac{1}{\pi\Delta_0}.
\label{pippardl}
\eeq
$\De_0\sim 1.764T_c$
is the gap at zero temperature \cite{Abuki:2001be}.
The Pippard length is the weak coupling (BCS) approximation of the pair
size at $T=0$, \ie, $\xi_p\cong\xi_c(T=0)$ in the BCS regime.
The inset of \Fig{fspecs}(1b) shows the magnitude of wavefunction
actually decays exponentially at large scale $r>\xi_c\equiv\xi_c(T_c)$;
note that the pair size is robust against the increase in temperature,
$\xi_c(T_c)\sim\xi_c(T=0)$ \cite{Abuki:2003ut}.
Also it is consistent with \cite{Abuki:2003ut} that the antiquark pair
correlation is negligible in the BCS regime.
How this situation changes as the coupling grows?
\Fig{fspecs}(2b) shows the wavefunctions $4\pi^2k_Fr\phi_\pm(r)$
in the BEC region.
We first notice that the wavefunction $\phi_+(r)$ has no oscillating
structure like that in the BCS.
This is easily anticipated by the fact that the in-medium bound state
forms in this regime and that the bound state in the $s$-wave channel 
is unique; so the ``wavefunction'' should not have any nodes in the
real space \cite{nozieres}.
This is in contrast to \cite{Matsuzaki:1999ww,Abuki:2001be} where the pair
structure at low density still has a oscillating structure;
this fact shows that the in-medium bound state is not formed there.
In this region, the Pippard length is not a good approximation of the
pair size. 
In this regime, the pair size is roughly approximated by the $s$-wave
scattering length $a_s$. This is because, neglecting matter effects,
the bound state wavefunction in the C.M. frame should behave
as $\sim e^{-r/\Sqrt{2}a_s}$ as $r\to\infty$.
We will confirm these later.
It is also notable that there is a slight deviation in the 
wavefunctions $\phi_+(r)$ and $\phi_-(r)$ in this regime reflecting
the explicit breaking of charge conjugation by finite $\mu$.
In the RBEC phase, however, $\phi_+$ and $\phi_-$ show almost the same
behaviour because of $\mu_c\ll T_c$ (see \Fig{fspecs}(3b)).

Finally, we discuss how the Fermi surface is destructed when the
attraction is increased. 
To see this, we have plotted the occupation numbers 
in \Fig{fspecs}(c1) $\sim$ (c3).
From these figures, we can clearly see that the Fermi surface gets
significantly broken once the in-medium bound state have appeared in the
spectrum.
Baryon number carrier in the BEC regime is almost the bosonic degrees of
freedom as we can see also from \Fig{Tcmuc}(b).
In the RBEC phase, the temperature is high $T_c\gg\mu_c$, and thus
fermions are again thermally active for the entropic reason we noted in
the previous section. 
Also, the quark-antiquark asymmetry is small because of ($\mu_c\ll T_c$).

\section{The evolution of the soft mode dynamics}\label{softmodes}
In this section, we construct the low energy effective theory for the
pair fluctuation transport and study how the static and dynamic nature
of the soft mode evolves as the system goes from the weak coupling BCS
to the RBEC.
In the context of a nonrelativistic BCS superconductor, the effective
theories for the fluctuating pair field both for (i) near $T_c$ and (ii)
near $T=0$ were derived in \cite{Abrahams} where a {\em diffusion} type
equation was found above $T_c$.
The dynamics of the spece-time variation of fluctuation
in color superconductivity above $T_c$ was first investigated in
\cite{Kitazawa:2005vr} where the authors found a 
{\em damped-oscillation} type equation within the linear level.
The effect of fluctuating pair fields to the specific heat was also
investigated in \cite{Voskresensky:2004jp} starting with the ansatz
that the pair fluctuation is a {\em propagating} mode with a
relativistic dispersion.
By extending \cite{Kitazawa:2005vr} along with the nonrelativistic
calculations \cite{melo93,Abrahams}, we will find the following three
points:
(i) The pair fluctuation in the BCS regime is a diffusive
mode but with small propagating piece \cite{melo93,Kitazawa:2005vr}.
(ii) In the BEC regime, the fluctuating pair field is of a propagating
mode but with a nonrelativistic dispersion \cite{melo93}.
(iii) The fluctuation is also propagating in the RBEC region and
the relativistic (an almost linear) dispersion is realized in a wide
kinematical region.

Let us start our analysis with first looking at the static part of 
the fluctuation. 
In the context of color superconductivity, this was first done in
\cite{Iida:2000ha} in the weak coupling BCS regime.
The long wavelength expansion of the pair susceptibility near $T=T_c$
leads to
\beq
  \left[\Gamma^{\rm Ren}_{\mu,T}(0,\bfm{P})\right]^{-1}%
  =a_0\frac{T-T_c}{T_c}+\frac{c}{4m}\bfm{P}^2+\cdots,
\eeq
where the mass-squared parameter $a_0$ and the stiffness
parameter (or diffusion constant) $c$ are defined by
\beq
\ba{rcl}
  a_0&=&T_c\frac{\partial}{\partial T}\left[\Gamma^{\rm
  Ren}_{\mu,T}(0,\bfm{0})\right]^{-1}\big|_{T=T_c},\\[2ex]
  c&=&4m\frac{\partial}{\partial\bfsm{P}^2}\left[\Gamma^{\rm
  Ren}_{\mu,T}(0,\bfm{P})\right]^{-1}\big|_{\bfsm{P}=\bfsm{0}}.\\[1ex]
\ea
\eeq
The explicit calculation yields \cite{Kitazawa}:
\begin{widetext}
\beq
\ba{rcll}
  a_0&=&\frac{1}{T_c}\int\frac{d\bfsm{q}}{(2\pi)^3}%
  \biggl[\frac{1}{\cosh^2\left(\frac{E_q-\mu}{2T_c}\right)}%
  +\frac{1}{\cosh^2\left(\frac{E_q+\mu}{2T_c}\right)}\biggl],&\\[2ex]
  \frac{c}{4m}&=&\int\frac{d\bfsm{q}}{(2\pi)^3}%
  \left[\frac{q^2{\rm sech}^2\left(\frac{E_q-\mu}{2T_c}\right)%
  \tanh\left(\frac{E_q-\mu}{2T_c}\right)}%
  {24T_c^2 E_q^2(E_q-\mu)} +
 (\mu\leftrightarrow-\mu)\right]&\quad\cdots{\mathcal O}%
  \big(\frac{\mu^2}{T_c^2}\big)\\[2ex]
  &&-{\mathcal P}\int\frac{d\bfsm{q}}{(2\pi)^3}%
  \left[\frac{(2E_q^2+m^2){\rm
  sech}^2\left(\frac{E_q-\mu}{2T_c}\right)}%
  {24 T_c E_q^3(E_q-\mu)}+ (\mu\leftrightarrow-\mu)\right]&%
  \quad\cdots{\mathcal O}%
  \big(\frac{\mu}{T_c}\big)\\[2ex]
  &&-{\mathcal P}\int\frac{d\bfsm{q}}{(2\pi)^3}%
  \left[\frac{(2E_q^2+m^2)(2E_q-\mu)%
 \tanh\left(\frac{E_q-\mu}{2T_c}\right)}%
  {6\mu E_q^3(E_q-\mu)^2}+ (\mu\leftrightarrow-\mu)\right]%
  &\quad\cdots{\mathcal O}\big(1\big).\\[1ex]
\ea
\label{ac}
\eeq
\end{widetext}
We note that $a_0$ has no UV divergence, while $c$ still has a
logarithmic divergence in the last term in \Eqn{ac} which is
sub-sub-dominant in the $(\mu/T_c)$-expansion;
accordingly, the integral is conversing in the weak coupling limit,
while it is not in the strong coupling.
We may simply regularize the divergence by sharp cutoff $\Lambda$.

In the extremely weak coupling BCS regime where $\mu_c\gg T_c$,
we can derive some analytic approximations.
First, we may safely ignore the contribution from antiquarks, and can
replace the momentum integral $\int\frac{d\bfsm{q}}{(2\pi)^3}$
by $N_0\int_{-\infty}^{\infty}dE_q$ with $N_0=\frac{\mu^2}{2\pi^2}$
being the density of state at $E_q=\mu$.
We then arrive at the approximations:
\beq
  a_0^{\rm BCS}=4N_0,\quad\frac{c^{\rm
  BCS}}{4m}=\frac{7\zeta(3)k_F^2}{12\pi^2\mu^2T_c^2}N_0.
\label{GLap}
\eeq
Because 
$\Gamma^{\rm Ren}_{\mu,T}(0,\bfm{P})\sim\frac{1}{\bfsm{P}^2%
+\frac{4m a_0}{c}t_r}$ 
near $T=T_c$, with $t_r=\frac{T-T_c}{T_c}$ being the reduced
temperature, the static disturbance of the system restores at the
``healing'' length
\beq
  \xi_{\rm GL}t_r^{-1/2},
\eeq
where the Ginzburg-Landau (GL) coherence length $\xi_{\rm GL}$ is
defined by
\beq
 \xi_{\rm GL}\equiv\Sqrt{\frac{c}{4m a_0}}.
\label{xiGL}
\eeq
The healing length diverges as $T\to T_c+0$ as a consequence of
the second order phase transition.
In the weak coupling BCS regime ($\mu\gg T_c$), the GL
coherence length is approximated by (substituting \Eqn{GLap}
into \Eqn{xiGL}),
\beq
\ba{rcl}
  \xi_{\rm GL}^{\rm BCS}&=&
  \frac{k_F}{E_F}\frac{\Sqrt{21\zeta(3)}}{12T_c}
  \cong0.739\xi_p,\\[1ex]
\ea
\label{uni2}
\eeq
a well-known relation between the Pippard length and the GL
healing length in the nonrelativistic BCS theory.

We now look at the dynamic part of the problem.
One must be careful in performing the low frequency expansion of the
dynamic pair susceptibility near $T_c$ because there arises a spectral
singularity in the kinematical region $\omega/P<v_F$ once the system
goes below $T_c$ \cite{Abrahams}; $v_F$ is the Fermi velocity of
quasi-quarks.
This singularity is attributed to the local absorption (or emission)
of the Goldstone mode (phonon) by thermally active quasi-quarks.
A simple expansion is allowed only for the region near the critical
temperature $T\sim T_c$ where the gap is small $\omega\gg\Delta(T)$.
In such region, we can naively perform the low frequency expansion
\cite{Abrahams} as
\beq
  \left[\Gamma^{\rm Ren}_{\mu,T}(\omega,\bfm{0})\right]^{-1}%
  =-d\omega-d_2\omega^2+\cdots,
\eeq
where $d$ and $d_2$ are defined by
\beq
\ba{rcl}
  d&=&-\frac{\partial}{\partial\omega}\left[\Gamma^{\rm
  Ren}_{\mu,T_c}(\omega,\bfm{0})\right]^{-1}\big|_{\omega=0},\\[2ex]
  d_2&=&-\frac{1}{2}\frac{\partial^2}{\partial\omega^2}\left[\Gamma^{\rm
  Ren}_{\mu,T_c}(\omega,\bfm{0})\right]^{-1}\big|_{\omega=0}.\\[1ex]
\ea
\eeq
These quantities are complex in the BCS region $(\mu>m)$, while they
become real in the BEC region $(\mu<m)$ because the imaginary part of
the dynamic pair susceptibility looks like
\beq
\ba{rcl}
{\rm Im}\left[\Gamma^{\rm Ren}_{\mu,T_c}(\omega,\bfm{0})\right]^{-1}%
&=&-\theta(\omega+2\mu>2m)\\[2ex]
&&\times\frac{(\omega+2\mu)\Sqrt{(\omega+2\mu)^2-4m^2}%
\tanh\left(\frac{\omega}{4T_c}\right)}{4\pi}.
\ea
\label{imag}
\eeq
Further explicit calculation leads
\beq
\ba{rcl}
  d&=&{\mathcal P}\int\frac{d\bfsm{q}}{(2\pi)^3}\left[%
      \frac{\tanh\left(\frac{E_q-\mu}{2T_c}\right)}%
      {(E_q-\mu)^2}%
      -\frac{\tanh\left(\frac{E_q+\mu}{2T_c}\right)}%
      {(E_q+\mu)^2}\right]\\[2ex]
   & &+i\theta(\mu>m)\frac{\mu\Sqrt{\mu^2-m^2}}{4\pi T_c},\\[2ex]
  d_2&=&{\mathcal P}\frac{1}{2}\int\frac{d\bfsm{q}}{(2\pi)^3}\left[%
      \frac{\tanh\left(\frac{E_q-\mu}{2T_c}\right)}%
      {(E_q-\mu)^3}%
      +\frac{\tanh\left(\frac{E_q+\mu}{2T_c}\right)}%
      {(E_q+\mu)^3}\right]\\[2ex]
   & &+i\theta(\mu>m)\frac{2\mu^2-m^2}{8\pi T_c\Sqrt{\mu^2-m^2}}.%
      \\[1ex]
\ea
\eeq
The weak coupling $(\mu\gg T_c)$ approximation again applies both
to $d$ and $d_2$ yielding the following analytic results.
\beq
\ba{rcl}
 d^{\rm BCS}&=&i N_0\frac{\pi}{2T_c},\\[2ex]
 d^{\rm BCS}_2&=&-N_0\frac{7\zeta(3)}{4\pi^2 T_c^2}%
  +iN_0\left(\frac{2\mu^2-m^2}{\mu^2-m^2}\right)\frac{\pi}{4\mu T_c}.
\ea
\label{dBCS}
\eeq
Note that $d^{\rm BCS}$ is pure imaginary reflecting the particle-hole
symmetry in the weak coupling limit \cite{Abrahams}.
This means that, to this order, the pair fluctuation obeys a
diffusion-type equation in the BCS phase \cite{Abrahams,melo93}.
In reality, there is a finite real part which is an order $(T_c/\mu)$
suppressed from the imaginary part though.
This real part arises from a particle-hole asymmetry and adds a small
propagating nature to the soft mode; as a consequence, the fluctuation
in the BCS region becomes a {\em damped-oscillation} mode
\cite{Kitazawa:2005vr} or an {\em overdamped} mode \cite{melo93}.
The coefficient of second-time derivative, $d_2$, is also non-vanishing.
This is another source of a propagating nature of the soft mode.

So far, we have concentrated on the kinematical aspect of the pair
fluctuation.
It is sometimes important to know about the interaction between soft
modes to investigate the transport properties of system. 
For example, the shear viscosity is proportional to the mean free path
$(l_{\rm MFP})$
of soft modes in the classical level. 
This requires an information about the collision between soft modes.
We here look at the interaction between soft modes.

The partition function is factorized as $Z_0 e^{-S_{\rm eff}}$
where $Z_0$ is the partition function for free quark gas,
and $S_{\rm eff}$ is the effective action for fluctuation.
We need to go beyond the gaussian approximation for $S_{\rm 
eff}$.
Up to the quartic order in $\Delta$, we have
\begin{widetext}
\beq
\ba{rcl}
 S_{\rm eff}[\Delta,\Delta^*]&=&%
 \dsp T\sum_{N,\eta}\int\frac{d\bfm{P}}{(2\pi)^3}%
 \left[\Gamma^{\rm Ren}_{\mu,T}(i\Omega_N,\bfm{P})\right]^{-1}%
 |\Delta_\eta(i\Omega_N,\bfm{P})|^2\\[2ex]
 &&\dsp+\frac{1}{4}\sum_{\eta,\xi}%
 \sum_{1,2,3}b_{1,2,3}\Delta_{\eta}(1)%
 \Delta_{\eta}(2)^*\Delta_{\xi}(3)\Delta_{\xi}^*(1-2+3)\\[2ex]
 &&\dsp+\frac{1}{4}\sum_{\eta,\xi}%
 \sum_{1,2,3}b^{'}_{1,2,3}\Delta_{\eta}(1)%
 \Delta_{\xi}(2)^*\Delta_{\xi}(3)\Delta_{\eta}^*(1-2+3).\\[1ex]
\ea
\eeq
\end{widetext}
Here numbers $\{1,2,3\}$ are labeling sets of frequency and momentum
$\{(i\Omega_{N1},\bfm{P}_1),(i\Omega_{N2},\bfm{P}_2),
(i\Omega_{N3},\bfm{P}_3)\}$,
and $\sum_{1,2,3}$ is a shorthand of
$T^3\sum_{N_1,N_2,N_3}\int\frac{d\bfsm{P}_1}{(2\pi)^3}%
\frac{d\bfsm{P}_2}{(2\pi)^3}\frac{d\bfsm{P}_3}{(2\pi)^3}$.
To derive the low energy effective action, 
we set $\Delta_\eta(i\Omega_N)=0$ for $N=\pm1,\pm2,\cdots$,
and only consider the fluctuation mode with the lowest Matsubara
frequency.
Thus $\Delta$ has no dependence on $\tau$.
Then the quadratic term of $S_{\rm eff}$ is casted into the form
$\beta\int d\bfm{x}f_{\rm eff}^{\rm kin}(\Delta(x),\Delta^*(x))$
with the kinetic part of the local effective (classical) free energy
density defined by
\beq
\ba{rcl}
 f_{\rm eff}^{\rm kin}&=&\sum_\eta\Delta_\eta(\bfm{x})^*\left[%
  a_T-\frac{c}{4m}%
  \nabla_{\bfsm{x}}^2+\cdots\right]\Delta_\eta(\bfm{x})\\[2ex]
  &=&\sum_{\eta} \left[a_T|\Delta_\eta(\bfm{x})|^2%
  +\frac{c}{4m}|\nabla_{\bfsm{x}}\Delta_\eta(\bfm{x})|^2\right].
\ea
\eeq
$a_T\equiv a_0\frac{T-T_c}{T_c}$
is the mass parameter and $c$ is the diffusion constant introduced
before.
The quartic term also can be written in the form
$\beta\int{d\bfm{x}}f^{\rm int}_{\rm eff}(\Delta,\Delta^*)$ 
with
\beq
\ba{rcl}
f_{\rm eff}^{\rm int}&=&\sum_{\eta,\xi}\frac{b_0}{2}%
 |\Delta_\eta(\bfm{x})|^2|\Delta_\xi(\bfm{x})|^2.
\ea
\eeq
where $b_{0}=b_{0,0,0}=b^{'}_{0,0,0}$ is calculated as
\begin{widetext}
\beq
\ba{rcl}
b_0&=&\dsp2T\sum_n\int\frac{d\bfm{q}}{(2\pi)^3}\tr%
\left[S_F(i\omega_n,\bfm{q})S_F(-i\omega_n,-\bfm{q})%
S_F(i\omega_n,\bfm{q})S_F(-i\omega_n,-\bfm{q})\right]\\[2ex]
  &=&\dsp\frac{1}{T}\int\frac{d\bfm{q}}{(2\pi)^3}%
     \Biggl[\frac{-(E_q-\mu)%
     +T\sinh\big(\frac{E_q-\mu}{T}\big)}{(E_q-\mu)^3%
     \big[1+\cosh\big(\frac{E_q-\mu}{T}\big)\big]}%
     +\frac{-(E_q+\mu)%
     +T\sinh\big(\frac{E_q+\mu}{T}\big)}{(E_q+\mu)^3%
     \big[1+\cosh\big(\frac{E_q+\mu}{T}\big)\big]}\Biggl]
\ea
\eeq
\end{widetext}
To derive this, we have taken care the lowest term omitting
the gradient terms. This corresponds to the soft limit of 
the two body scattering where all the incoming and outgoing 
momenta vanish.
The weak coupling BCS ($\mu\gg T_c$) approximation again applies
to $b_0$ resulting in
\beq
  b_0^{\rm BCS}=N_0\frac{7\zeta(3)}{2\pi^2 T_c^2}.
\eeq

Collecting all the above results, we arrive at the following expression
for the local (but coarse-grained) Ginzburg-Landau (GL) free energy
functional 
($F_{\rm eff}=\int d{\bfm{x}}(f_{\rm eff}^{\rm kin}+f_{\rm eff}^{\rm int})$) 
which describes the long wavelength (static) excitation of soft mode:
\begin{widetext}
\beq
\ba{rcl}
 F_{\rm eff}[\Delta,\Delta^*]%
  &=&\int d\bfm{x}\left[\sum_{\eta}\left[a_T|\Delta_\eta(\bfm{x})|^2%
  +\frac{c}{4m}|\nabla_{\bfsm{x}}\Delta_\eta(\bfm{x})|^2\right]%
  +\sum_{\eta,\xi}\frac{b}{2}|\Delta_\eta(\bfm{x})|^2|\Delta_\xi(\bfm{x})|^2\right].\\[1ex]
\ea
\label{TDGLf}
\eeq
On the other hand, we have already derived the time dependence of the
excitation at long time scale. 
Combining the nonlinear term in $\Delta(t,x)$, we obtain the following
 dynamic equation for the fluctuation transport:
\beq
  \left[-i d\partial_t+d_2\partial_t^2%
  +a_T-\frac{c}{4m}\nabla_{\bfsm{x}}^2%
  +b_0\sum_{\xi}|\Delta_\xi(t,\bfm{x})|^2\right]\Delta_\eta(t,\bfm{x})=0.
\eeq
\end{widetext}
It can be written in the compact and familiar form:
\beq
  \left[-id\partial_t+d_2\partial_t^2\right]\Delta_\eta(t,\bfm{x})%
  =-\frac{\delta F_{\rm
  eff}[\Delta,\Delta^*]}{\delta\Delta^*_{\eta}(t,\bfm{x})}.
\label{Eqm}
\eeq
As for extremely low frequency modes with 
$(\omega\ll{\rm min.}\,(T_c,|m-\mu|))$, 
we may simply ignore the second term (including $d_2$) in the left hand
side of the above equation.

In the BCS regime, $d$ is imaginary dominant and is pure imaginary in
the weak coupling limit as shown in \Eqn{dBCS}.
In this regime, \Eqn{Eqm} describes how the fluctuation about the
thermal equilibrium relaxes; 
the thermodynamic restoring force to the equilibrium is then 
given by $-\frac{1}{d^{\rm BCS}}\frac{\delta F_{\rm eff}[\Delta,\Delta^*]}%
{\delta\Delta^*_{\eta}(\bfm{x})}$.
By rescaling the field by $\Psi(t,\bfm{x})=\Sqrt{c}\Delta(t,\bfm{x})$,
we find the standard expression for the time-dependent
Ginzburg-Landau (TDGL) equation in the BCS regime:
\beq
  \left[\frac{\Im d}{c}\partial_t+\frac{a_T}{c}%
  -\frac{\nabla_{\bfsm{x}}^2}{4m}+\frac{b_0}{c^2}%
  \sum_\xi|\Psi_\xi(t,\bfm{x})|^2\right]\Psi_\eta(t,\bfm{x})=0.%
\eeq
The pair fluctuation is diffusive as in the nonrelativistic case
\cite{Abrahams,melo93}; ignoring the nonlinear term, the mode with 
wavenumber $k$ decays exponentially 
\beq
\Psi_\xi(t,k)\sim e^{-t/\tau_k}
\eeq
with $\tau_k$ being the relaxation time
\beq
 \tau_k=\frac{\Im d}{a_T}\frac{1}{1+\frac{ck^2}{4m a_T}}%
 \cong\frac{\pi}{8T_c}\frac{1}{t_r+k^2\xi_{\rm GL}^{2}}.
\eeq
The relaxation time diverges as $k\to0$ and $T\to T_c+0$ reflecting
the critical slowing down. 
The TDGL equation for fluctuating diquarks is derived in the
linear level \cite{Kitazawa}.

\begin{figure}[tp]
  \includegraphics[width=0.3\textwidth,clip]{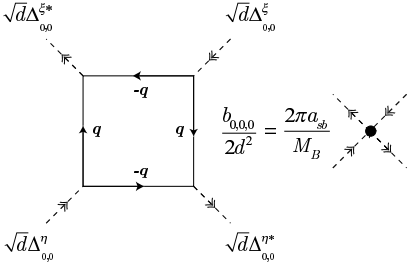}
  \caption[]{%
  The microscopic process which contributes the boson-boson
  repulsive force in the BEC region. 
  The thin line expresses propagation of fermion,
  while the external dashed line represents incoming/outgoing 
  boson. $\Sqrt{d}$ is the wavefunction renormalization.
  The arrow indicated on the line represents the flow of baryon
  charge.
  }
  \label{bosonint}
\end{figure}

Now we discuss the fluctuation transport property in the BEC regime.
In the BEC regime where $\mu<m$ is realized, $d$ in turn becomes
pure real as shown in \Eqn{imag}.
Again by ignoring $d_2$ term looking at the dynamics of sufficiently
low energy excitation, and by rescaling the field as
$\Psi(t,\bfm{x})=\Sqrt{d}\Delta(t,\bfm{x})$,
we arrive at
\beq
  \left[-i\partial_t+\frac{a_T}{d}-\frac{c\nabla_{\bfsm{x}}^2}{4m d}%
  +\frac{b_0}{d^2}\sum_\xi|\Psi_\xi(t,\bfm{x})|^2\right]\Psi_\eta(t,\bfm{x})=0.
\eeq
It is obvious that this is the Gross-Pitaevskii (GP) equation
\cite{Pitaevskii,Gross1,Gross2} which describes
the low energy excitation above the Bose-Einstein condensate after
identifying
\beq
  \mu_b\equiv\dsp-\frac{a_T}{d},\quad%
  M_b\equiv\dsp2m\,\frac{d}{c},\quad%
  \frac{4\pi a_{sb}}{M_b}=\dsp\frac{b_0}{d^2},
\label{Bmass}
\eeq
where $\mu_b$, $M_b$, and $a_{sb}$ are the effective chemical potential,
the effective boson mass, and the effective scattering length
for $s$-wave boson-boson scattering.
In Fig.~\ref{bosonint}, we show the schematic Feynman graph which
contributes to $a_{sb}$.
When the temperature goes below $T_c$, $\mu_b$ becomes positive
indicating the instability to formation of the Bose-Einstain condensate.

We see that the fluctuation becomes propagating in the BEC regime as in
the nonrelativistic fermion system \cite{melo93}.
This is because a gap $(2m-2\mu)$ appears between the continuum
excitation and $\omega=0$ due to the formation of bound state.
Therefore, the low energy fluctuation about the equilibrium cannot 
decay in this region; the fluctuation with wavenumber $k$ can decay
only via the nonlinear term, \ie, the scattering among soft modes.

\begin{figure*}[tp]
 \begin{minipage}{0.49\textwidth}
  \includegraphics[width=0.9\textwidth,clip]{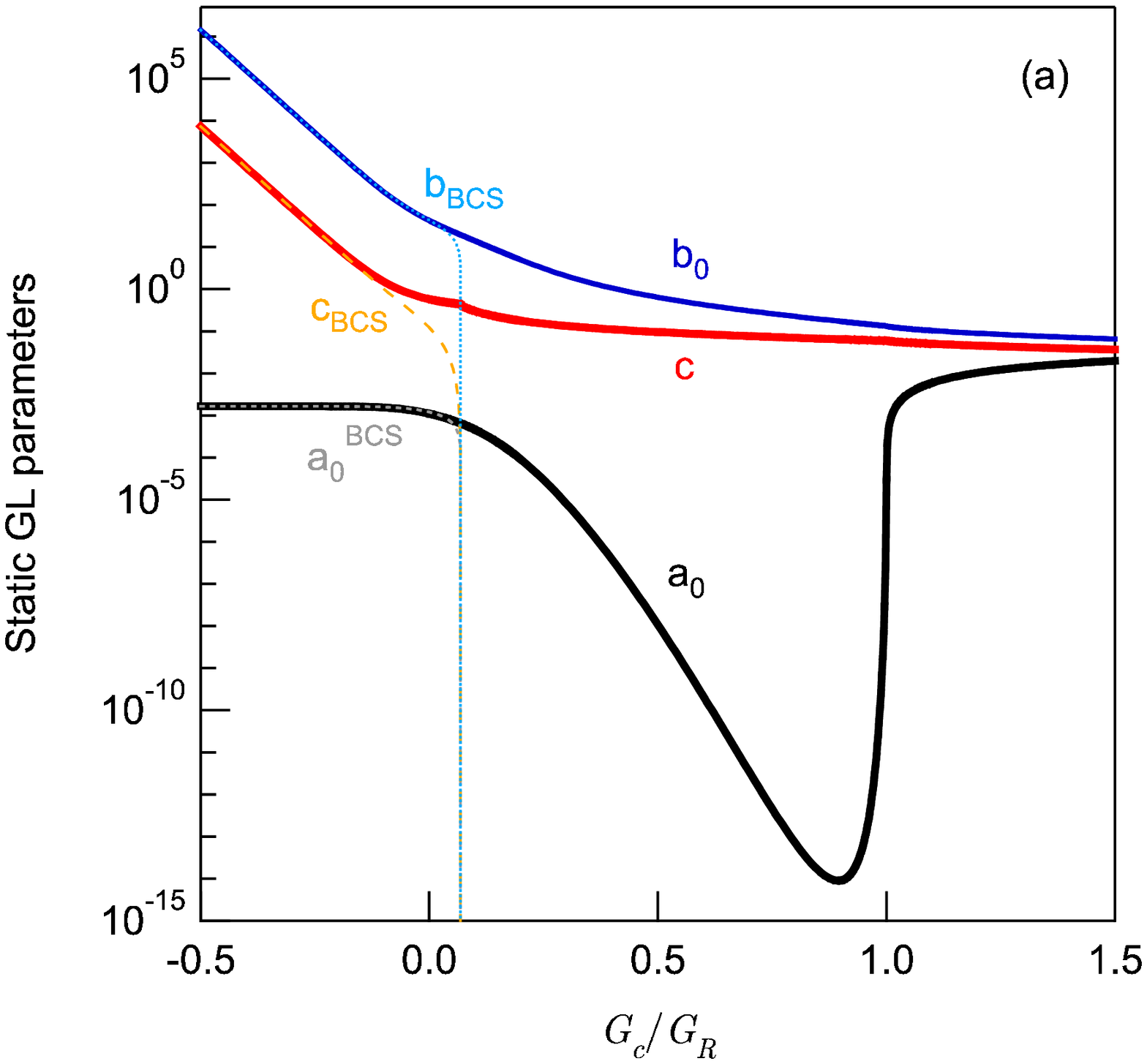}
 \end{minipage}%
 \hfill%
 \begin{minipage}{0.49\textwidth}
  \includegraphics[width=0.9\textwidth,clip]{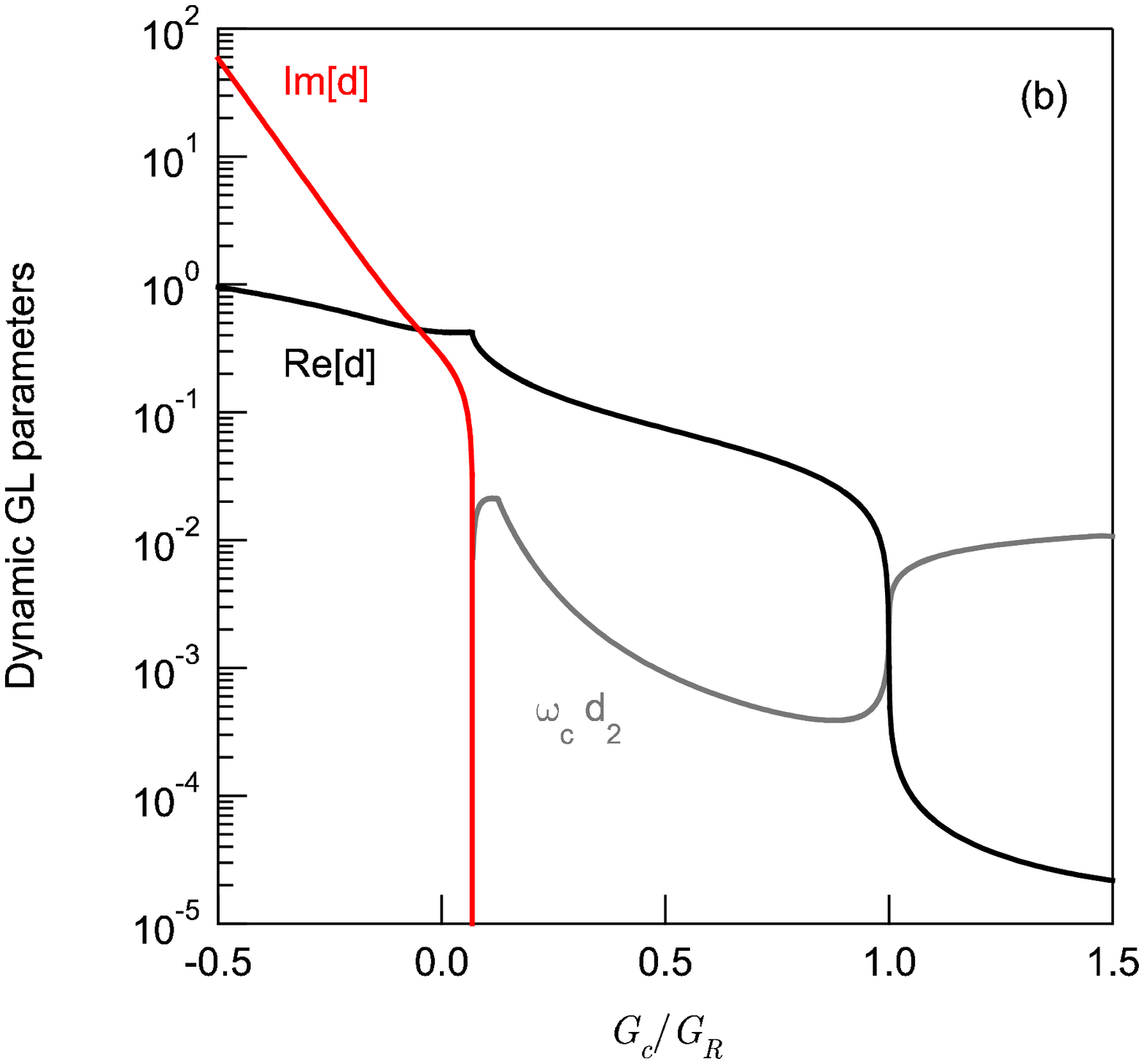}
 \end{minipage}%
  \caption[]{%
  {\bf(a)}~~The static GL parameters as a function of $G_R^{-1}$. 
  $a_0$, $c$, and $b$
  are the mass-squared parameter, the diffusion 
  constant, and the interaction parameter. 
  $a_0^{\rm BCS}$, $c^{\rm BCS}$, and $b^{\rm BCS}$
  are the weak coupling analytic approximations $(\mu_c\gg T_c)$
  of $a_0$, $c$, and $b$.
  {\bf(b)}~~The dynamic GL parameters. $d$ ($d_2$)
  is the coefficient for the first (second) time-derivative of
  the TDGL equation. 
  $\omega_c\equiv{\rm min.}(T_c,|\mu_c-m|)$
  is the limit energy such that the low energy GL expansion
  is valid for $\omega\ll\omega_c$.
  }
  \label{GLparams}
\end{figure*}

We now study the situation $|d\omega|\ll d_2\omega^2$;
we will see that, in the RBEC phase, there actually exists such an
``intermediate'' frequency regime $d/d_2\alt|\omega|$
where the expansion is still valid $(\omega\ll\omega_c\equiv{\rm
min.}\,(T_c,|\mu-m|))$.
In such regime, we find $d_2$ becomes pure real and positive.
By rescaling field $\Psi(t,\bfm{x})=\Sqrt{d_2}\Delta(t,\bfm{x})$, 
we obtain the relativistic Gross-Pitaevskii (RGP) equation (or simply
the Klein-Gordon equation with $\Phi^4$ interaction)
\cite{Fukuyama:2005jq}:
\beq
  \left[\partial_t^2-v_s^2\nabla_{\bfsm{x}}^2%
  +M^2+\lambda\sum_\xi|\Psi_\xi(t,\bfm{x})|^2\right]\Psi_\eta(t,\bfm{x})%
  =0,
\eeq
where $v_s$, $M^2$ are the velocity and the effective mass of soft
mode. $\lambda$ represents the two body repulsive force between soft modes. 
They are
\beq
  v_s^2\equiv\frac{c}{4m d_2},\quad M^2\equiv\frac{a_T}{d_2},%
  \quad\lambda\equiv\frac{b_0}{d_2^2}.
\label{vs}
\eeq
When the temperature goes below $T_c$, $M^2$ becomes negative.
The soft mode then becomes tachyonic without nonlinear terms, 
which signals the instability to the formation of nonzero Bose-Einstein
condensate $\langle\Psi(0,\bfm{0})\rangle\ne 0$.

To see which region of $(\omega,k)$-space is governed by the RGP (or GP)
equation, we look at the dispersion relation of the fluctuation in the
(R)BEC region.
The dispersion is given by the solution of
\beq
  -d\omega-d_2\omega^2+\frac{ck^2}{4m}+a_T=0,
\eeq
At the critical temperature $(a_T=0)$, we find
\beq
 \omega_k=\frac{d}{2d_2}\left(\Sqrt{1+\frac{cd_2}{md^2}k^2}-1\right).
\label{fdisp}
\eeq
For a sufficiently small wavenumber $k\ll\Sqrt{\frac{md^2}{cd_2}}$, 
\beq
 \omega_k\sim\frac{k^2}{2M_b},
\label{nrdisp}
\eeq
where $M_b$ is defined in \Eqn{Bmass}. 
This is the dispersion of nonrelativistic particle with velocity
$v=k/M_b$.
In the opposite case, the dispersion is approximated as
\beq
 \omega_k\sim v_s k,
\label{rldisp}
\eeq
with $v_s$ defined by \Eqn{vs}.
This is the dispersion of phonon with velocity $v_s$.
We note that as $d_2$ becomes large such that $d_2\gg d^2 m/c$, 
the momentum region where the dynamics of fluctuation is described by
the RGP equation becomes wide. 
This is the case when we increase the attraction between quarks as we
will see later.

In \Fig{GLparams}(a) and (b) show the numerical results for the
static and dynamic GL parameters from the weak coupling BCS regime
to the ultra-strong coupling RBEC regime.
The weak coupling approximation for these parameters are also indicated
by several thin dashed lines.
One can see the drastic change going from the BCS to the RBEC
particularly in the dynamic GL parameters.
From these results, we shall discuss the character change of the soft
mode, and transport properties.

\subsection{How does the static property of the soft mode 
evolve with the crossover?}
\begin{figure}[tp]
  \includegraphics[width=0.45\textwidth,clip]{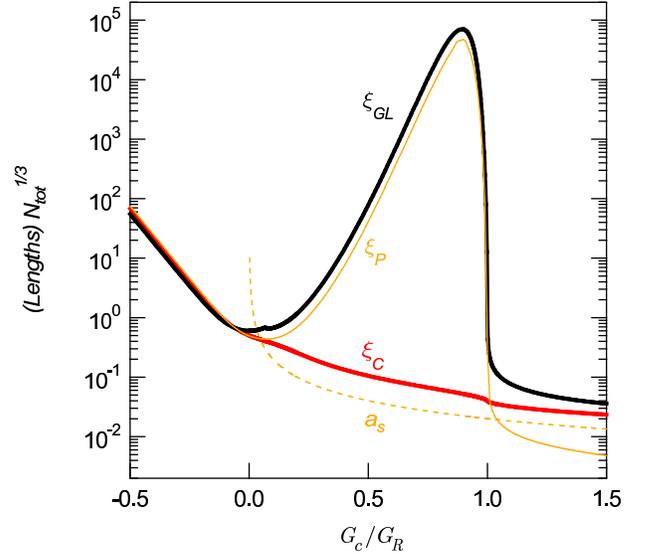}
  \caption[]{%
  The GL coherence length $\xi_{\rm GL}$ and the Cooper pair
  size $\xi_c$ evaluated on $(\mu_c(G_R),T_c(G_R))$.
  The Pippard length is shown by the line indicated by $\xi_p$. $a_s$
  is the scattering length for $s$-wave quark-quark scattering. 
  These are plotted in the unit of the mean inter-fermion 
  distance $N_{\rm tot}^{-1/3}$.
  }
  \label{size}
\end{figure}

We first focus on the GL coherence length \Eqn{xiGL}.
As noted, the static disturbance to the system restores at this healing
length.
Also the anomalous heat capacity $\delta C_{\rm a}$ per unit volume is
controlled by this length as
\beq
  \frac{\delta C_{\rm a}}{V}=\frac{N_c(N_c-1)}{2}%
  \frac{1}{8\pi\xi_{\rm GL}^3}t_r^{-1/2}.
\eeq
Furthermore, this quantity is known to appear in the intensity of 
the diverging conductivity at $T=T_c$ \cite{Aslamazov,Maki,Larkin}.
\beq
  \sigma_{\rm AL}\sim \frac{g^2}{\xi_{\rm GL}}t_r^{-1/2}.
\eeq
where $g$ is the gauge coupling.

In \Fig{size}, we show the GL coherence length $\xi_{\rm GL}$ as a
function of $G_R^{-1}$.
The Cooper pair size $\xi_c(T_c)$ and the Pippard length $\xi_p$ are
also shown.
In the weak coupling region, all these scales exhibit almost the same
behaviour. 
In particular, the weak coupling ``universal relation'' $\xi_{\rm
GL}/\xi_p=0.74$ (see \Eqn{uni2}) is satisfied in good accuracy.
However, once one goes into the BEC region, these scales start to
deviate. 
First, the pair size $\xi_c$ becomes smaller than the mean inter-fermion
distance, which supports the picture of independent bosons.
In this regime, the pair size $\xi_c$ is approximated by
the scattering length $a_s$ rather than $\xi_p$ as noted before.
In contrast, the GL coherence length $\xi_{\rm GL}$ takes a minimum
between the BCS and BEC regimes as nonrelativistic calculation \cite{melo97}.
On the other hand, the behaviour of $\xi_{\rm GL}$ is approximated
by the Pippard length $\xi_p\sim v_F/\De_0$ which is defined in
the weak coupling regime though.
This fact suggests that, the gap parameter not only has the role of the
order parameter but also plays a physical role to determine the response
to the static disturbance. 

\vspace*{1ex}
\noindent
{\bf Ginzburg-Levanyuk region:}~%
The above derived low energy effective theories may fail in describing
the dynamics of soft modes when $T$ is very close to $T_c$, \ie, 
they cannot be applied in the {\em critical region}.
In this critical region the thermodynamic quantities diverge
with the anomalous power law.
We can estimate this region by Ginzburg-Levanyuk criterion
\cite{ref:Leva59,ref:Ginz60}.
We first define the order parameter averaged in the 
sphere $V_{T}=4\pi(\xi_{\rm GL}/\Sqrt{t_r})^3$,
a characteristic size of static fluctuation:
\beq
  \Psi_\eta=\frac{1}{V_T}\int_{V_T} d\bfm{x}%
  \Delta_\eta(\bfm{x}).
\eeq
Then we estimate the magnitude of the thermal average of
fluctuation in the gaussian approximation.
\beq
\ba{rcl}
 \langle{\Psi}_\eta^*\Psi_\xi\rangle_{\rm th}&=&%
 \frac{\delta_{\eta\xi}}{V_T}\int_{V_T}%
 d\bfm{x}\langle\Delta^*_1(\bfm{x})\Delta_1(\bfm{0})%
 \rangle_{\rm th}\\[2ex]
 &=&\delta_{\eta\xi}\frac{m T}{c}\frac{1-1/e}{\pi\xi_{\rm GL}}\Sqrt{t_r}.
\ea
\eeq
If $a_T\langle\Psi^*_\eta\Psi_\eta\rangle_{\rm th}$ becomes smaller
than the quartic term $\frac{b}{2}|\Psi_\eta|^2|\Psi_\xi|^2$, then
the mean field approximation fails due to higher terms in the soft mode
action; this condition is
\beq
  t_r=\frac{T-T_c}{T_c}%
  \alt\frac{N_c(N_c-1)}{2}\frac{b_0 T_c(1-1/e)}{8a_0^2\xi_{\rm GL}^3}.
\eeq
When the temperature is in the above range, the Ginzburg-Landau
approximation fails to describe the anomalous divergence of the
thermodynamic quantities like the heat capacity
\cite{Kitazawa:2005vr}.
We must note that the critical region estimated by above formula 
is not small in the RBEC regime, which means we need to go beyond
the quartic approximation in order to describe the soft mode
there in the quantitative level.

\subsection{What about the dynamic nature of fluctuation?}
\begin{figure*}[tp]
 \begin{minipage}{0.49\textwidth}
  \includegraphics[width=0.9\textwidth,clip]{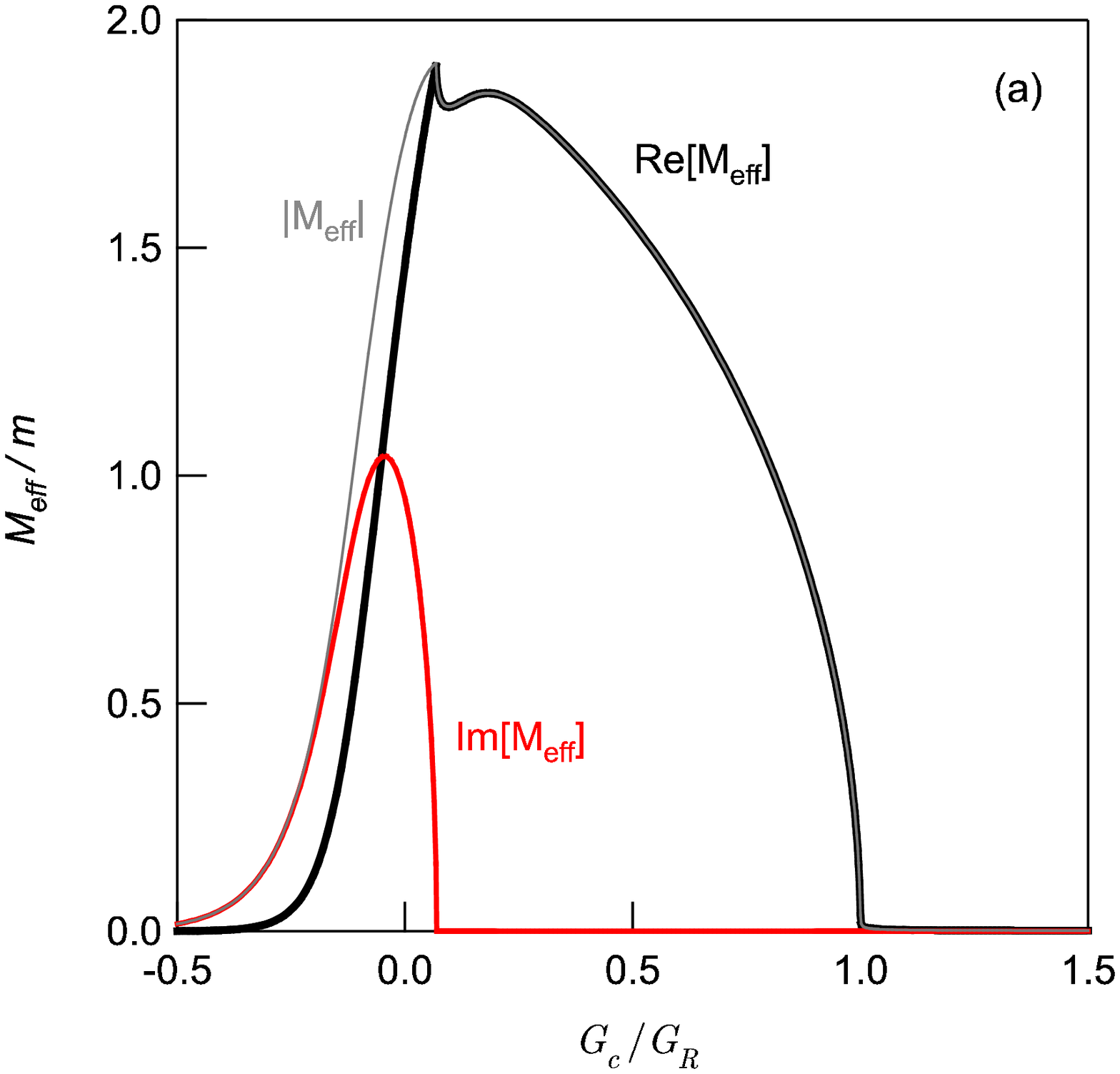}
 \end{minipage}%
 \hfill%
 \begin{minipage}{0.49\textwidth}
  \includegraphics[width=0.9\textwidth,clip]{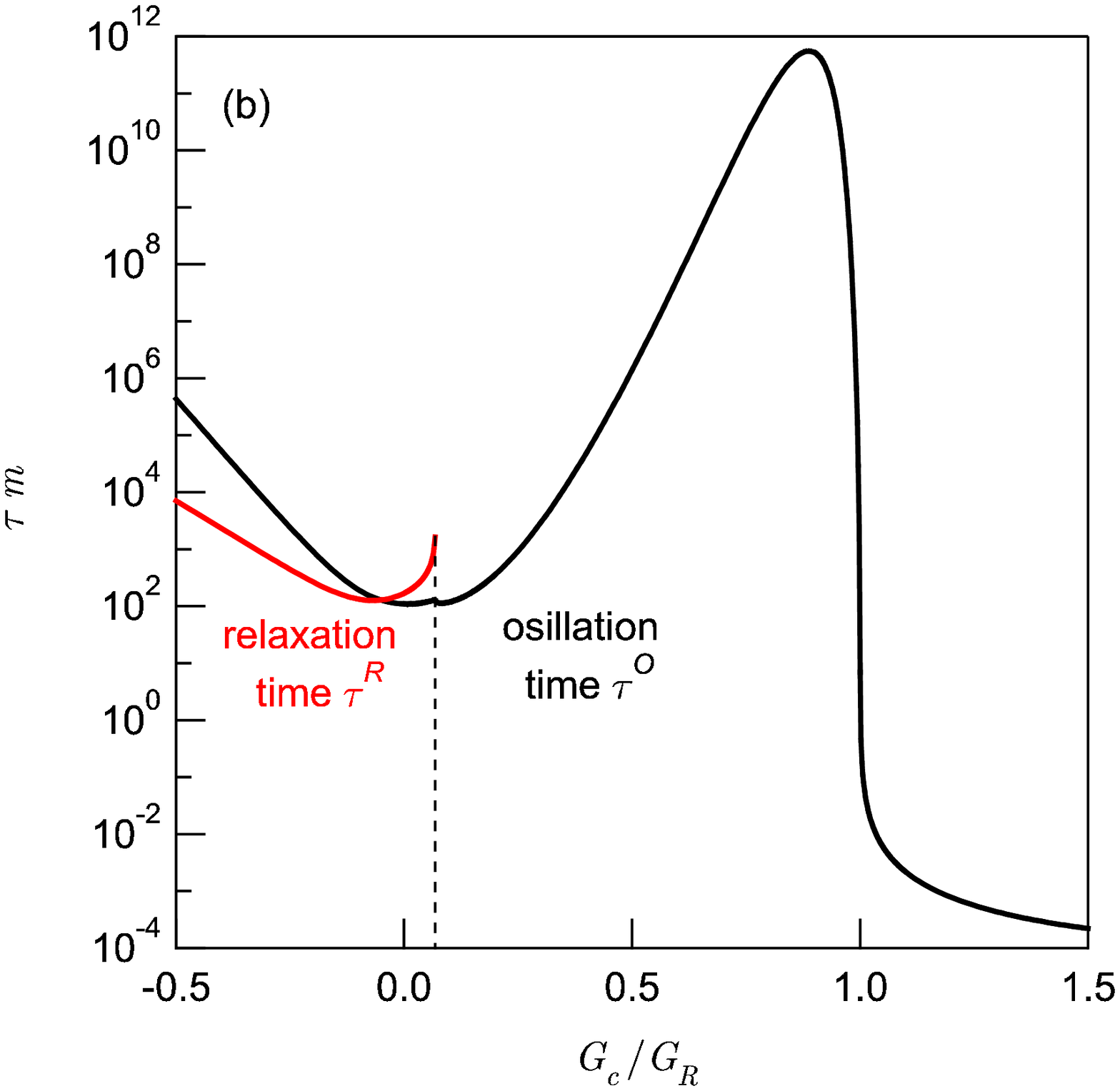}
 \end{minipage}%
  \caption[]{%
  {\bf(a)}~~The complex effective mass as a function of $G_R$, which
  determines the time dependence of the fluctuation with finite $k$ at
  $T\cong T_c$.
  {\bf(b)}~~The oscillation time $\tau^{O}$ and the relaxation 
  time $\tau^{R}$ for the fluctuation in the long wavelength limit.
  }
  \label{dynamicresponces}
\end{figure*}
We now look at the dynamic property.
We first ignore the effect of $d_2$ looking at the low frequency regime.
Then the complex pole ($\omega_{\rm pole}$) of the dynamic susceptibility
in the TDGL approximation should be given by
\beq
\ba{rcl}
  \omega_{\rm pole}&=&\dsp\frac{ck^2}{4m d}+\frac{a_T}{d}
        =\dsp\frac{k^2}{2M_{\rm
	eff}}-i\frac{1}{\tau^{O} t_r^{-1}}%
	+\frac{1}{\tau^{R} t_r^{-1}}.
\ea
\label{effmass}
\eeq
where the effective (complex) mass $M_{\rm eff}$, the relaxation
time $\tau^{R}$, and the oscillation time $\tau^{O}$ are defined by
\beq
\ba{rcl}
  M_{\rm eff}&=&\dsp 2m\frac{d}{c},\quad%
  \tau^R=\frac{|d|^2}{a_0\,\Im d},\quad%
  \tau^O=\frac{|d|^2}{a_0\,\Re d}.\\[2ex]
\ea
\eeq
In the BEC region, $M_{\rm eff}$ is real and coincides with effective
boson mass $M_b$ defined in \Eqn{Bmass}. 
For the stability that the fluctuation about the thermal equilibrium is
not tachyonic, $\Im\omega_{\rm pole}<0$ should be satisfied. 
This requires both $T\agt T_c$ and $\Im d\agt0$, and our numerical
result shows that the latter is actually satisfied
(see \Fig{GLparams}(b)).

The fluctuation with a wavenumber $k$ behaves as
\beq
  \Delta_\eta(t,k)\sim \exp\left(-i\frac{tk^2}{2M_{\rm
  eff}}-i\frac{t}{\tau^{R} t_{r}^{-1}}+\frac{t}{\tau^{O}
  t_r^{-1}}\right).
\label{decay}
\eeq
For $T\cong T_c$, the fluctuation looks like
$\De_\xi(t,k)|_{T=T_c}=e^{-t/\tau^D_k}\sin(t/\tau^P_k+{\rm const.})$ 
where $\tau_k^D=\frac{2|M_{\rm eff}|}{k^2}\frac{1}{\Im M_{\rm eff}}$ 
and $\tau_k^P=\frac{2|M_{\rm eff}|}{k^2}\frac{1}{\Re M_{\rm eff}}$ 
are the diffusion and propagating times.
Thus the complex effective mass determines whether the fluctuation
with a finite momentum $k$ is diffusive or propagating.
In contrast, $\tau^R$ and $\tau^O$ determine how the long wavelength
limit of the fluctuation behaves for $T\agt T_c$. 
In this case, the fluctuation looks like as $\Delta_\eta(t,0)\sim
e^{-t/\tau^{R}t_r^{-1}}\sin(t/\tau^{O}t_r^{-1}+{\rm const.})$
where $\tau^{R}$ and $\tau^O$ are the relaxation and oscillation times. 

\Fig{dynamicresponces}(a) shows the complex effective mass $M_{\rm eff}$
as a function of $G_R$, and (b) shows the relaxation time $\tau^{R}(G_R)$
and the oscillation time $\tau^{O}(G_R)$. 
We can see from figure (a) that the fluctuation is diffusive
($\tau^P_k\gg\tau^D_k$) in the BCS regime,
while in the BEC regime, the pair fluctuation is of propagating 
($\tau^P_k\ll\tau^D_k$) because of the reality of boson \cite{melo93}.
In contrast to the nonrelativistic calculation \cite{melo93,Haussmann},
$|M_{\rm eff}|\ne 2m$
in the strong coupling limit.
It decreases due to the binding effect as the coupling becomes stronger. 
Also the long wavelength limit of the pair excitation is
of overdamped in the BCS side ($\tau^R\ll\tau^O$) while it is 
oscillating mode in the BEC side ($\tau^R=\infty$) because of $\Im
d=0$. This is again attributed to the bound state gap.

\subsection{Relativistic Gross-Pitaevskii equation}
\begin{figure*}[tp]
 \begin{minipage}{0.49\textwidth}
  \includegraphics[width=0.9\textwidth,clip]{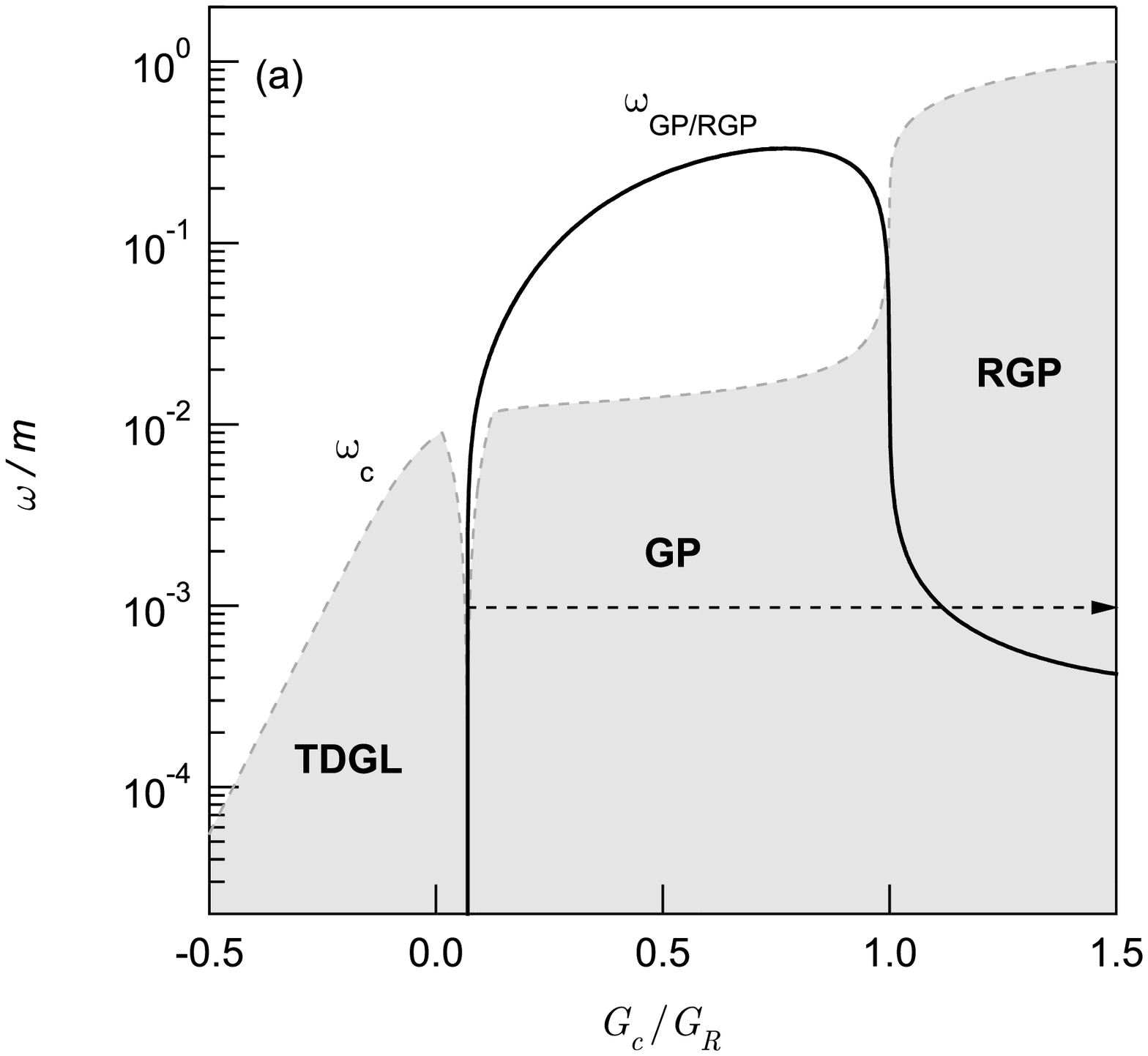}
 \end{minipage}%
 \hfill%
 \begin{minipage}{0.49\textwidth}
  \includegraphics[width=0.9\textwidth,clip]{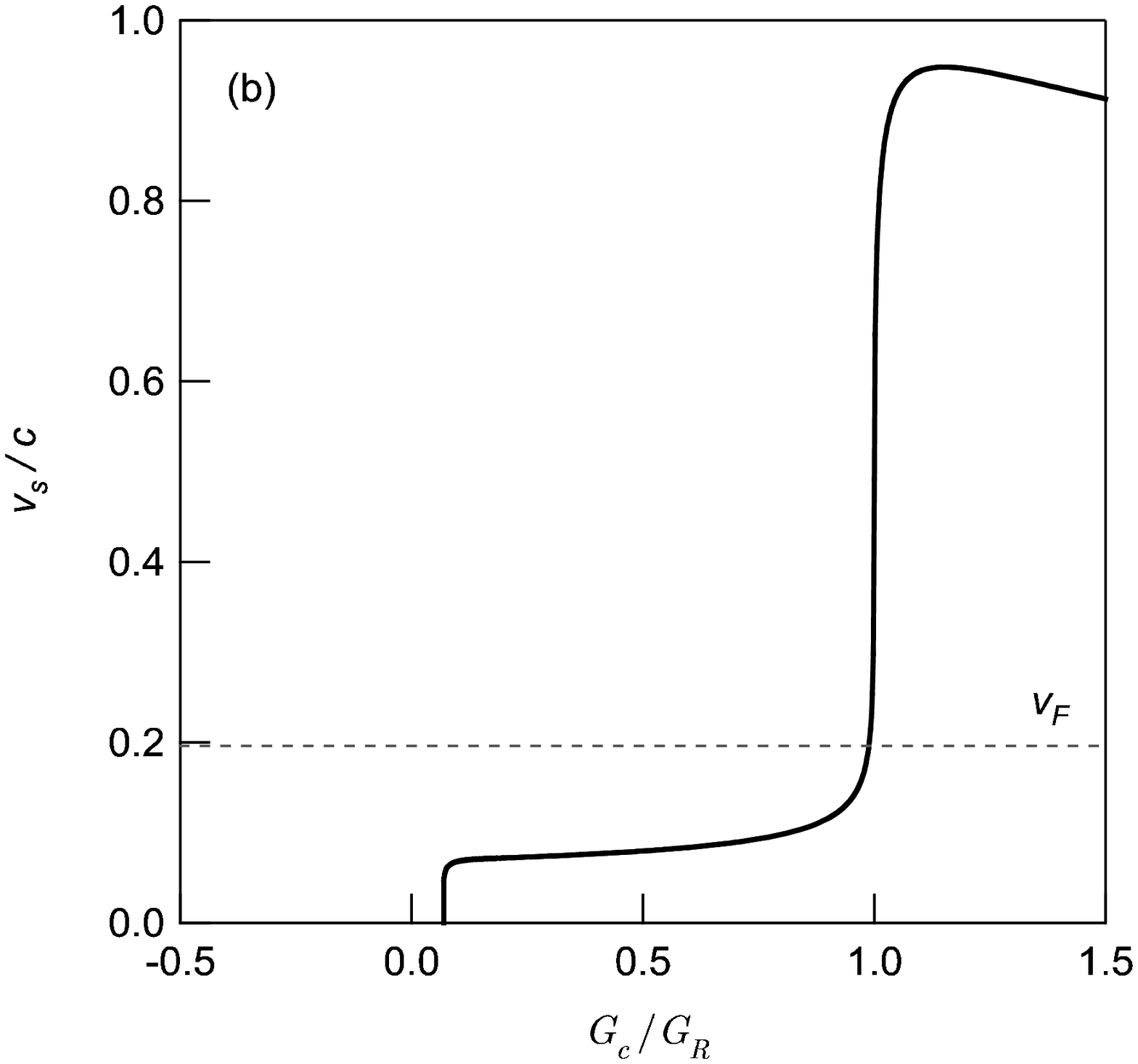}
 \end{minipage}%
  \caption[]{%
  {\bf(a)} The three regions where the fluctuation is described by TDGL,
  GP, and RGP equations.
  The low energy expansion is allowed for the shaded area 
  where $\omega<\omega_c\equiv{\rm min.}(T_c,|\mu_c-m|)$.
  {\bf(b)} The group velocity of the fluctuation (boson), 
  $d\omega_k/dk$,
  evaluated along the light dashed arrow in (a). 
  }
  \label{TDGL}
\end{figure*}
We have shown that there may exist the three distinct-$(\omega,k)$
regions where the fluctuation is described by the TDGL, GP, and RGP
equations.
\Fig{TDGL}(a) shows where in the $\omega$-region these three
equations govern the dynamics of soft mode.
The shaded area satisfies the 
condition $\omega<\omega_c\equiv{\rm min.}(T_c,|\mu_c-m|)$ where the
low energy expansion is valid.
As in the nonrelativistic case \cite{melo93}, the validity
breaks down in the very vicinity of the BCS-BEC crossover due
to $|\mu_c-m|\to 0$. The solid line indicated by $\omega_{\rm GP/RGP}$ 
divides the $\omega$-space into two pieces, one in which the 
GP equation applies, and the other where the RGP equation applies.
We determined this boundary by
\beq
  \omega_{\rm GP/RGP}=\frac{(\Sqrt{2}-1)d}{2d_2},
\eeq
which is evaluated by equating $\frac{cd_2}{md^2}k^2$ in \Eqn{fdisp} to
unity; if $\omega\gg\omega_{\rm GP/RGP}$ the dispersion of the
fluctuation is approximated by \Eqn{rldisp}, and in the opposite
case with $\omega\ll\omega_{\rm GP/RGP}$, 
it is approximated by the nonrelativistic dispersion, \Eqn{nrdisp}.
We see that the $\omega$-region where the RGP equation describes
the fluctuation rapidly grows when the system goes into the RBEC region.
This is obviously because boson is light in the RBEC region.

\Fig{TDGL}(b) shows the group velocity of the fluctuation
mode, \ie, $v_s=\frac{d\omega_k}{dk}$ (with $\omega_k$ defined by
\Eqn{fdisp}) evaluated along the dashed arrow in figure (a).
The velocity of fluctuation is smaller than the Fermi velocity $v_F$
and is almost independent of $G_R$ in the BEC region.
It rapidly grows at $G_c/G_R\sim1$ where the system goes into the RBEC
phase; the speed $v_s$ becomes comparable to the speed of light; 
this is also consistent with the fact that the boson is light in
this regime.

\subsection{Interaction between soft modes}
\begin{figure}[tp]
  \includegraphics[width=0.45\textwidth,clip]{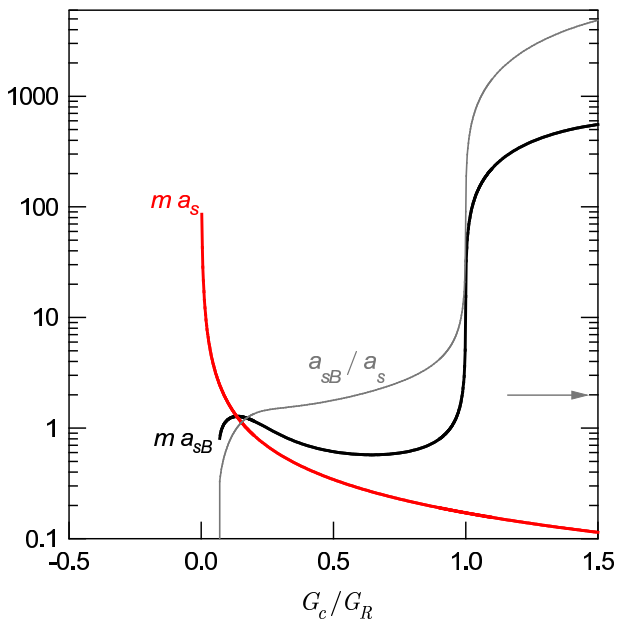}
  \caption[]{%
  The fermion-fermion scattering length $a_s=mG_R/4\pi$, 
  and the boson-boson scattering length $a_{sb}$.
  The arrow indicates $2$, the strong coupling limit which the ratio
  $a_{sb}/a_{s}$
  approaches in the nonrelativistic calculation \cite{melo93}.
  }
  \label{bint}
\end{figure}
As we have discussed, there are multi-body interaction among soft
modes. 
To the lowest order beyond the gaussian approximation, we found
the repulsive two-body interaction which causes a binary collision. 
From \Eqn{Bmass}, the scattering length for this collision is given by
\beq
 a_{sb}=\frac{mb_0}{2\pi cd},
\label{bsl}
\eeq
which determines the scattering cross section in the soft limit $(s\to
4M_b^2)$ in the C.M. frame as
\beq
 \sigma_b=4\pi a_{sb}^2=\frac{m^2 b_0^2}{\pi c^2d^2}.
\eeq

In Fig.~\ref{bint}, we show the fermion-fermion scattering 
length $m a_s$ and the boson-boson scattering length $m a_{sb}$ as a
function of $G_R^{-1}$. 
Also its ratio $a_{sb}/a_s$ is depicted.
When $G_R^{-1}$ is increased beyond the BCS/BEC boundary, 
the boson scattering length first decreases, \ie, 
the weaker boson-boson repulsion in the stronger fermion-fermion
attraction \cite{melo93}. 
This is because the boson-boson interaction is microscopically induced
only via the dissociation of boson in the intermediate state as shown
in Fig.~\ref{bosonint};
it becomes harder to dissociate the boson in the stronger coupling.
When $G_R^{-1}$ is further increased and the RBEC is approached, the
boson-boson scattering length rapidly increases, 
and the nonrelativistic limit value of $a_{sb}/a_{s}$, $2$
\cite{melo93}, is significantly exceeded \footnote{%
  In contrast to the approach taken in \cite{melo93},
  the study of 4-body Schr\"odinger equation produces $a_B\sim 0.6 a_s$
  near the unitarity \cite{petrov}.
  In addition, the renormalization group approach yields the similar
  value also in the strong coupling regime \cite{ohashi}.
  Thus it is clear that, for the quantitative understanding of the dimer
  interactions, we need to go beyond the current approximation taking 
  into account how higher energy 2-boson processes renormalize
  the low energy scattering length. We defer this to future study.
}.
This is because the temperature is large ($T_c>2\mu_c$) in the RBEC
regime, which favors dissociation. 

\subsection{Smooth change of transport properties}
We here discuss some of the transport properties on the basis of 
the dynamic equations obtained in the previous section.
In the nonrelativistic superconductor, it is known that the fluctuating
pair fields affect the electric conductivity above $T_c$
\cite{Aslamazov,Maki,Larkin}.
In the context of two-flavor color superconductivity, it is noted 
in \cite{Kitazawa:2001ft} that these soft modes may affect not only
transport coefficients but also the dilepton spectrum.

We here focus on the shear viscosity near the critical temperature and
derive approximate formulas for the viscosities in three regions.
To do this, we employ the semi-classical Kubo's formula and also the
Boltzmann equation for soft modes in the sprit that the short range
quantum effects are already integrated out in the low-energy
coefficients in the soft mode action.
For this to be correct, soft modes should be only dilutely
distributed in the phase space such that they can be described by
the classical effective theory.
Technically, this means that we only take the ``thermal average'' of
the soft mode distribution in the Kubo's formula.
We will make a short statement regarding the validity of this treatment
in the end of this section.

\begin{figure}[tp]
  \includegraphics[width=0.38\textwidth,clip]{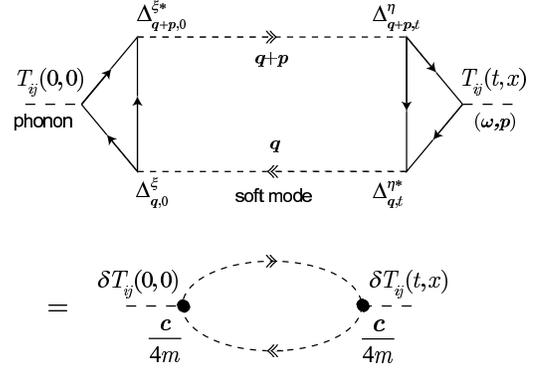}
  \caption[]{%
  The microscopic process which contributes to the shear viscosity
  in the BCS region ($\mu_c>m$). 
  The same graph expresses the
  Aslamazov-Larkin type correction to the conductivity if the
  vertex $T_{ij}(t,\bfm{x})$ is replaced by the paramagnetic current 
  \cite{Kitazawa:2005vr}.
  In our TDGL approach, a phonon couples to two soft modes by
  the TDGL coefficient ($c/4m$).
  }
  \label{Ablamasov}
\end{figure}

\vspace*{1ex}
\noindent
{\bf Shear viscosity due to the soft mode 
diffusion in the BCS regime:}~%
We now look at the precursory soft mode contribution to the shear
viscosity, $\eta_s$, in the BCS regime. 
In the rotationally symmetric system considering here, the shear
viscosity is formally given by the quantum Kubo's formula
\cite{Hosoya:1983id},
\beq
 \eta=\lim_{\omega\to0}\frac{1}{\omega}\int
 d\bfm{x}\int_0^{\infty}dte^{i(\omega+i\delta)t}%
 \langle[\hat{T}_{xy}(\bfm{x},t),\hat{T}_{xy}(\bfm{0},0)]\rangle,
\label{kubo}
\eeq
where $\hat{T}_{ij}$ denotes the traceless part of the energy-momentum
tensor.
First thing we have to do is to derive the classical approximation
($\omega\ll T$) of the Kubo's formula \Eqn{kubo}. 
From the spectral representation of the real part of
\Eqn{kubo}, we find for $\omega\ll T$,
\beq
\ba{rcl}
  \Re\eta%
  &\cong&\dsp\beta\int d\bfm{x}%
  \int_0^\infty%
  dt\,\langle\hat{T}_{xy}(t,\bfm{x})\hat{T}_{xy}(0,\bfm{0})\rangle.
\ea
\eeq
Note that we have no longer the commutation in the bracket. 
We now replace $\hat{T}_{xy}$ by the classical stress tensor and the
expectation by the thermal average looking only at the contribution 
from soft modes (with $\omega\ll T$) to the viscosity $\eta_s$.
For this purpose, we first extract the soft mode contribution to
the stress-momentum tensor which we denote by $\delta T_{ij}$. 
The stress arising from soft modes is
\beq
\ba{rcl}
  \delta T_{ij}(t,x)&=&-\delta_{ij}f_{\rm eff}(\Delta,\Delta^*)\\[2ex]
  &&+\frac{c}{4m}\left[%
  \partial_i\Delta_\eta^*(t,x)\partial_j\Delta_\eta(t,x)%
  +(i\leftrightarrow j)\right]
\ea
\eeq
This is followed by calculation of the free energy shift $\delta F_{\rm
eff}=-\int d\bfm{x}\delta T_{ij}\frac{\partial}{\partial x_j}u_i(x)$
under the small distortion $x_i\to x_i'=x_i+u_i(\bfm{x})$.
The shear viscosity from fluctuating pair fields is then given by
\begin{widetext}
\beq
\ba{rcl}
 \eta_s%
  &=&{\dsp\beta}\dsp\lim_{\bfsm{p}\to\bfsm{0}}%
     \int_0^\infty dt\frac{1}{Z_{\rm eff}}\int%
     {\mathcal D}\Delta%
     {\mathcal D}\Delta^*%
     \,e^{-\beta F_{\rm eff}(\Delta,\Delta^*)} %
  \delta T_{xy}(t,\bfm{p})\delta T_{xy}(0,-\bfm{p})\\[2ex]
  &=&\dsp{\beta}\left(\frac{c}{4m}\right)^2%
     \lim_{\bfsm{p}\to\bfsm{0}}%
     \int\frac{d\bfm{\bfm{q}}}{(2\pi)^3}(2q_xq_y+q_xp_y+q_yp_x)^2%
     \int_0^\infty dt%
     \langle\Delta^\eta_{\bfsm{q}+\bfsm{p},t}%
     \Delta^{\xi*}_{\bfsm{q}+\bfsm{p},0}\rangle_{\rm th}%
     \langle\Delta^{\eta*}_{\bfsm{q},t}\Delta^\xi_{\bfsm{q},0}%
     \rangle_{\rm th}.\\[2ex]
\ea
\label{ooo}
\eeq
\end{widetext}
Here $\langle\cdots\rangle_{\rm th}$ denotes the thermal
average at $t=0$, \ie, 
$\frac{1}{Z_{\rm eff}}\int{\mathcal D}\Delta{\mathcal D}\Delta^*(\cdots)$.
The diagrammatic interpretation for the above formula is given
in Fig.~\ref{Ablamasov}.
To proceed further, we use the standard ansatz that the time evolution
of the fluctuation is determined by the classical TDGL equation in the
gaussian approximation;
\beq
  \dsp\Delta_{\bfsm{q},t}=U_{\bfsm{q},t}\Delta_{\bfsm{q},0}%
 =e^{-i\frac{ct}{4md}(\bfsm{q}^2+\xi_{\rm GL}^{-2}t_r)}\Delta_{\bfsm{q},0}.
\eeq
The gaussian approximation to $F_{\rm eff}[\Delta,\Delta^*]$ is also
 essential in order for the replacement of the expectation of $\Delta^4$
 by the product of the two body Green's functions in \Eqn{ooo} to be valid.
By taking the identity
$\langle\Delta_{\bfsm{q},0}^{\xi}\Delta_{\bfsm{q},0}^{\eta*}\rangle_{\rm
th}=\frac{4m  T}{c}\frac{\delta_{\eta\xi}}{\bfsm{q}^2+\xi_{\rm
GL}^{-2}t_r}$ into account, we obtain
\beq
\ba{rcl}
 \eta_s&=&\dsp N_c(N_c-1){T}\frac{4m\,|d|^2}{c\,\Im[d]}%
  \int^{q_c}\!\!\frac{d\bfm{\bfm{q}}}{(2\pi)^3}\frac{q_x^2q_y^2}{(\bfm{q}^2%
 +\xi_{\rm GL}^{-2}t_r)^3}\\[2ex]
 &\to&%
  \dsp\frac{N_c(N_c-1)}{30\pi^2}\frac{T_c\tau^R}{\xi_{\rm GL}^2/k_F},%
  \quad\mbox{(as $T\to T_c,\,q_c=k_F$)}.
\ea
\eeq
where we have introduced the UV cutoff $q_c$ and set $q_c=k_F$. 
Unlike the conductivity which diverges as $T\to T_c$ 
like $\frac{g^2}{\xi_{\rm GL}\Sqrt{t_r}}$ \cite{Aslamazov,Maki}, 
the shear viscosity is regular.
Also the shear viscosity is proportional to the soft mode relaxation
time; the transport coefficients are always proportional to the
relaxation time (mean free time) of the quasi-particle according to
the relaxation time approximation to the Boltzmann equation.
Because of the divergence of the relaxation time $\tau^R\to\infty$
at the BCS-BEC crossover point $\mu=m$, this formula fails to
describe the shear viscosity in the BEC regime. 
We thus try the other approach to derive the shear viscosity in the BEC
region.

\vspace{1ex}
\noindent
{\bf Shear viscosity via boson binary collision in the BEC regime:}~%
In the BEC regime, the soft fluctuation about the equilibrium cannot
decay via the quantum diffusion because of the bound state gap.
In reality, however, the soft mode with momentum $k$ decays
via the nonlinear term in the GP equation, \ie, the binary collision of
soft modes.
Here, we simply apply the result from the classical Boltzmann equation
for dilute gas of a particle with a hard sphere $r=\Sqrt{2}a_{sb}$.
\cite{Reichl,Chapman,BLTZ},
\beq
\ba{rcl}
 \eta_s&=&\frac{5\pi}{32\Sqrt{2}}n_{\rm soft}\langle M_{\rm
 eff}v\rangle_{\rm
 th}l_{\rm mfp}\\[2ex]
 &\sim&0.35\times n_{\rm soft}\langle M_{\rm
 eff}v\rangle_{\rm th}l_{\rm mfp}.\\[1ex]
\ea
\eeq
$n_{\rm soft}$ 
is the soft mode density, and $l_{\rm mfp}$ is the mean free path.
$\langle M_{\rm eff}v\rangle_{\rm th}$
is the average momentum of thermally excited soft modes. 
These are given by
\beq
\ba{rcl}
 l_{\rm mfp}&=&\dsp\frac{1}{n_{\rm soft}\sigma_{sb}}=\frac{\pi
 c^2d^2}{n_{\rm soft}m^2b_0^2},\\[3ex]
 \langle M_{\rm eff}v\rangle_{\rm th}%
 &=&\dsp\lambda_T^{-1}=\Sqrt{\frac{M_{\rm eff}T}{2\pi}}%
 =\Sqrt{\frac{mdT}{c\pi}}.\\[1ex]
\ea
\eeq
Combining the above results, we arrive at the formula
\beq
 \eta_s=\frac{5\Sqrt{2}}{256}\frac{1}{\lambda_T a_{sb}^2}%
 =\frac{5\pi^{3/2}}{32\Sqrt{2}}\frac{c^{3/2}d^{5/2}\Sqrt{T}}{m^{3/2}b_0^2}.
\label{etabec}
\eeq
This equation relates the low energy coefficients
in the GP equation to the shear viscosity.
We note that this formula is valid only for the classical (dilute)
limit, \ie, $n_{\rm soft}a_{sb}^3,n_{\rm soft}\lambda_T^3\ll1$, 
where $\lambda_T=\Sqrt{\frac{2\pi}{M_{\rm eff}T}}$ is the thermal
de-Broglie length, the quantum radius at $T$.

\vspace{1ex}
\noindent
{\bf
Shear viscosity via binary collision between light bosons
in the RBEC phase:}~%
The nonrelativistic formula we have derived above fails in the 
RBEC regime because of the lightness of boson.
The qualitative behaviour of the shear viscosity is 
\cite{Jeon:1994if,Jeon:1995zm}
\beq
\ba{rcl}
  \eta_s&\sim& l_{\rm mfp}\bar{v}\langle\ep+P\rangle_{\rm th}
  \sim l_{\rm mfp}\bar{v}T\langle s\rangle_{\rm th},
\ea
\eeq
where $s$ denotes the entropy density, and $\bar{v}$ is the average
velocity of soft modes. 
In the above formula, we ignored the chemical potential dependent part
of the pressure, \ie, $P=\mu n+Ts-\ep\to Ts-\ep$ with $\ep$ being the
energy density.
The relativistic thermal de-Broglie length is given by \cite{Yan}
\beq
  \lambda_{RT}=\frac{\pi^{2/3}}{T},
\eeq
and therefore, the soft mode dynamics is described by the RGP equation we
derived in the previous section. $\bar{v}$ 
and the scattering cross section for a collision between soft
modes is estimated as
\beq
  \bar{v}=\Sqrt{\frac{c}{4md_2}},\quad%
  \sigma_{sb}(s\sim T^2)\sim\frac{\lambda^2}{T^2}\sim\frac{b_0^2}{T^2d_2^4},
\eeq
with $\lambda$ being the $\Delta^4$ coupling (see \Eqn{vs}).
We obtain
\beq
  \eta_s\sim\frac{\bar{v}}{\sigma_{sb}}\frac{T s}{n_{\rm soft}},
\eeq
where we adopted the simplified notation $s=\langle s\rangle_{\rm th}$
for the equilibrium entropy density. $n_{\rm soft}$ again denotes the
density of thermal soft modes.
We further simplify the equation by estimating $s/n_{\rm soft}$ in
the present gaussian approximation for the soft mode action.
The thermodynamic potential coming from soft modes is
\begin{widetext}
\beq
\ba{rcl}
 \delta\Omega&=&\frac{N_c(N_c-1)}{4}T\sum_N\int\frac{d\bfsm{P}}{(2\pi)^3}%
 \sum_{\sigma=\pm}\log\left(-d_2(i\Omega_N+\sigma\mu)^2%
 +\frac{cP^2}{4m}+a_T\right)\equiv\delta\Omega_++\delta\Omega_-,\\[2ex]
\ea
\label{rbecs}
\eeq
where $\delta\Omega_\pm$ 
represents the bosonic (antibosonic) contribution to the partition function:
\beq
\ba{rcl}
  \delta\Omega_\pm=\frac{N_c(N_c-1)}{4}T\sum_N\int\frac{d\bfsm{P}}{(2\pi)^3}%
 \log\Big(d_2\Omega_N^2%
 +\Big(\Sqrt{\frac{cP^2}{4m}+a_T}+\sigma\Sqrt{d_2}\mu\Big)^2\Big).
\ea
\eeq
\end{widetext}
By differentiating the difference $(\delta\Omega_+-\delta\Omega_-)$ with
respect to $\mu$, we find the following formula for the soft mode
density in the limit $T\to T_c$ and $T_c\gg\mu$,
\beq
\ba{rcl}
 n_{\rm soft}&=&\left(\frac{d_2m}{c}\right)^{3/2}%
	     \frac{16\zeta(3)}{\pi^2}\frac{N_c(N_c-1)}{2}T_c^3.
\ea
\eeq
Also, we take the same limit in \Eqn{rbecs} to find the
Stephan-Boltzmann pressure
\beq
\ba{rcl}
 \delta\Omega=\frac{N_c(N_c-1)}{2}\frac{\pi^2}{45}%
 \left(\frac{4m}{c}\right)^{3/2}d_2^2T_c^4.
\ea
\eeq
Then by differentiating this with respect to $T_c$, we find
\beq
\ba{rcl}
 s&=&\langle s\rangle_{\rm th}=\frac{N_c(N_c-1)}{2}\frac{4\pi^2}{45}%
 \left(\frac{4m}{c}\right)^{3/2}d_2^2T_c^3.
\ea
\eeq
Combining all the above, we obtain the following parametric dependence
of the shear viscosity
\beq
 \eta_s\sim\Sqrt{d_2}\bar{v}\frac{T_c}{\sigma_{sb}}.
\eeq
Then, adjusting the prefactor to the result of the relativistic
Boltzmann equation \cite{BLTZ}, we end up with
\beq
\eta_s=\frac{3}{10\pi}\Sqrt{\frac{c}{4m}}\frac{T_c}{\sigma_{sb}}%
\sim\frac{3}{20\pi}\frac{c^{1/2}d_2^4}{m^{1/2}b_0^2}T_c^3.
\eeq

\begin{figure*}[thp]
 \begin{minipage}{0.49\textwidth}
  \includegraphics[width=0.95\textwidth,clip]{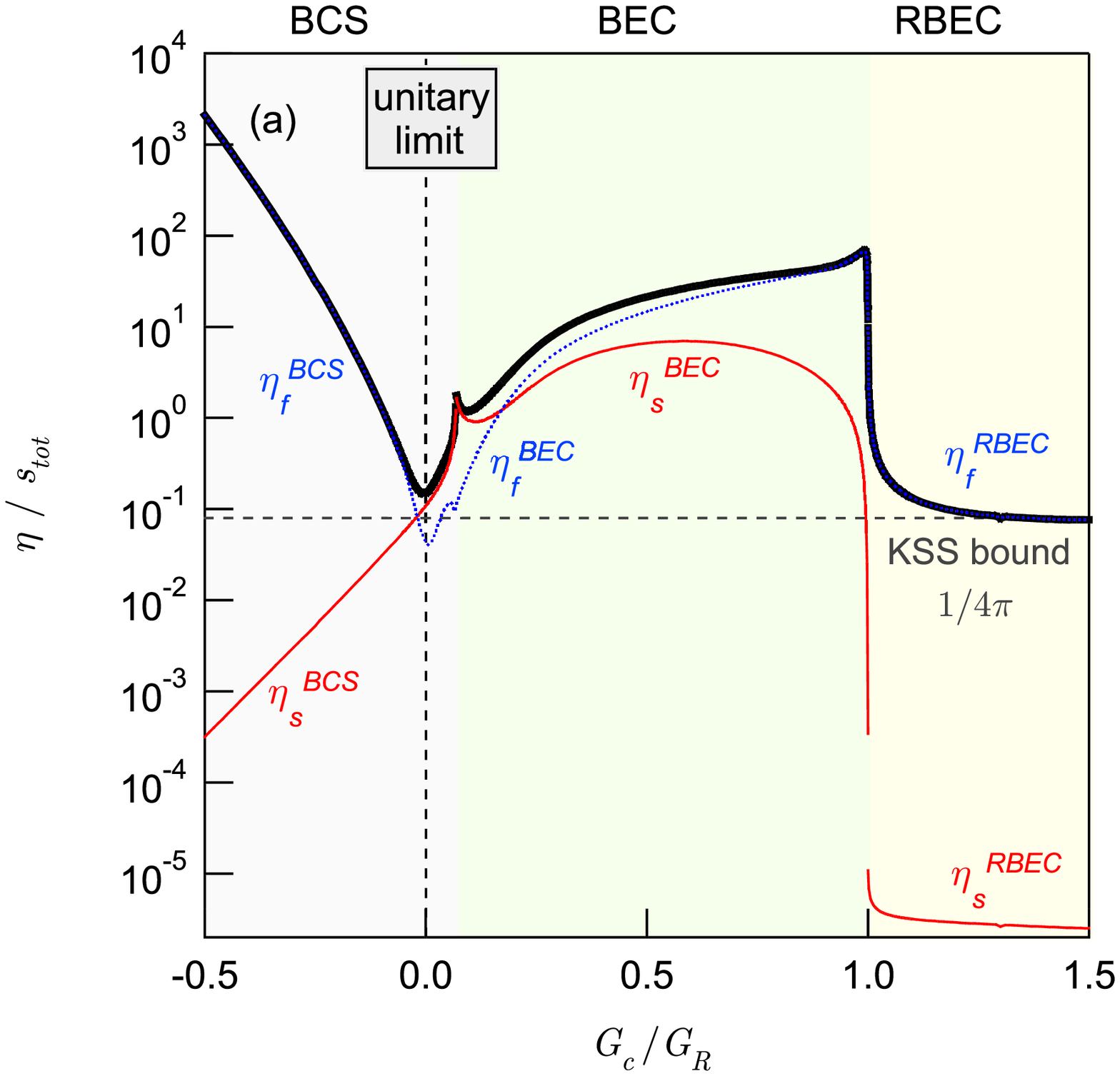}
 \end{minipage}%
 \hfill%
 \begin{minipage}{0.49\textwidth}
  \includegraphics[width=0.95\textwidth,clip]{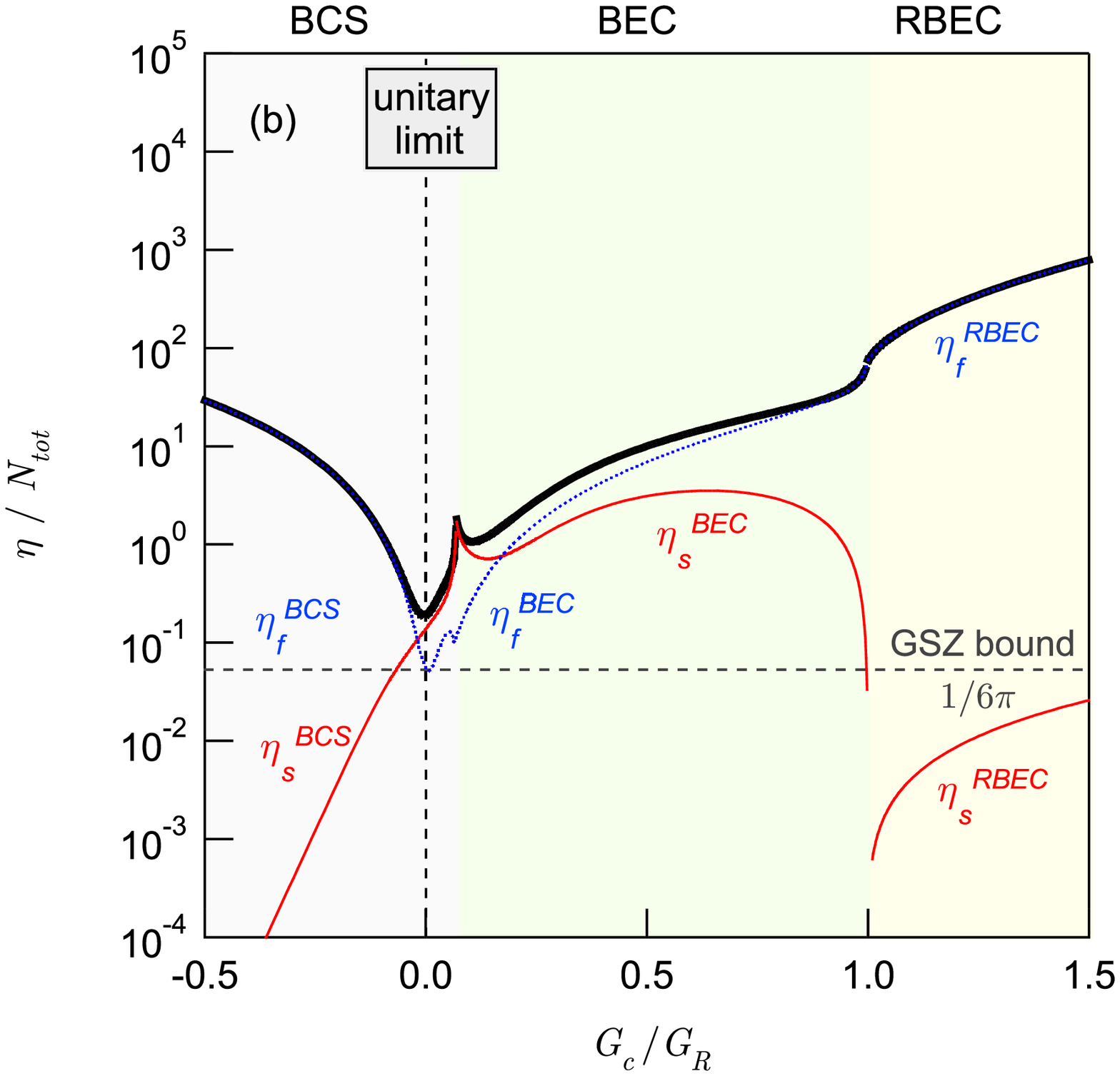}
 \end{minipage}%
  \caption[]{%
  {\bf(a)}~The shear viscosity to entropy ratios from soft modes,
  $\eta_s^{\rm BCS}$, $\eta_s^{\rm BEC}$, and $\eta_s^{\rm RBEC}$ as a
  function of four-fermion coupling $G_R^{-1}$ (thin solid lines).
  The dashed lines indicated by $\eta_f^{\rm BCS}$ $\eta_f^{\rm BEC}$,
  and $\eta_f^{\rm RBEC}$ correspond to the shear viscosities from
  fermions in BCS/BEC and RBEC regimes.
  The total viscosity, the sum of the fermion- and boson- originated
  viscosities, are shown by bold solid line.
  The ``KSS bound'', ${1}/{4\pi}$, is shown by the horizontal dashed
  line. 
  At the unitary limit, the cross section for the low energy $s$-wave
  fermion scattering diverges as $\sim 1/k^2$ with $k$ being the relative
  momentum.
  {\bf (b)}~The shear viscosity to quark number ratios. 
  The dashed line indicated by ``GSZ bound'' is the minimum bound,
  $1/6\pi$,
  proposed in \cite{Gelman:2004fj}.
  }
  \label{viscos}
\end{figure*}

Summarizing our results, shear viscosities arising from soft modes
in the three regimes are
\beq
\ba{rcl}
\eta_s({\rm 
BCS})&=&\dsp\frac{4}{5\pi^2}\frac{m|d|^2k_FT_c}{c\Im d},\\[3ex]
\eta_s({\rm
BEC})&=&\dsp\frac{5\pi^{3/2}}{32\Sqrt{2}}\frac{c^{3/2}d^{5/2}\Sqrt{T_c}}%
{m^{3/2}b_0^2},\\[3ex]
\eta_s({\rm
RBEC})&=&\dsp\frac{3}{20\pi}\frac{c^{1/2}d_2^4}{m^{1/2}b_0^2}T_c^3.\\[2ex]
\ea
\eeq
The system has fermion's degrees of freedom as well as soft modes.
As for the viscosity from fermion-fermion binary collision, we estimate
that in the nonrelativistic region by $\eta_f=\frac{5\pi\bar{k}}{32\Sqrt{2}}%
\frac{1+\bar{k}^2a_s^2}{4\pi a_s^2}$ with $\bar{k}$ being the
average momentum of fermions,
and that in the RBEC regime by $\eta_f=\frac{3}{40\pi^2}\frac{T_c^3}{m^2a_s^2}$.
We also evaluate $\bar{k}$ simply by $\bar{k}=\Sqrt{{\rm
max.}\left(\mu^2-m^2,\frac{mT_c}{2\pi}\right)}$, \ie, we take the larger
of (a) the matter momentum due to the Pauli-blocking, and (b) the
thermal momentum.

In \Fig{viscos}, we have examined the above-derived formulas for 
shear viscosities in three regimes. 
\Fig{viscos}(a) shows the shear viscosity to entropy density ratio
as a function of $G_R^{-1}$.
This quantity is related to the sound attenuation length by
$\Gamma_s=\frac{\frac{4}{3}\eta+\zeta}{s_{\rm tot}T}$
with $\zeta$ being the bulk viscosity.
For a hard sphere particle gas, $\zeta\ll\eta$ both for
nonrelativistic $T\ll m$ and relativistic $T\gg m$ situations 
\cite{Reichl,Chapman,BLTZ}.
For the fluid-dynamic picture to be valid, this length must be
much smaller than the time scale of the fluid-dynamic evolution of
system.
From the figure, we can see that in the weak coupling BCS regime, the
total viscosity is dominated by contribution from the fermion binary
scattering, and that from the soft mode diffusion is negligible %
\footnote{%
We here estimated the total viscosity just by adding up the fermionic
viscosity and the bosonic viscosity. 
In the kinetic theory, the viscosity of the gas mixture takes
much more complicated form even in the first approximation of
the Chapman-Enskog equation where the deviation of
the distribution function is expanded up to the first order in the
Sonine's polynomials. Let us here quote the result of \cite{Chapman}
(see Eq. 12.5.I);
$\eta_{\rm mix}=\frac{\frac{n_1}{n_2}(\frac{2}{3}+A\frac{m_1}{m_2})%
+\frac{n_2}{n_1}(\frac{2}{3}+A\frac{m_2}{m_1})+\frac{E}{2\mu_1}%
+\frac{E}{2\mu_2}+\frac{4}{3}-2A}%
{\frac{1}{\mu_1}\frac{n_1}{n_2}(\frac{2}{3}+A\frac{m_1}{m_2})%
+\frac{1}{\mu_2}\frac{n_2}{n_1}(\frac{2}{3}+A\frac{m_2}{m_1})%
+\frac{E}{2\mu_1\mu_2}+\frac{4A(m_1+m_2)^2}{3Em_1m_2}}$ with
$n_1$ $(n_2)$
being the number density of each particle
in the gas mixture, $\mu_1$ $(\mu_2)$ 
being the viscosity of the individual gas, and 
$m_1$ $(m_2)$
being the mass of each species.
$A$ and $E$
depend on the microscopic detail of interaction between cross species,
and in most cases, $E$ is inversely proportional to $\sigma_{12}^2$
where $\sigma_{12}$ is the collisional cross section between cross species.
Then, if there is no interaction between cross species, $E$ is $\infty$
and the above formula for $\eta_{\rm mix}$ reduces
to the simple sum of each viscosity, \ie, $\lim_{E\to\infty}%
\eta_{\rm mix}=\eta_1+\eta_2$.
In our case, it is simply assumed that there is no interaction between
boson and fermion as a first step.
Of course, further investigations are needed to judge if this treatment
is quantitatively appropriate.
}.
As the coupling $G_R^{-1}$ becomes larger, $\eta_f^{\rm BCS}$
monotonically decreases with the behaviour $\sim a_s^2\sim1/|G_R|^2$;
because the cross section for the fermion binary scattering monotonically
increases and this effect on the shear viscosity overcomes the effect
of the decrease in the average momentum $\bar{k}$.
In contrast, $\eta_s^{\rm BCS}$ monotonically increases because
the relaxation time of the soft mode, $\tau_RT_c$, becomes large.
As the unitary limit $G_R^{-1}=0^-\,(a_s=-\infty)$ is 
approached, $\eta_f^{\rm BCS}$ rapidly decreases and crosses the ``KSS
bound'' ($1/4\pi$), the minimum bound of $\eta/s$ conjectured out by
Kovtun, Son and Starinets \cite{Kovtun:2004de}.
However owing to the viscosity due to soft modes, the sum of the
viscosity ratios, $\eta_s+\eta_f$, does not lower the bound.

When the crossover point $G_c/G_R\sim0.068$ where $\mu_c=m$ is closely
approached, $\eta_s^{\rm BCS}$ rapidly diverges due to the
divergent relaxation time ($\tau_R=\infty$) at $\mu_c=m$.
As noted, however, this is not physical for the following two reasons:
(i) The soft mode with $k$ actually relaxes via the non-linear
interaction caused in the TDGL action. 
(ii) The low energy expansion fails in the vicinity of $\mu_c=m$ 
because of the absence of expansion scales.
In short, the singularity in the crossover point is attributed both
to our gaussian approximation to the soft mode relaxation, and to the
breakdown of the low energy expansion \cite{melo93}.

In summary, we found the following facts.
(1)~Soft mode may play important role to the shear viscosity
   near the unitary regime.
   In fact, the model calculation performed here shows that
   it makes the minimum bound for shear viscosity proposed
   in \cite{Kovtun:2004de} remain valid. 
   On the other hand, it is completely dominated by fermionic
   processes in the weak coupling BCS regime.
(2)~Shear viscosity takes almost minimum near the unitary limit
   in the present model calculation.
   It is close to the minimum bounds \cite{Kovtun:2004de,Gelman:2004fj}.
   It is also noteworthy that the bulk viscosity vanishes in the
   unitary limit \cite{Son:2005tj}.   
   These suggest that the intermediate of the BCS/BEC crossover 
   is rather liquid-like as conjectured in \cite{Gelman:2004fj}.

Finally we must note that our ansatz that the short range quantum
correlation is completely incorporated in the coefficients in 
the low energy soft mode action may not be good approximation in 
the BEC and RBEC regimes.
In fact, the ratio of the thermal de-Broglie length of soft mode to the
average distance between soft modes is not smaller than the unity,
taking around $2\sim4$.
Further theoretical investigations are indeed needed in future to
clarify how secondary quantum corrections affect the shear viscosity.

\section{The density dependence and the quantum fluctuation}\label{densquan}
We here again look at some static aspects of crossover.
In Sec.~\ref{a}, we ask how the characteristics of the critical
temperature as a function of $G_R^{-1}$ depends on quark number density
which we have fixed so far.
We then discuss whether quark matter with the standard choice of diquark
coupling exhibits the crossover.
In Sec.~\ref{b}, we look at how large the effect of quantum fluctuation
in the ``boson sea'', which we have neglected so far.

\subsection{The density dependence; what about the ultra-relativistic
 limit of the crossover?}\label{a}
\begin{figure}[tp]
  \includegraphics[width=0.45\textwidth,clip]{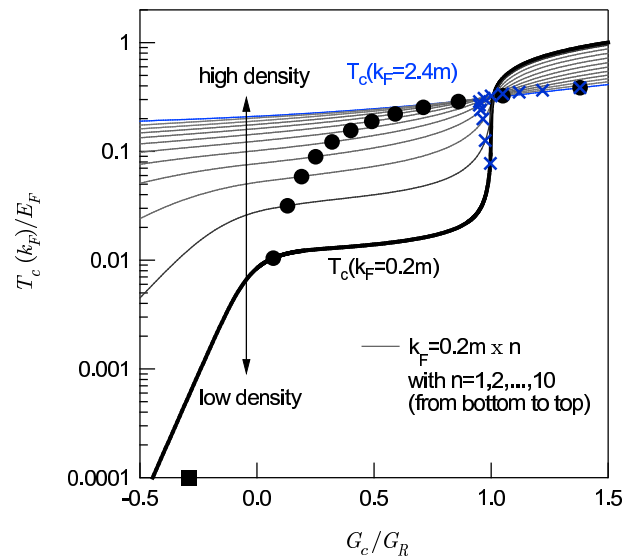}
  \caption[]{%
  The $k_F$-dependence of the critical temperature as a function
  of $G_c/G_R$. From bottom to top, the Fermi momentum increases
  as $k_F=0.2m,0.4m,\cdots,2.4m$.
  The large point located on each critical line represents the 
  BCS/BEC crossover point where $\mu_c=m$ holds, while the large cross
  corresponds to the BEC/RBEC boundary which we estimated by
  $T_c=2\mu_c$.
  The large square put on the horizontal axis indicates the standard
  choice of the diquark attraction, \ie, $\frac{G}{4}\sim\frac{3}{4}G_s$
  with $G_s=2.17/\Lambda^2$ set to reproduce the dynamical quark
  mass $M_q=400\MeV$ in vacuum.
  }
  \label{kFdeps}
\end{figure}
So far we have worked with a fixed Fermi momentum $k_F=0.2m$.
We now study how the BCS/BEC/RBEC crossover will be affected by the
change of quark density.
In \Fig{kFdeps}, we show the $k_F$-dependence of $T_c$ as a function
of the four-Fermi coupling, $G_R^{-1}$.
As the system becomes denser, the critical temperature increases
monotonically \footnote{%
One may think this statement is
incorrect because the critical temperature seems to be shifted downward
in the RBEC regime.
This is not true; the density dependent normalization, $E_F$, 
is making $T_c$ at higher $k_F$
look smaller than those at lower densities.}.
At the same time, the BCS/BEC crossover boundary ($\mu_c=m$) indicated
by the large point shifts to a larger coupling.
As noted before, this is because the Pauli-blocking prevents 
the formation of bound boson in medium.
In addition, the characteristic change in the shape of
($T_c$, $G_R^{-1}$)-relation gets somewhat smeared and seems to vanish for
$k_F\agt m$.
Moreover, a typical diquark coupling strength, $\frac{3}{4}$ of the
scaler coupling $G_s\Lambda^2=2.17$ chosen so that the dynamical quark
mass $400\MeV$ is reproduced at vacuum, is located below the unitary
coupling $G_R^{-1}=0$ (see the large square on the horizontal axis).
From these points, we conclude that quark matter only with light flavors
hardly exhibits the BCS-BEC crossover.
We need a somewhat exotic situation where (i) the in-medium constituent
quark mass is comparable to $\mu$ and (ii) the $(qq)$ attraction between
constituent quarks is stronger than that in the $(q\bar{q})$ channel in
order for the BCS-BEC crossover to be realized in possible quark matter
inside compact stars.

\subsection{How does the quantum fluctuation affect the crossover?}\label{b}
We here examine how large the quantum fluctuation we have ignored so
far can be.
As we discussed in Sec.~\ref{specs}, the gaussian fluctuation consists
of two parts, \ie, the Nozi{\`e}res--Schmit-Rink correction
\beq
\ba{rcl}
 \Omega_{\rm NSR}(\mu,T)&=&-\frac{N_c(N_c-1)}{2}%
     \int_{-\infty}^{\infty}\!\frac{d\omega}{\pi}%
     \frac{d\bfsm{P}}{(2\pi)^3}\\[2ex]
  &&\times\tilde{f}_B(\omega)\delta^{\rm Ren}_{\mu,T}(\omega,\bfm{P}),\\[2ex]
\ea
\eeq
and the ``quantum fluctuation'' which remains finite as $T\to 0$ 
in the presence of finite $\mu$
\beq
\ba{rcl}
 \Omega_{\rm qfl}(\mu,T)&=&%
 -\frac{N_c(N_c-1)}{2}\int_{-\infty}^{\infty}\!\frac{d\omega}%
     {\pi}\frac{d\bfsm{P}}{(2\pi)^3}\\[2ex]
 &&\times\frac{\eps(\omega)}{2}\left[\delta^{\rm Ren}_{\mu,T}(\omega,\bfm{P})%
     -\delta^{\rm Ren}_{0}(\omega,\bfm{P})\right].
\ea
\label{qfl}
\eeq
$\omega$%
-integral in $\Omega_{\rm qfl}$ is finitem, while the remaining momentum
integral is quadratically divergent and thus we need to introduce a
new three momentum cutoff $\Lambda_B$.
In principle, there exists no a priori relation between the
fermionic cutoff $\Lambda$ and the bosonic cutoff $\Lambda_B$.
But it is reasonable that $\Lambda_B<\Lambda$ because the short range
quantum effects are already taken in part to the bosonic effective
lagrangian.
Here we shall try two choices, $\Lambda_B=0.2\Lambda$
and $\Lambda_B=0.3\Lambda$.

Before going into discussion of numerical results,
let us little closely look at the structure of quantum fluctuation
to see its physical meaning.
In order to grab an intuition into this term, we try to estimate the
quantum fluctuation by the bound state approximation for the in-medium
phase shift; that is
\beq
\ba{rcl}
 \delta_{\mu,T}(\omega,\bfm{P})&=&%
 \pi\theta(\omega>E_{B\bfsm{P}}^{\mu,T}-2\mu),\\[2ex]
 &&-\pi\theta(\omega<-E_{\bar{B}\bfsm{P}}^{\mu,T}-2\mu),\\[1ex]
\ea
\eeq
where $E^{\mu,T}_{B\bfsm{P}}$ and $E^{\mu,T}_{\bar{B}\bfsm{P}}$
are the boson and antiboson energy dispersions.
Then substituting this into \Eqn{qfl} yields
\beq
\ba{rcl}
 \Omega_{\rm qfl}(\mu,T)&=&\frac{N_c(N_c-1)}{2}%
 \int^{\Lambda_B}\!\!\!\frac{d\bfsm{P}}{(2\pi)^3}
 \bigg[\frac{E_{{B}\bfsm{P}}^{\mu,T}+E_{\bar{B}\bfsm{P}}^{\mu,T}}{2}
 -E_{B\bfsm{p}}^0\bigg],\\[2ex]
\ea
\label{qfl2}
\eeq
where the three momentum cutoff $\Lambda_B$ is introduced.
Now it is clear that this correction is due to the vacuum fluctuation
(quantum {\em Casimir pressure} for boson).
It differs from fermionic one by minus sign.
For example, the chiral condensation energy at $T=0$ can be written by
the integral of the zero-point energy
shift as
\beq
\ba{rcl}
  \Omega_{\chi\rm SB}&=&-2N_cN_f\int^{\Lambda}%
  \frac{d\bfsm{p}}{(2\pi)^3}%
  \bigg[{E_{M\bfsm{p}}-E_{0\bfsm{p}}}\bigg].\\[2ex]
\ea
\eeq
We conclude that the quantum fluctuation is interpreted as
the ``{\it vacuum}'' energy coming from the in-medium
shift of the boson (antiboson) dispersion.
In the BCS regime, the above argument will be slightly modified
because there is no stable boson in the spectrum.
However, even in the BCS regime, it can be understood in the same way;
it is the vacuum fluctuation coming from the in-medium spectral shift.
 
Differentiating \Eqn{qfl2} with respect to $\mu$ results in the following
expression for the number density from the quantum fluctuation.
\beq
\ba{rcl}
 N_{\rm qfl}&=&-\frac{N_c(N_c-1)}{2}\frac{1}{2}\int^{\Lambda_B}\!\!\!%
 \frac{d\bfsm{P}}{(2\pi)^3}\biggl[%
 \frac{\partial E^{\mu,T}_{B\bfsm{P}}}{\partial\mu}
 +\frac{\partial E^{\mu,T}_{\bar{B}\bfsm{P}}}{\partial\mu}\biggr].\\[2ex]
\ea
\label{qfl3}
\eeq
Let us recall here that in the bound state approximation, the
number density coming from the ordinary thermal fluctuation is given by
\Eqn{effectivenumber} 
where we see that the ``effective charge'' of (anti)boson
deviates from $(\pm)2$.
From these points, we can see that a small 
fraction ${\partial E_{B(\bar{B})\bfsm{P}}^{\mu,T}}/{\partial\mu}$
of quark number charge in the thermal fluctuation {\it escapes} to the
vacuum sector.
In the nonrelativistic situation where $m\gg(|\mu-m|,T_c)$,
we can ignore the antiboson contribution to fluctuations.
In addition, the effective charge shift is of order
$(P^2/M_B^2)(\partial M_B/\partial\mu)$, 
which is negligible by definition of nonrelativity, \ie, $(P^2/2M_B\ll M_B)$.
This argument justifies the ordinary nonrelativistic treatment.
Also at high temperature $T\agt N_{\rm tot}^{1/3}/M_B$, the thermal
fluctuation dominates over the quantum fluctuation because
of the Bose-enhancement in $N_{\rm NSR}$, \ie, $f_B(\omega)\sim T/\omega$
for $\omega\ll T$, and we can safely ignore the quantum fluctuation. 
Thus, we expect that the effect of the quantum fluctuation
becomes less significant with going from the weak coupling BCS regime to
the RBEC regime.

\begin{figure}[tp]
  \includegraphics[width=0.45\textwidth,clip]{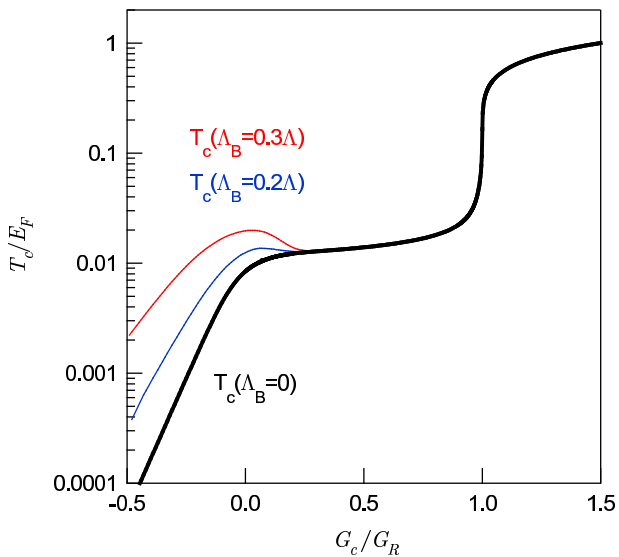}
  \caption[]{%
  The critical temperature versus the four-Fermi coupling.
  The bold curve shows the result without the quantum fluctuation,
  while two thin lines are the results when the quantum fluctuation
  is incorporated with two choices for cutoff, \ie, $\Lambda_B=0.2\Lambda$
  (down) and $\Lambda_B=0.3\Lambda$ (up).
  }
  \label{vac}
\end{figure}

In \Fig{vac}, we show how the $(T_c, G_R^{-1})$-relation is affected by
the incorporation of the quantum fluctuation.
We solved the number equation
\beq
\ba{rcl}
 \dsp N_cN_F\frac{k_F^3}{3\pi^2}&=&\dsp N_{\rm MF}(\mu,T)%
 +N_{\rm NSR}(\mu,T)%
 -\frac{\partial\Omega_{\rm qfl}}{\partial\mu},\\[1ex]
\ea
\eeq
together with the Thouless condition \Eqn{thl} in $(\mu,T)$
to obtain the critical line $(\mu_c(G_R),T_c(G_R))$.
The bold line is the result without the quantum fluctuation,
while two thin lines show the results with two different choices 
for a cutoff $\Lambda_B$ in $\Omega_{\rm qfl}$; the lower (upper) line
corresponds to $\Lambda_B=0.2\Lambda$ ($0.3\Lambda$).
We conclude that the quantum fluctuation greatly affects the critical
temperature in the BCS regime.
This is a feature which is specific in the {\rm relativistic} BCS state
where the condition $m\gg(|\mu-m|,T_c)$ does not hold.
As anticipated, the effect of quantum fluctuation becomes smaller
for the (R)BEC regime where the critical temperature is relatively high 
($T_c\agt N_{\rm tot}^{1/3}/M_B$).

Let us finally explain why the quantum fluctuation {\em increases} 
the critical temperature in the BCS regime.
The quark number from the vacuum fluctuation is {\it negative}
as seen in \Eqn{qfl3}.
Then, to keep the quark number unchanged, the quark number from the mean
field contribution $N_{\rm MF}(\mu,T)$ should increase.
To realize this on the {\it Thouless line} in $(\mu,T)$-plane, 
the line $(\mu,T_c(\mu))$ along which the Thouless criterion is satisfied, 
the temperature should also increase because $T_c(\mu)$ is an 
increasing function as well as $N_{\rm MF}(\mu,T_c(\mu))$.
So we have to go higher $\mu$ (and $T_c(\mu)$) in the $(\mu,T)$%
-plane in order to keep the total quark number unchanged.

\section{Summary and Outlook}\label{concluding}
We have made an extensive analysis on the static and dynamic aspects of
the BCS/BEC crossover in a relativistic superfluid.
We first developed the relativistic formulation of the
Nozi{\`e}res--Schimit-Rink framework for a Dirac fermion interacting
with a four-Fermi point attraction. 
By carrying out the regularization of the dynamic pair susceptibility
using the low energy expansion of the T-matrix in vacuum, we have
performed a systematic extension of the nonrelativistic
Nozi{\`e}res--Schimit-Rink (NSR) scheme \cite{nozieres} to a
relativistic fermion system.
Although we aimed particularly at a possible crossover in quark matter,
our framework is general being not restricted to quark matter,
and may be used for other relativistic fermion systems such as a
possible neutrino superfluidity \cite{Kapusta:2004gi} inside compact
stars at the early stage of their thermal evolution.
We have found that in the relativistic case, there is an additional
source of fluctuation which we have called the ``quantum fluctuation'',
and that our approach is consistent with the traditional
Nozi{\`e}res--Schmit-Rink framework when $(m\gg|\mu-m|,T_c)$
where quantum fluctuation is absent.

We have shown by the numerical calculation that three physically
distinct regimes appear successively when the attraction is increased:
In the weak coupling regime, the system is in the BCS phase where
the critical temperature is exponentially smaller than the Fermi
energy.
When the coupling is increased beyond the unitary limit $a_s=\pm\infty$,
the system gradually goes into the ordinary BEC phase where the increase
of the critical temperature is little suppressed because of the
appearance of bound states in the spectrum;
in this regime, the increase in the attraction results mainly 
in the stabilization of boson.
However, the critical temperature still shows a gentle increase
due to the mass shift of in-medium bosons, which is highly in contrast
to the nonrelativistic case.
When the attractive coupling is increased further beyond the
critical coupling for the Majorana mass formation in vacuum, 
the system goes into the relativistic BEC (RBEC) phase \cite{kapusta}
where the thermodynamics is dominated by antibosons, antifermions as
well as fermions and bosons.
To make physics of these three phases transparent, we discussed how the
spectral function, occupation numbers, Cooper pair size, and entropy
density behave as a function of attraction.

We have also clarified the physical meaning of ``quantum fluctuation''
and have shown numerically that it greatly affects the critical
temperature in the BCS regime, which is also in contrast to the
nonrelativistic BCS phase.

In addition to the study of the static aspects of the crossover, we have
also studied the evolution of soft mode dynamics.
By carrying out the low-energy and long-wavelength expansion
of the dynamic pair susceptibility, we have seen how the effective
theory for the dynamics of fluctuating pair field changes throughout
the crossover.
We saw that, in the weak coupling BCS regime, the dynamics of soft modes
is described by the time-dependent Ginzburg-Landau (TDGL) theory. 
Accordingly, the pair fluctuation is an overdamped mode, which
dissociates into two fermions in the Fermi sea \cite{Abrahams,melo93}.
When the system goes into the BEC regime crossing the unitary limit,
the TDGL theory is taken over by the Gross-Pitaevskii (GP) theory
as in the nonrelativistic calculation \cite{melo93}. 
In this regime, the low energy pair fluctuating field becomes
a propagating mode which cannot decay into fermions because of the
bound state gap.
When the coupling is further increased and the system goes into
the RBEC phase, the relativistic Gross-Pitaevskii (RGP) theory
in turn describes a wide kinematical region. In this region, the
velocity of the fluctuation is large and the repulsive force between
them is also large.

Based on these low energy effective theories, we discussed how the shear
viscosity behaves throughout the BCS/BEC crossover.
In the spirit that the short range quantum correlation is already taken
into the low-energy coefficients in the effective theories, and also
on the basis of the ansatz that soft modes are dilutely distributed in
the phase space,
we estimated the shear viscosity using the {\it classical} Kubo's
formula and the Boltzmann equation.
We have found that the viscosity in the BCS regime
comes from the diffusion process of soft modes, while it comes from the
binary scattering between soft modes in the (R)BEC regime.
Our model calculation indicated that the soft mode contribution is
important near the unitary limit; the shear viscosity to
the entropy density ratio takes the minimum in the unitary regime, but
the minimum bound of the viscosity proposed by Kovtun, Son and Starinets
\cite{Kovtun:2004de} is kept owing to soft modes.

Unfortunately, as the density is increased $(k_F\agt m)$, the
Pauli-blocking effect against the formation of in-medium bosons, 
pushes the BCS/BEC crossover boundary to a larger attraction, 
and makes the crossover characteristics of the critical temperature
somewhat ambiguous.
Also a typical diquark coupling which is usually adopted in the
literature of color superconductivities
\cite{Abuki:2004zk,Ruster:2005jc} is located well below the unitary point.
Based on these points, we have concluded that the following ``exotic''
conditions are required to have the diquark BEC
\cite{Nishida:2005ds,Nawa:2005sb}
in quark matter core in compact stars; (i)~the attraction
between quarks is much larger than expected perturbatively
\cite{Nakamura:2004ur}, and
(ii)~the in-medium fermion mass is larger than the perturbative hard
dense loop calculation \cite{braaten92} and must be comparable to
chemical potential.

Investigating the transport properties for various candidates of the
color superconducting phase of quark matter \cite{Manuel:2004iv} as
well as the normal nuclear matter \cite{Itoh79}, or exploring the
BCS/BEC crossover with paying respect to more realistic conditions of
quark matter remains an important issue.
It may also be interesting to extend our framework so as to take into
account the feedback of soft modes to the fermion propagation in medium,
and to ask how the fermion is modified near the critical temperature
\cite{Kitazawa:2005pp}.
The self-consistent T-matrix theory \cite{Haussmann} or the relativistic
version of the Brueckner-Hartree-Fock theory \cite{BHF} will be one of
the best ways to study the in-medium fermion property affected by soft
modes.

\noindent
\begin{acknowledgments}
This work is motivated in part by my previous study
 \cite{Nishida:2005ds} performed in collaboration with Y.~Nishida to
 whom I am very grateful.
The first part of this paper is greatly indebted to fruitful
 discussions with him.
I have also been greatly benefited from discussions with
 M.~Kitazawa, T.~Kunihiro and Y.~Nemoto, which I acknowledge here.
I am thankful to J.~P.~Blaizot, K.~Itakura, E.~Kolomeitsev,
 G.~Nardulli, K.~Ohnishi, M.~Ruggieri, and K.~Tsumura for helpful
 comments, discussions, and informing me of some references on this
 topic. 
I also thank T.~Hatsuda for useful comments on pair size.
The numerical calculations were carried out on Altix3700 BX2 at YITP
 owned by Kyoto University.
This work was supported in part by a Grant-in-Aid for the 21st Century 
   COE ``Center for Diversity and Universality in Physics''.
\end{acknowledgments}

\vspace*{1em}

\appendix
\section{Imaginary part of the dynamic pair susceptibility}
The imaginary part of the dynamic pair susceptibility can be
calculated as
\beq
\ba{lll}
 &&\Im\chi_{\mu,T}(\omega+i\delta,\bfm{P})\\[2ex]
 &&\quad=\theta\big[z>\Sqrt{4m^2+P^2}\big]f_{qq}^{\mu,T}(z,P)\\[2ex]
 &&\quad\quad-\theta\big[|z|\le P\big]f_{\rm LD}^{\mu,T}(z,P)\\[2ex]
 &&\quad\quad-\theta\big[z<-\Sqrt{4m^2+P^2}\big]f_{\bar{q}\bar{q}}^{\mu,T}(z,P),
\ea
\eeq
where we have defined a variable $z=\omega+2\mu$. 
We can show the identity
$f_{\bar{q}\bar{q}}^{\mu,T}(z,P)=f_{qq}^{-\mu,T}(-z,P)$.
The explicit forms for $f_{qq}^{\mu,T}$ and $f_{\bar{q}\bar{q}}^{\mu,T}$
can be found as
\beq
\ba{rcl}
  f_{qq}^{\mu,T}(z,P)&=&-\frac{\Sqrt{z^2-P^2}}{4\pi}\Sqrt{z^2-P^2-4m^2}\\[2ex]
  &&+\frac{z^2-P^2}{2\pi P}T\ln%
  \left[\frac{1+e^{\frac{\omega}{2T}+\frac{P}{2T}%
  \Sqrt{\frac{z^2-P^2-4m^2}{z^2-P^2}}}}%
  {1+e^{\frac{\omega}{2T}-\frac{P}{2T}%
  \Sqrt{\frac{z^2-P^2-4m^2}{z^2-P^2}}}}\right],\\[3ex]
  f_{\rm LD}^{\mu,T}(z,P)&=&\frac{P^2-z^2}{2\pi P}%
  \omega\\[2ex]
  &&+\frac{P^2-z^2}{2\pi P}T\ln\left[%
  \frac{1+e^{-\frac{\omega}{2T}+\frac{P}{2T}%
  \Sqrt{\frac{z^2-P^2-4m^2}{z^2-P^2}}}}%
  {1+e^{\frac{\omega}{2T}+\frac{P}{2T}%
  \Sqrt{\frac{z^2-P^2-4m^2}{z^2-P^2}}}}\right].\\[3ex]
\ea
\eeq
$f_{\rm LD}$ 
is present only for $T\ne0$, and is called as the Landau damping,
\ie, the Cherenkov radiation (absorption) of soft mode from (to) the
thermally excited quarks.
We can easily check the identity $\Im\chi_{\mu,T}(0,\bfm{P})=0$
which states that the quasi-quark excitation on the Fermi surface is
stable.

\end{document}